\documentclass[12pt]{article}
\usepackage{amsmath}
\usepackage{multirow}
\usepackage{amsfonts}
\usepackage{amssymb,graphics,psfrag}
\usepackage{array,epsfig,multirow,stmaryrd,graphicx}
\usepackage{comment}

\usepackage{hyperref}

\def\hybrid{\topmargin -20pt    \oddsidemargin 0pt
        \headheight 0pt \headsep 0pt
        \textwidth 6.25in      
        \textheight 9 in      
        \marginparwidth .875in
        \parskip 5pt plus 1pt
          \jot = 1.5ex
  }
\hybrid
\numberwithin{equation}{section}
\numberwithin{table}{section}

\newcommand{\beq}{\begin{equation}}
\newcommand{\eeq}{\end{equation}}
\newcommand{\be}{\begin{equation}}
\newcommand{\ee}{\end{equation}}
\newcommand{\bea}{\begin{eqnarray}}
\newcommand{\eea}{\end{eqnarray}}
\newcommand{\ben}{\begin{eqnarray*}}
\newcommand{\een}{\end{eqnarray*}}               
\newcommand{\ba}{\begin{aligned}}
\newcommand{\ea}{\end{aligned}}
\newcommand{\bt}{\begin{tabular}}
\newcommand{\et}{\end{tabular}}
\newcommand{\bc}{\begin{center}}
\newcommand{\ec}{\end{center}}

\newcommand{\cC}{\mathcal{C}}
\newcommand{\cD}{\mathcal{D}}

\newcommand{\cK}{\mathcal{K}}
\newcommand{\cN}{\mathcal{N}}

\newcommand{\cF}{\mathcal{F}}
\newcommand{\cI}{\mathcal{I}}
\newcommand{\cJ}{\mathcal{J}}
\newcommand{\cR}{\mathcal{R}}

\newcommand{\cV}{\mathcal{V}}

\newcommand{\cM}{\mathcal M}

\newcommand{\I}{\text{Im}}
\newcommand{\R}{\text{Re}}

\newcommand{\ib}{{\bar\imath }}
\newcommand{\jb}{{\bar\jmath }}

\newcommand{\bbZ}{\mathbb{Z}}
\newcommand{\bbR}{\mathbb{R}}

\newcommand{\bbP}{\mathbb{P}}

\newcommand{\nn}{\nonumber}

\newcommand{\cref}{{\bf [check ref]}}

\newcommand{\tr}{\mathrm{tr}\, }

\psfrag{n1}{$\nu_1$}
\psfrag{n2}{$\nu_2$}
\psfrag{n1'}{$\nu_1'$}
\psfrag{n2'}{$\nu_2'$}
\psfrag{n9}{$\nu_9$}
\psfrag{n10'}{$\nu_{10}'$}
\psfrag{t1}{$\tilde{\nu}_1$}
\psfrag{t2}{$\tilde{\nu}_2$}
\psfrag{t9}{$\tilde{\nu}_9$}
\psfrag{t1'}{$\tilde{\nu}_1'$}
\psfrag{t2'}{$\tilde{\nu}_2'$}
\psfrag{t10'}{$\tilde{\nu}_{10}'$}

\newcommand{\goa}{\tilde{\mathfrak{a}}^{(0)}  }
\newcommand{\gob}{\tilde{\mathfrak{b}}^{(1)} }
\newcommand{\goc}{\tilde{\mathfrak{c}}^{(0)} }

\def\blfootnote{\xdef\@thefnmark{}\@footnotetext}
\long\def\symbolfootnote[#1]#2{\begingroup%
\def\thefootnote{\fnsymbol{footnote}}\footnote[#1]{#2}\endgroup}

\begin{document}

\baselineskip=15pt

\begin{titlepage}
\begin{flushright}
\parbox[t]{1.08in}{MPP-2011-136}
\end{flushright}

\begin{center}

\vspace*{ 1.2cm}

{\large \bf Six-dimensional (1,0) effective action of F-theory  \\[.1cm] 
             via M-theory on Calabi-Yau threefolds}

\vskip 1.2cm

\begin{center}
 {Federico Bonetti and Thomas W.~Grimm \footnote{bonetti,\ grimm\ \textsf{at}\ mppmu.mpg.de}}
\end{center}
\vskip .2cm

{Max-Planck-Institut f\"ur Physik, \\
F\"ohringer Ring 6, 80805 Munich, Germany} 

 \vspace*{1cm}

\end{center}

\vskip 0.2cm
 
\begin{center} {\bf ABSTRACT } \end{center}

The six-dimensional effective action of
F-theory compactified on a singular elliptically fibred 
Calabi-Yau threefold is determined by using an M-theory lift.
The low-energy data are derived by comparing a circle reduction 
of a general six-dimensional $(1,0)$ gauged supergravity theory with 
the effective action of M-theory on the resolved Calabi-Yau threefold.
The derivation includes six-dimensional tensor multiplets for which the (anti-) self-duality constraints
are imposed on the level of the five-dimensional action. The vector sector 
of the reduced theory is encoded by a non-standard 
potential due to the Green-Schwarz term 
in six dimensions. This Green-Schwarz term also contains higher curvature couplings 
which are considered to establish the full map between anomaly coefficients and geometry.
F-/M-theory duality is exploited by moving to the five-dimensional 
Coulomb branch after circle reduction and integrating out massive vector multiplets 
and matter hypermultiplets. The associated fermions then generate additional 
Chern-Simons couplings at one-loop. Further couplings involving 
the graviphoton are induced by quantum corrections due to 
excited Kaluza-Klein modes. On the M-theory side integrating out massive 
fields corresponds to resolving the singularities of the Calabi-Yau threefold,
and yields intriguing relations between six-dimensional 
anomalies and classical topology. 

\vskip 0.2cm

\hfill {December, 2011}
\end{titlepage}

\tableofcontents

\newpage



\section{Introduction}

The study of effective theories arising in string compactifications is of 
crucial importance both from a conceptional as well as phenomenological point of view.
It is now believed that there is a vast landscape of four-dimensional  
effective theories with minimal or no supersymmetry arising in string theory,
but it is an open problem to systematically characterize these theories \cite{Reports,Denef:2008wq}.
A systematic study becomes more
tractable in compactifications to higher dimensions and with more supersymmetry.
Highly supersymmetric compactifications have a more constrained
effective theory, and arise from restricted classes of candidate
string constructions. In the maximally
supersymmetric case the theory and compactification geometry 
are in fact almost unique.

An intermediate scenario is provided by six-dimensional (6d) $(1,0)$
supergravity theories \cite{Taylor:2011wt}.
While there are constraints both from supersymmetry and
anomalies in this dimension, the moduli space of these theories still
permits a rich
structure and is not fixed by the symmetries. The $(1,0)$ multiplets
in the spectrum are
the gravity multiplet, a number of tensor and vector multiplets, as
well as neutral and matter
hypermultiplets.
A special complication arises from the fact 
that in six dimensions the $(1,0)$ two-form tensors in the tensor multiplets
and the gravity multiplet
obey duality constraints. The two-form in the gravity multiplet 
has a self-dual field strength, while the two-forms in the tensor 
multiplets will admit an anti-self-dual field strength.
This fact  makes it hard to give a Lagrangian formulation for the
dynamics of these forms. While such formulations exist \cite{hep-th/9611100}, 
we will take a different route in this work. Our
6d actions will be formulated as pseudo-actions which yield equations
of motions for the tensor fields which still need to be additionally 
restricted by imposing the self- and anti-self-duality
constraints \cite{hep-th/9703075,Riccioni:1997ik,arXiv:1012.1818,Samtleben:2011fj}. Moreover, our computations will proceed by first
determining a five-dimensional (5d) action for which
these conditions can be consistently imposed on the level of the
action.  We will see that this 
transdimensional treatment is natural in
connecting compactifications of F-theory and M-theory. 
Recently, a transdimensional treatment was suggested 
to study the M5-brane action with self-dual 
non-Abelian two-forms \cite{Douglas:2010iu,Lambert:2010iw,arXiv:1104.4040}. 

In the last years a systematic study of six-dimensional $(1,0)$ supergravity
theories has been undertaken to study the consistency conditions imposed
by quantum gravity~\cite{Taylor:2011wt}. In 6d there are
gravitational, gauge as well as mixed anomalies.
These impose constraints on the number of multiplets, and link the
matter spectrum to the anomaly
coefficients; see e.g.~\cite{Green:1984sg,Sagnotti:1992qw,Schwarz:1995zw}.
A fruitful starting point has
been to ask for a realization of these supergravity theories as a
compactification of F-theory on Calabi-Yau threefolds \cite{Vafa:1996xn,Morrison:1996na,hep-th/9604097,Sadov:1996zm,KumarTaylor,arXiv:1110.5916,arXiv:1111.2351}. These threefolds
have to be elliptically fibred with a base space being a K\"ahler twofold.
At the loci in the base where the elliptic fibre becomes singular, the
dilaton-axion, parameterizing the
complex structure of the elliptic fibre, indicates the existence of
seven-brane sources. These
seven-branes wrap complex curves in the base. Two seven-branes can
intersect at points at which strings ending on different branes
yield new massless matter hypermultiplets in the effective theory.  
This gives the embedding of six-dimensional gauge theories with
matter in a general geometric framework. Note that 
in order to obtain non-Abelian gauge groups the 
elliptic Calabi-Yau threefold $Y_3$ has to be singular itself. 
For Calabi-Yau singularities localized along a single seven-brane 
divisor one can infer the gauge group $G$ at co-dimension one in the base.
The singularity enhancements at co-dimension two in the
base predict the representations of matter fields \cite{Bershadsky:1996nh,hep-th/9606086,arXiv:1106.3563}.

To study the 6d $(1,0)$ effective action arising by compactifications
of F-theory on an elliptically fibred Calabi-Yau threefold, we 
take a detour via M-theory. Our analysis will be analogous to the 4d/3d treatment 
of F-theory on Calabi-Yau fourfolds presented in \cite{arXiv:1008.4133,Grimm:2011fx}, 
but will be more refined and use the enhanced constraints 
of 6d $(1,0)$ supersymmetry and anomalies.
M-theory and F-theory on the same 
Calabi-Yau manifold are connected by a certain limit which 
shrinks the elliptic fibre in M-theory and grows an extra dimension 
required to the match with F-theory. The simplest physical description of 
this limit is provided by considering M-theory on a two-torus. Shrinking the 
size of the torus, one ends up in a Type IIA string set-up on a small circle. 
Performing a T-duality along this circle leads to a Type IIB string 
compactifications on a large circle. Indeed, if the torus shrinks to zero size, 
the Type IIB set-up grows an extra dimension. One can extend this limit 
adiabatically to elliptic fibrations. Furthermore, also branes and flux sources 
can be traced through this duality. We recall more details on this 
duality and the geometry of elliptically fibred Calabi-Yau 
threefolds in section \ref{F-theoryin6d}.

Since we want to determine 
the characteristic data of the 6d F-theory effective 
action, we start with a rather general 6d $(1,0)$ pseudo-action with 
a non-Abelian gauge group $G$, and a generalized Green-Schwarz term
to cancel 6d anomalies \cite{hep-th/9703075, Riccioni:1997ik,arXiv:1012.1818}. The self-duality of the tensors 
is imposed on the level of the equations of motion. We perform 
the Kaluza-Klein reduction on a circle, and derive an actual 
5d effective action for the Kaluza-Klein zero-modes in section \ref{6d5dsection}.
We show how the self-duality can now be imposed  
on the action level, and
determine the characteristic data of the 5d $\cN=2$ theory. 
In particular, we find that the kinetic terms of the 5d 
vectors are encoded by a real function $\cN^{\rm F}$, which 
is homogeneous of degree three. It is interesting to point out 
that it contains a non-polynomial term which is not allowed in 
a standard 5d $\cN=2$ supergravity theory. This correction is 
induced by the fact that the 6d theory contained a classically 
non-gauge invariant Green-Schwarz term to cancel 6d one-loop anomalies. 
Our findings are then interpreted as counterterms in five dimensions, following 
the suggestion of \cite{hep-th/9604097}. In order to prepare the ground for the 
comparison with the M-theory reduction, it will be crucial to comment on 
the modifications when moving to the Coulomb branch of the 5d gauge theory.
Furthermore, also higher curvature terms are required in 6d for anomaly 
cancellation, and we provide a partial dimensional reduction which will 
be compared with the M-theory result.

To determine the 6d characteristic data in terms of the geometric 
data of the compactification threefold, we also determine the 5d  
M-theory effective action in section \ref{Mtheory_on_threefold}. The derivation 
is performed on a fully 
resolved Calabi-Yau threefold $\tilde Y_3$. This implies that the 
5d gauge theory will be in the Coulomb branch, and all M2-brane 
states wrapped on the elliptic fibre and the resolution cycles 
will be massive. The resulting 5d $\cN=2$ action has already been known in the 
literature \cite{Cadavid:1995bk}. Also higher curvature corrections have 
been dimensionally reduced from eleven to five dimensions \cite{Antoniadis:1997eg}. 
It was shown in \cite{Antoniadis:1997eg} that the second Chern class of the 
threefold $\tilde Y_3$ determines 5d higher curvature couplings of the form 
$A \wedge {\rm tr} \cR \wedge \cR$, with $A$ being a 5d vector and $\cR$ 
being the 5d curvature two-form. 

In the comparison of the general 6d/5d reduction with 
the M-theory reduction in section \ref{F-theory-lift},
we argue that the latter does not only contain the classical 
terms but also certain one-loop corrections. 
We identify the F-theory limit of M-theory which leads to 
a perfect match of the classical terms and allows us to 
extract all characteristic data for the 6d $(1,0)$ theory
in terms of the geometry of the resolved Calabi-Yau threefold $\tilde Y_3$.
This includes the geometric data determining the classical metrics on the 
6d vector, tensor and hypermultiplet moduli spaces.
Including a comparison of the dimensionally reduced 
higher curvature terms, we also infer the discrete data 
determining the 6d Green-Schwarz term and hence encode 6d anomalies. 
Our results confirm more indirect arguments 
using the Chern-Simons action of seven-branes \cite{Sadov:1996zm,GrimmTaylor}. Furthermore, 
our results agree with the analysis of 6d anomalies presented in \cite{KumarTaylor}.

Remarkably, we identify several terms in the 5d M-theory reduction on 
$\tilde Y_3$ which do not arise in the classical 6d/5d reduction. We argue using 
\cite{Witten:1996qb,Intriligator:1997pq} that 
this is due to the fact that there are one-loop corrections in the 6d/5d
reduction which arise form two sources: (1) in going to the Coulomb branch, 
there are massive charged hypermultiplets, and massive vector multiplets containing 
the W-bosons which have to be integrated out, (2) in the dimensional reduction there are massive 
Kaluza-Klein modes for all 6d multiplets. In particular, we argue that 
massive fermions running in the loop
generate constant corrections to the 
5d Chern-Simons terms of the from $A \wedge F \wedge F$, with $F$ being the 5d gauge fields.
Integrating out
5d massive Kaluza-Klein modes of 6d chiral fields
also generates 
one-loop Chern-Simons couplings $A^0 \wedge F^0 \wedge F^0$ and $A^0 \wedge \text{tr}(\cR \wedge \cR)$ 
for the 5d vector zero-mode $A^0$ arising from the 
reduction of the 6d gravity multiplet. Both coefficients 
depend on the number of 6d tensor multiplets. The comparison with the 
M-theory result is expected to yield the 6d anomaly conditions.


\section{F-theory in six dimensions} \label{F-theoryin6d}

This section is devoted to a brief account on the F-theory set-up, 
applied to the construction of 6d models. Firstly, we 
recall the basics of F-/M-theory duality and make some comments about the Type IIB picture 
of F-theory vacua. Secondly, we develop a minimal mathematical toolkit to describe 
elliptically fibred Calabi-Yau threefolds. 

\subsection{F-theory via M-theory} \label{FviaM}

F-theory \cite{Vafa:1996xn} is a twelve-dimensional geometric framework introduced to capture some crucial 
non-perturbative aspects of Type IIB vacua in presence of seven-branes. One of the most efficient ways to extract 
information about F-theory vacua is given by duality with M-theory compactifications \cite{Denef:2008wq}. Since we will
follow this strategy throughout the paper, let us briefly review some basic material about F-/M-theory duality 
and its application to the study of 6d vacua.

Consider M-theory on the product manifold
\begin{equation}
 \mathcal{M}_{11} = T^2 \times B_2 \times \mathbb{R}^{1,4}\ ,
\end{equation}
where $T^2$ is a two-torus, $B_2$ is a K\"ahler manifold of complex dimension two, and 
$\mathbb{R}^{1,4}$ is 5d Minkowski spacetime. The metric on the torus can be written as
\begin{equation} \label{torus-metric}
 ds^2_{T^2} = \frac{v^0}{\I\, \tau} \left[ (d x_A + \R\, \tau d x_B)^2 + (\I\, \tau)^2 d x_B^2 \right] \; .
\end{equation}
Here $x_A, x_B$ are real coordinates with period 1, $\tau$ is the complex structure parameter, and $v^0$ is the volume 
of the torus. The canonical coordinates $x_A, x_B$ parameterize the two one-cycles which we 
name  $A$- and $B$-cycle, respectively. 
Upon compactification along a small $A$-cycle, M-theory reduces to Type IIA string theory. An application 
of T-duality acting on the $B$-cycle results in Type IIB string theory on the background
\begin{equation}
 \mathcal{M}_{10} = S^1 \times B_2 \times \mathbb{R}^{1,4}\ ,
\end{equation}
where the circle $S^1$ corresponds to the $B$-cycle. Note that in this duality the 
complex structure parameter $\tau$ in \eqref{torus-metric} is identified with dilaton-axion $\tau = C_0 + i e^{-\phi}$, 
where $C_0$ is the RR scalar and $\phi$ is the dilaton.
In this way, $SL(2,\mathbb{Z})$ modular invariance of Type IIB is interpreted as the $SL(2,\mathbb{Z})$ 
reparameterization symmetry of the complex structure parameter of $T^2$. Furthermore, in the 
limit of vanishing $v^0$, the size of the compact $S^1$ becomes infinite, thus 
leading effectively to Type IIB on 
\begin{equation}
 \mathcal{M}'_{10} = B_2 \times \mathbb{R}^{1,5} \;,
\end{equation}
where $\mathbb{R}^{1,5}$ denotes 6d Minkowski spacetime. The present discussion can be generalized 
to the case in which $\mathcal{M}_{11}$ is a $T^2$-fibration over $B_2 \times \mathbb{R}^{1,4}$, repeating the argument 
fibrewise. We require $T^2$ to depend holomorphically on the complex coordinates of $B_2$. 
More precisely, the 11d background can be written as
\begin{equation}
 \mathcal{M}_{11} = Y_3 \times \mathbb{R}^{1,5}\ ,
\end{equation}
where $Y_3$ is an elliptic fibration with zero-section over the base $B_2$ with fibres being possibly singular 
elliptic curves.  
In order to preserve a fraction of supersymmetry, $Y_3$ must be a Calabi-Yau manifold. 
In summary we are thus led to consider elliptically 
fibred Calabi-Yau threefolds. We introduce some basic facts about their geometry in 
subsection \ref{elliptic_basics}.

Carrying out the duality program outlined above, we end up with a Type IIB vacuum with non-trivial dilaton-axion 
profile $\tau$ varying along $B_2$. As a consequence general F-theory vacua do not admit a perturbative description
in terms of of fundamental strings and D-branes. The fundamental objects of F-theory are 
$(p,q)$-strings and $(p,q)$-branes, which are $SL(2,\bbZ)$-generalizations of the fundamental 
strings and D-branes. A particular role is played by $(p,q)$ seven-branes which magnetically couple 
to $\tau$. This allows to treat them geometrically. In fact, space-time filling seven-branes are located
at co-dimension one loci in $B_2$ at which the elliptic fibre becomes singular. More precisely, a $(p,q)$ seven-brane
can be found at a point where the $(p A+qB)$-cycle collapses. In the following we will include stacks of such 
seven-branes which admit a non-Abelian gauge-theory on their worldvolume. 

Despite the non-perturbative nature of general F-theory vacua, the connection with M-theory will allow us 
to restrict to a low-energy supergravity framework. More precisely, we will compute 
a 6d effective action for F-theory vacua through the following steps:
\begin{enumerate} 
 \item computation of the 5d $\cN=2$  action resulting from Kaluza-Klein reduction on a circle of a general 6d $(1,0)$ supergravity action;
       move to the 5d Coulomb branch of the gauge theory;
 \item computation of the 5d $\cN=2$ effective action of M-theory on a resolved elliptically fibred Calabi-Yau threefold;
 \item comparison between the results and determination of the characteristic data which specify the 6d $(1,0)$ effective action.
\end{enumerate}

In carrying out this program we will restrict to the zero-modes in both the 6d/5d reduction as well 
as in the M-theory reduction. In the 6d/5d-reduction the Kaluza-Klein modes will become 
light in the decompactification limit and restore the dependence of the supergravity fields on all 6d coordinates. 
Moreover, in the M-theory reduction additional M2-brane modes become relevant in the F-theory limit due to the vanishing size 
of the $T^2$-fibre. Both contributions will be neglected when working with 5d massless modes only. 
However, the duality outlined above suggests that the massive 
5d corrections of both sides can also be matched. The crucial
observation which we will use in our work, is that the 
functional dependence of the characteristic data of the supersymmetric actions on the fields should 
be already captured by the zero-modes. This allows us to carry out the above program and indeed determine the 6d effective 
action of an F-theory compactification on a Calabi-Yau threefold. In addition, we find that also certain 
one-loop corrections can be matched under this duality and are crucial to complete the 
picture.

\subsection{Elliptically fibred Calabi-Yau threefolds} \label{elliptic_basics}

As explained in the previous subsection, we want to consider elliptically fibred Calabi-Yau threefolds. 
To this end, it is useful to recall the Weierstrass description of a $T^2$ as a complex curve inside the 
weighted projective space $\mathbb{P}_{2,3,1}$, as discussed e.g.~in \cite{Morrison:1996na,Denef:2008wq}.
In this ambient space the $T^2$ is given by the equation
\begin{equation}
 y^2 = x^3 + f x z^4 + g z^6 \;,
\end{equation}
modulo the identification $(x,y,z) \equiv (\mu^2 x, \mu^3 y, \mu z)$ for all  $\mu \in \mathbb{C} \setminus \{0\}$.
If $f,g$ are complex constants we are describing a specific elliptic curve. 
In order to describe an elliptic fibration over $B_2$, we have to promote $f,g$ 
to sections of the line bundles $-4K, -6K$ respectively, where $K$ denotes the canonical line bundle on $B_2$. 
Locally $f,g$ can be given as polynomials 
in some holomorphic coordinates of $B_2$.

In order to describe seven-branes we have to find the loci where the elliptic fibre degenerates.
This happens at points on $B_2$ where the discriminant 
\begin{equation}
 \Delta = 4f^3 + 27g^2
\end{equation}
vanishes. Let us denote by $[\Delta]$ the two-form cohomology class Poincar\'e dual in $B_2$ to the divisor 
given by $\Delta=0$. It has been shown by Kodaira that this class $[\Delta]$ 
must be related to the canonical class $[K] =- c_1(B_2)$ of the base $B_2$ via
\begin{equation}
 -12[K] = [\Delta]\ ,
\end{equation}
in order for the total space $Y_3$ to be a Calabi-Yau manifold. 
Generally speaking, a singularity in the fibration may or may not yield 
a singularity of the whole Calabi-Yau threefold. 
We are thus led to represent $[\Delta]$ as
\begin{equation} \label{eq:Delta_decomposition}
 [\Delta] = \sum_A \nu_A [S_A] + [\Delta']
\end{equation}
where $[S_A]$ are the Poincar\'e dual two-form classes of the irreducible, effective divisors $S_A$ 
on which the Calabi-Yau threefold develops a singularity, while $[\Delta']$ is the residual 
class associated to singularities of the fibration which leave the total space smooth. 
Singularities of the Calabi-Yau threefold along $S_A$ corresponds to stacks of 
seven-branes on $S_A$ which admit a non-Abelian gauge theory on their world-volume. 
Possible gauge groups can be classified looking at the possible singularities
which occur in $Y_3$ \cite{Morrison:1996na,Bershadsky:1996nh,hep-th/9606086,arXiv:1106.3563}. The 
constants $\nu_A$ are related to group-theoretical invariants.
The divisor $\Delta'$ 
is wrapped by a single seven-brane with no massless gauge bosons on its world-volume.  
 Furthermore, if $[\Delta ']$ and some of the $[S_A]$'s have non-vanishing intersection, 
singularity enhancements take place, which give rise to charged matter in the Type IIB picture.

In order to perform dimensional reduction of M-theory, it is necessary to resolve the 
Calabi-Yau threefold $Y_3$ if it is singular. This amounts to find a smooth Calabi-Yau threefold 
$\tilde{Y}_3$ and a map $\tilde{f} : \tilde{Y}_3 \rightarrow Y_3$
such that singular loci on $Y_3$ are preimages through $\tilde{f}$ of so-called exceptional 
divisors on $\tilde{Y}_3$. This can be done in a canonical way, both if the singularity locus
 is a point and if it is a smooth curve \cite{Morrison:1996na,Bershadsky:1996nh,hep-th/9704097}. 
The resolution procedure can be given the following physical description in the F-/M-theory 
duality picture. The non-Abelian gauge theory with group $G_A$ living on the unresolved stack of 
seven-branes at $[S_A]$ goes to its 5d Coulomb branch in the resolved space, with Abelian gauge group
$U(1)^{\mathrm{rank}(G)}$. Indeed, in M-theory M2-branes wrapping the $\bbP^1$-fibres of the 
exceptional divisors encode the degrees of freedom of 
vectors that are massive in the Coulomb branch and become massless
as the exceptional divisors are shrunk to zero size.

In the remaining part of this section we collect some results about divisors and intersection
numbers of an elliptically fibred Calabi-Yau threefold.\footnote{Full $SU(3)$-holonomy is always understood.} Let us start by 
considering the case of a smooth
threefold $Y_3$. On such a space there is a natural set of divisors which 
span $H_4(Y_3,\bbR)$. Firstly, one has the section of the fibration which is homologous to the 
base $B_2$. Secondly, there is the set of vertical divisors $D_\alpha$ which are 
obtained as $D_\alpha = \pi^{-1}(D_\alpha^{\rm b})$, where $D_\alpha^{\rm b}$ is 
a divisor of $B_2$ and  $\pi$ is the projection to the base $\pi: Y_3 \rightarrow B_2$.
For these smooth elliptic fibrations one has $h^{1,1}(B_2)=h^{1,1}(Y_3)-1$ such divisors.
Let $\omega_0,\omega_\alpha$ be the two-form cohomology classes Poincar\'e dual to $B_2,D_\alpha$.
It is useful to record some facts concerning intersections of divisors for smooth elliptic fibrations. Due to the 
fibration structure one has
\begin{equation}
D_\alpha \cap D_\beta \cap D_\gamma = 0 \;.
\end{equation}
We also introduce the matrix $\eta_{\alpha \beta}$ by defining
\begin{equation}
  \eta_{\alpha \beta}=D^{\rm b}_\alpha \cap D^{\rm b}_\beta  = B_2 \cap D_\alpha \cap D_\beta\ .
\end{equation}
Note that $\eta_{\alpha \beta}$ is a non-degenerate symmetric matrix with mostly minus Lorentzian signature $(1,h^{1,1}(B_2)-1)$.
Finally, let us recall the cohomological identity \footnote{We will be slightly sloppy with the notation in the following, since 
we do not explicitly indicate that certain quantities, e.g.~the first Chern class $c_1(B_2)$, have to be pulled 
back from $B_2$ to the Calabi-Yau threefold.} 
\begin{equation} \label{omega0squaredidentity}
 \omega_0 \wedge \omega_0 + c_1(B_2) \wedge \omega_0 = 0 \;.
\end{equation}
We also introduce the vector $K^\alpha$ by expanding the canonical class $[K]$ in a basis two-forms dual to vertical divisors as
\begin{equation} \label{eq:K_expansion}
 [K] = K^\alpha \omega_\alpha \;.
\end{equation}
Some basic formulas for the base $B_2$ of $Y_3$ will be useful later.
The Euler number $\chi(B_2)$ and the integral of $c^2_1(B_2)$ can be generally evaluated as
\beq \label{chic12_base}
  \chi(B_2) = \int_{B_2} c_2(B_2) = 2 + h^{1,1}(B_2)\ , \qquad \int_{B_2} c^2_1(B_2) = K^\alpha K^\beta \eta_{\alpha \beta} = 10 - h^{1,1}(B_2)\ , 
\eeq
where we have used $c_1(B_2) = - K^\alpha \omega_\alpha$, and the fact that $h^{1,0}(B_2)=h^{2,0}(B_2)=0$
for a base of a Calabi-Yau manifold.

Let us now take into account a singular Calabi-Yau threefold $Y_3$ and its resolution $\tilde Y_3$. For the sake of simplicity, 
we will restrict ourselves to the case of a single seven-brane stack, thus omitting the sum over 
index $A$ in (\ref{eq:Delta_decomposition}), $[\Delta] = \nu [S] + [\Delta ']$. Let $D_i$ be the exceptional divisors 
introduced by resolving the singularity. The index $i$ runs from 1 to $\mathrm{rank}(G)$. The cohomology class 
Poincar\'e dual to $D_i$ is denoted $\omega_i$. Furthermore, let us expand the divisor $S$ wrapped by the stack 
of branes in a basis two-forms dual to vertical divisors as
\begin{equation}
 [S] = C^\alpha \omega_\alpha \;.
\end{equation}
Note that, after resolution, this is replaced by 
\begin{equation}
 [\hat{S}] = C^\alpha \omega_\alpha + a^i \omega_i \;,
\end{equation}
where $a^i$ are the Dynkin numbers characterizing the Dynkin diagram of $G$.\footnote{Note that after 
singularity resolution also (\ref{eq:K_expansion}) is modified by the addition of non-trivial $\omega_i$ terms. 
Nonetheless, these terms do not affect the following discussion on intersection numbers, thanks to identities
(\ref{eq:intersection_identities_2})} Exceptional divisors enjoy the following properties:
\begin{align} \label{eq:intersection_identities_2}
 B_2 \cap D_i &= 0 \nn \\
 D_\alpha \cap D_i \cap D_j &= -C_{ij} \; B_2 \cap D_\alpha \cap S \nn \\
 D_\alpha \cap D_\beta \cap D_i &= 0  \;,
\end{align}
where $C_{ij}$ is the Cartan matrix of the group $G$.

We are now in a position to summarize all intersection numbers on the resolved Calabi-Yau threefold $\tilde Y_3$. We have
found a cohomology basis $\{ \omega_0, \omega_\alpha, \omega_i \}$ which can be denoted collectively as $\{\omega_\Lambda\}$.
Intersection numbers are defined as
\begin{equation} \label{def-intersectionnumbers}
 \mathcal{V}_{\Lambda \Sigma \Theta} = \int_{\tilde Y_3} \omega_\Lambda \wedge \omega_\Sigma \wedge \omega_\Theta \;. 
\end{equation}
Identities and properties listed above imply that intersection numbers must satisfy
\begin{align} \label{ell_intersections}
  \mathcal{V}_{000} &= \eta_{\alpha\beta} K^\alpha K^\beta   &  \mathcal{V}_{0i\Lambda} & = 0 \\
 \mathcal{V}_{00\alpha} &= \eta_{\alpha\beta} K^\beta  &   \mathcal{V}_{\alpha i j} & = - C_{ij} \eta_{\alpha \beta} C^\beta \nn \\
 \mathcal{V}_{0\alpha\beta} &= \eta_{\alpha\beta}  &   \mathcal{V}_{\alpha \beta i} & = 0  \nn \\
 \mathcal{V}_{\alpha \beta \gamma} &= 0 \;, \nn
\end{align}
where $\Lambda = 0,\alpha,j$. As far as $\mathcal{V}_{ijk}$ is concerned, in general it is non-vanishing, 
but otherwise unconstrained by our discussion so far. These intersection numbers arise from 
intersecting exceptional divisors. In fact, as we will discuss below, they will be linked 
to group-theoretical factors depending on the charged matter content of the gauge theory.


\section{Circle compactification from six to five dimensions} \label{6d5dsection}

In this section we discuss the circle reduction of a 
general 6d $(1,0)$ supergravity theory. After reviewing 
some foundational material about 6d supergravities with a simple non-Abelian 
gauge group in subsection \ref{6dgeneralities}, the details of the 
dimensional reduction are presented in subsection \ref{KKreduction} supplemented 
by appendix \ref{appendix_6dto5d}. We emphasize the treatment of 
self-dual two-forms, and describe both the reduction 
of the non-Abelian gauge theory and its broken phase relevant 
in the match with M-theory. The 5d action is brought into canonical $\cN=2$ form in 
subsection \ref{canonica5daction}. We point out an intriguing generalization of the $\cN=2$ formalism
which captures the full reduced action. In subsection \ref{6dhighercurvature} 
certain higher order curvature corrections are reduced which 
carry crucial information about gravitational 6d anomalies.

\subsection{Generalities on 6d (1,0) supergravity} \label{6dgeneralities}

In this subsection we review some basic facts about the spectrum and the dynamics of a 
generic 6d supergravity model with $(1,0)$ supersymmetry, corresponding to 8 real supercharges.  
Massless states in six dimensions are classified by representations of the little group
 $SO(4) \cong SU(2) \times SU(2)$ and are therefore labelled by a couple of integer or half-integer spins, 
$(j_L, j_R)$. Four different kinds of supersymmetric multiplets can be constructed, restricting to spin less or equal to two \cite{Schwarz:1995zw}.
We list them following the chirality conventions which are more common in the 6d supergravity
literature, cf.~e.g.~\cite{hep-th/9703075}: 
\begin{itemize}
\item gravity multiplet: $(1,1) \oplus 2 (\tfrac{1}{2},1) \oplus (1,0)$, i.e.~the graviton, 
one Weyl \footnote{An equivalent formulation  makes use of a $SU(2)$ doublet of Weyl 
left-handed gravitini ($SU(2)$ is the automorphism group of the supersymmetry algebra), 
supplemented by a symplectic Majorana condition. Similar remarks apply to all other fermions. 
This explains why this model is sometimes referred to as $\mathcal{N}=2$ in the literature.} 
left-handed gravitino, one self-dual two-form;
\item vector multiplet: $(\tfrac{1}{2}, \tfrac{1}{2}) \oplus 2(\tfrac{1}{2},0)$, i.e.~one vector 
and one Weyl left-handed gaugino;
\item tensor multiplet: $(0,1) \oplus 2(0,\tfrac{1}{2}) \oplus (0,0)$, i.e.~one anti-self-dual two-form, 
one Weyl right-handed tensorino, one real scalar;
\item hypermultiplet: $2(0,\tfrac{1}{2}) \oplus 4(0,0)$, i.e.~one Weyl right-handed hyperino and two complex scalars.
\end{itemize}
A general model features one gravity multiplet, $n_V$ vector multiplets, $n_H$ hypermultiplets, 
$n_T$ tensor multiplets. It is well known that the (anti-)self-duality condition is incompatible 
with a na\"ive Lagrangian formulation, because the usual kinetic term for two-forms vanishes 
identically once it is taken into account. In the special case $n_T=1$, the anti-self-dual two-form 
from the gravity multiplet and the self-dual two-form from the tensor multiplet can be combined into 
a two-form without any self-duality property, and the standard Lagrangian formulation applies. 
Nonetheless, a set of consistent, supersymmetric, two-derivative, classical equations of motion is 
known for arbitrary $n_T$ \cite{hep-th/9703075}. We can still derive them from variation of a suitable 
functional of the fields (called pseudo-action), provided that the self-duality condition is imposed 
after computation of functional derivatives. In this paper, all 6d actions are to be interpreted 
in this weak sense.\footnote{This formalism is usually applied to Type IIB supergravity in ten 
dimensions to deal with the self-dual four-form in the RR sector.}

We will always restrict ourselves to the bosonic content of the model, and adopt notations 
described below. First of all, we denote all 6d two-forms collectively as $\hat{B}^\alpha$, 
where $\alpha=1,...n_T+1$.\footnote{Later on we will identify $n_T+1=h^{1,1}(B_2)$ in the duality to
M-theory. This provides the match of the indices of the present
section with the ones of section \ref{elliptic_basics}.} The scalars coming from the $n_T$ tensor multiplets parameterize the quotient
\begin{equation}
SO(1,n_T)/SO(n_T) \;.
\end{equation}
It is customary to describe this coset scalar manifold by means of a vielbein formalism. We refer the 
reader to e.g.~\cite{hep-th/9703075} for a detailed account.
For our present discussion we need only to recall that a constant $SO(1,n_T)$ metric
$\Omega_{\alpha\beta}$ is introduced, along with a set of $n_T+1$ scalar fields $j^\alpha$.
The metric $\Omega_{\alpha\beta}$ has mostly minus Lorentzian signature $(1,n_T)$,
and the scalars $j^\alpha$ are subject to the constraint
\begin{equation} \label{j_squared}
  \Omega_{\alpha \beta} {j}^\alpha {j}^\beta = 1 \;.
\end{equation}
Moreover, the scalar manifold is endowed with another non-constant, positive definite metric 
$g_{\alpha\beta}$, which is given in terms of $\Omega_{\alpha\beta}, j^\alpha$ by
\begin{equation} \label{g_equals_2jj}
g_{\alpha\beta} = 2j_\alpha j_\beta - \Omega_{\alpha \beta} \ ,
\end{equation}
where $j_{\alpha} = \Omega_{\alpha \beta} j^\beta$. This metric is needed to write down 
the (anti)-self-duality condition for $\hat{B}^\alpha$ in a $SO(1,n_T)$ covariant way, as
we will see in equation \eqref{6d_self}.

As far as vectors are concerned, in this section we consider a supergravity model with simple gauge group $G$. 
Let $\mathfrak{g}$ be the Lie algebra of $G$. We denote the $\mathfrak{g}$-valued gauge one-form by $\hat{A}$, 
and matrix multiplication will always be understood. 
Moreover, we use anti-Hermitian generators, and the expression for the non-Abelian field strength two-form reads
\begin{equation}
\hat{F} = d\hat{A} + \hat{A}\wedge \hat{A} = d\hat{A} + \tfrac{1}{2} [\hat{A},\hat{A}] \;.
\end{equation}
Let us recall the definition of the Chern-Simons three-form
\begin{equation}
\hat{\omega}^{\mathrm{CS}} = \tr \left( \hat{A} \wedge d\hat{A} + \tfrac{2}{3}\hat{A} \wedge \hat{A} \wedge \hat{A} \right)
\end{equation}
where the trace is taken in a suitable representation of $\mathfrak{g}$. More details
about our normalization for gauge traces can be found in appendix \ref{appendix_anomalies}.
It is also useful to point out two key properties of the Chern-Simons three-form,
\begin{equation}
\delta \hat{\omega}^{\mathrm{CS}} = \tr  d\hat{\lambda} \wedge d\hat{A}\;,  \qquad
d \hat{\omega}^{\mathrm{CS}} = \tr \hat{F} \wedge \hat{F} \;.
\end{equation}

Next, let us make some remarks about the hyper sector.
Each hypermultiplet contains four real scalars, and therefore we use the 
notation $q^U$ ($U=1,...,4 n_H$). These scalar fields can be considered as real 
coordinates for a quaternionic manifold, whose metric we write as $h_{U V}$.
The geometric structures of quaternionic manifolds have been studied intensively, see 
e.g.~\cite{Ferrara:1989ik,Andrianopoli:1996cm}. Since our main focus will be on the 
tensor and vector multiplet structure, we will refrain from giving 
a detailed account of these results here. 
However, in the following we will need to consider some aspects of 
charged hypermultiplets. The only piece of information relevant to our discussion 
is the 6d covariant derivative, which reads schematically  
\begin{equation}
 \hat\cD q^U = d q^U + \hat A^I (T^{\bf R}_{I}q)^U \;,
\end{equation}
where the index $I$ runs over all 
generators of the gauge group $G$, and $T^{\bf R}_I$ are the group generators acting on
the scalars $q^U$ in the representation ${\bf R}$. Several examples of gauged 6d $(1,0)$ supergravities 
are known. We refer the reader to \cite{IC-84-218, hep-th/0504033, hep-th/0508172, hep-th/0512019} and references therein for a detailed account on the subject.

Finally, gravitational degrees of freedom are described by means of
the vielbein formalism.  
The analogue of the one-form gauge connection $\hat{A}$ is provided by the $\mathfrak{so}(1,5)$-valued 
spin connection one-form $\hat{\omega}$, determined by the vielbein through the usual torsionless condition
\begin{equation}
d\hat{e} + \hat{\omega} \wedge \hat{e} = 0 \;,
\end{equation}
where matrix multiplication is understood. If $\hat{\ell}$ is a $\mathfrak{so}(1,5)$-valued zero-form 
which we interpret as infinitesimal parameter of a local Lorentz transformation, we have
\begin{equation}
\delta \hat{\omega} = d\hat{\ell} + [\hat{\omega}, \hat{\ell}] \;.
\end{equation}
The correct covariant field strength is the curvature two-form $\hat{\mathcal{R}}$, which is 
constructed out of the spin connection according to
\begin{equation}
\hat{\mathcal{R}} = d\hat{\omega} + \hat{\omega} \wedge \hat{\omega} \;,
\end{equation} 
and is related to the components of the 6d Riemann tensor $\hat{R}^{\hat\lambda}_{\phantom{A}\hat\tau \hat\mu\hat\nu}$ by
\begin{equation} \label{curvature_form_vs_tensor}
\hat{\mathcal{R}}^{\hat a}_{\phantom{A}\hat b} = \tfrac{1}{2} \hat{e}^{\hat a}_{\hat\lambda} \hat{e}^{\hat\tau}_{\hat b} 
\hat{R}^{\hat\lambda}_{\phantom{A}\hat\tau \hat\mu\hat\nu} \; d\hat x^{\hat\mu} \wedge d\hat x^{\hat\nu} \;, \qquad \hat a,\hat b,=0,...,5\;.
\end{equation}
We also define a gravitational Chern-Simons three-form
\begin{equation}
\hat{\omega}^{\mathrm{CS}}_{\mathrm{grav}} = \tr \left( \hat{\omega} \wedge d\hat{\omega} + 
\tfrac{2}{3}\hat{\omega} \wedge \hat{\omega} \wedge \hat{\omega} \right) \;.
\end{equation}
This definition implies the identities
\begin{equation}
\delta \hat{\omega}^{\mathrm{CS}}_{\mathrm{grav}} = \tr  d\hat{\ell} \wedge d\hat{\omega} \; , \qquad
d \hat{\omega}^{\mathrm{CS}}_{\mathrm{grav}} = \tr \hat{\mathcal{R}} \wedge \hat{\mathcal{R}} \;.
\end{equation}
Note that the right hand side of the last equation is proportional to a characteristic class build from the curvature two-form.
In general, the proportionality constant is fixed by the requirement that suitable integrals of
such classes take integer values. This standard normalization is achieved by inserting a factor of $(2\pi)^{-1}$
for each occurrence of the curvature two-form $\hat \cR$ specified by \eqref{curvature_form_vs_tensor}. In order to improve readability,   
we will never write down these factors of $(2\pi)^{-1}$ in the following.  Similar remarks apply to the 5d curvature two-form introduced in section \ref{KKreduction}.

As we have seen above, the spectrum of a general 6d (1,0) supergravity model contains chiral fermions and (anti)-self-dual two-forms.
As a result, gauge, gravitational, and mixed anomalies may appear once one-loop effects are taken into account. 
Nonetheless, a generalization of 10d Green-Schwarz mechanism, 
due to Sagnotti \cite{Green:1984sg, Sagnotti:1992qw, Sadov:1996zm}, 
can be implemented to generate consistent, anomaly-free theories: it is reviewed concisely 
in appendix \ref{appendix_anomalies}. 
Let us just recall now that, under suitable conditions on the matter content of the model, 
the anomaly polynomial factorizes,
\begin{equation}
\hat{I}_8 = \tfrac{1}{2} \Omega_{\alpha \beta} \hat{X}^\alpha_4 \wedge \hat{X}^\beta_4 \;,
\end{equation}
where
\begin{equation}
\hat{X}^\alpha_4 = \tfrac{1}{2} a^\alpha \tr \hat{\mathcal{R}} \wedge \hat{\mathcal{R}} + 
2 b^\alpha \tr \hat{F} \wedge \hat{F} \;.
\end{equation}
If this is the case, even if $\hat{I}_8$ is non-vanishing, anomalies can be counterbalanced 
by adding the so-called Green-Schwarz term to the action,
\begin{equation} \label{Green-Schwarz-term}
\hat{S}^{\mathrm{GS}} = - \tfrac{1}{2} \int_{\cM_6} \Omega_{\alpha\beta}  \hat{B}^\alpha \wedge \hat{X}^\beta_4 \;.
\end{equation}

In order for this generalized Green-Schwarz mechanism to work, we have to 
assign the following non-trivial transformation rules to the fields of the model:
\begin{align}
\delta \hat{A} &= d\hat{\lambda} + [\hat{A}, \hat{\lambda}] \\
\delta \hat{B}^\alpha &= d\hat \Lambda^\alpha - \tfrac{1}{2} a^\alpha \tr \hat{\ell} d\hat{\omega} 
-2 b^\alpha \tr \hat{\lambda} d\hat{A} \label{6d_B_rule} \;.
\end{align}
In the second equation, $\Lambda^\alpha$ is a collection of one-forms which are the parameters of 
the usual Abelian gauge invariance of two-form potentials. The correct, gauge-invariant field 
strength three-form for $\hat{B}^\alpha$ turns out to be
\begin{equation}
\hat{G}^\alpha = d\hat{B}^\alpha + \tfrac{1}{2} a^\alpha \hat{\omega}^{\mathrm{CS}}_{\mathrm{grav}}
 + 2 b^\alpha  \hat{\omega}^{\mathrm{CS}} \;,
\end{equation}
and satisfies a non-standard Bianchi identity,
\begin{equation} \label{6d_bianchi}
d\hat{G}^\alpha = \hat{X}^\alpha_4 \;.
\end{equation}
The self-duality constraint for the two-forms is written in terms of the three-form field strengths as
\begin{equation} \label{6d_self}
g_{\alpha \beta} \hat{*} \hat{G}^\alpha =   \Omega_{\alpha \beta} \hat{G}^\beta \;,
\end{equation}
where $g_{\alpha\beta}$ is the positive-definite, non-constant metric introduced in \eqref{g_equals_2jj}.

We are now in a position to write down the pseudo-action for 6d $(1,0)$ 
supergravity with simple gauge group $G$. 
Its purely bosonic terms relevant to us are given by 
\begin{align} \label{6d_action}
\hat{S}^{(6)} = \int_{\cM_6} &+ \tfrac{1}{2} \hat{R} \hat{*} 1 
-h_{U V} \hat \cD q^U \wedge * \hat \cD {q}^{V} - \tfrac{1}{4} g_{\alpha \beta}
 \hat{G}^\alpha \wedge \hat{*} \hat{G}^\beta -\tfrac{1}{2} g_{\alpha\beta}
 dj^\alpha \wedge \hat{*} dj^\beta   \nn\\
& 
 - 2 \Omega_{\alpha\beta} {j}^\alpha b^\beta \tr \hat{F} \wedge 
\hat{*} \hat{F} - \tfrac{1}{2} \Omega_{\alpha \beta} \hat{B}^\alpha \wedge \hat{X}^\beta_4 - \hat V \hat{*}1\;.
\end{align}
In the second line, $\hat V$ is a potential generated by gauging the hypermultiplet 
scalars $q^U$. Its explicit form can be found e.g.~in \cite{hep-th/0512019}, but will not 
be crucial for our discussion. Recall that this action has to be supplemented by the duality constraint \eqref{6d_self} imposed on the 
level of the equations of motion. Note that the second-order equation obtained through 
variation of $\hat{B}^\alpha$ is equivalent to the exterior derivative of (\ref{6d_self}) 
thanks to (\ref{6d_bianchi}). This action contains a two-derivative part which 
yields the equations of motion discussed in
\cite{Riccioni:1997ik}. We included in \eqref{6d_action} one additional higher derivative 
term which is the generalized Green-Schwarz term \eqref{Green-Schwarz-term} required 
for 6d anomaly cancellation.

It is appropriate to point out that the Green-Schwarz term is a possible source of non gauge-invariance 
of this classical action. Indeed, one computes
\begin{equation}
\delta \hat{S}^{(6)} =  \tfrac{1}{2}  \int_{\cM_6} \Omega_{\alpha \beta} \left( \tfrac{1}{2} 
a^\alpha \tr \hat{\ell} d\hat{\omega} + 2b^\alpha \tr \hat{\lambda} d\hat{A}  \right) \wedge \hat{X}^\beta_4 \;,
\end{equation}
which in general is not just a surface contribution. It is precisely this failure 
of gauge invariance at tree-level which cancels one-loop anomalies. We summarize the anomaly conditions 
in appendix \ref{appendix_anomalies}. 
For completeness let us point out that there is a simple special case where the action 
is already classically gauge invariant. It is enforced by the conditions 
\beq \label{classical_gaugeinvariance}
\Omega_{\alpha \beta} a^\alpha a^\beta  = 0\ , \qquad
\Omega_{\alpha \beta} a^\alpha b^\beta = 0 \ , \qquad
\Omega_{\alpha \beta} b^\alpha b^\beta =0 \;.
\eeq
These conditions on $a^\alpha,b^\alpha$ can be related to the spectrum of fields, in particular the charge matter content, 
through the anomaly cancellation conditions \eqref{eq:hv}-\eqref{eq:bij-condition} of appendix \ref{appendix_anomalies}. As we argue in section \ref{F-theory-lift}, 
the match between the F-theory set-up and the M-theory compactification is simpler in this special case.

\subsection{Kaluza-Klein reduction on the circle} \label{KKreduction}

Let us now study the supergravity model outlined above on a background with one compact 
spatial dimension, i.e.~with topology $\mathbb{R}^5 \times S^1$. Degrees of freedom along 
the circle can be analysed in terms of their Fourier expansion, giving rise to an infinite 
tower of Kaluza-Klein modes. As discussed above, we restrict ourselves to zero-modes only. 

Our metric Ansatz reads
\beq \label{KK_metric} 
d\hat s^2_{(6)} = \tilde{g}_{\mu\nu} dx^\mu dx^\nu + r^2 Dy^2\ ,\qquad Dy = dy -A^0\ ,
\eeq
where $A^0 = A^0_\mu dx^\mu$, and all 5d field are independent of the coordinate $y$ along $S^1$. A twiddle is used to 
stress that this form of the metric gives rise to a non-canonically normalized action for 
gravity, so that a Weyl rescaling has to be performed. As usual, $A^0$ is a 5d vector 
with Abelian $U(1)$ symmetry $A^0 \rightarrow A^0 + d\chi$ coming from $S^1$ diffeomorphisms 
$y \rightarrow y + \chi$, which leave the derivative $Dy$ invariant. The field strength of $A^0$ reads
\begin{equation}
 F^0 = dA^0 \; .
\end{equation}
It is useful to write down the Kaluza-Klein Ansatz for the metric in the vielbein formalism, too.
Up to local Lorentz transformations, we can take 
\begin{equation} \label{KK_vielbein}
\hat e^a = \tilde e^a_\mu dx^\mu \ ,\qquad \hat e^5 = r\, Dy \ ,
\end{equation}
where $Dy$ is given in \eqref{KK_metric}, and $\tilde e^a_\mu,a=0,\ldots,4$ is the 5d vielbein (independent of $y$) before Weyl rescaling.

Let us now turn to the  one-forms and two-forms, and take into account zero-modes only. In order to get 5d 
fields which are uncharged under the aforementioned $U(1)$ symmetry, we expand all fields on $Dy$ defined in \eqref{KK_metric}. 
To begin with, we set
\begin{equation} \label{KK_vector}
 \hat{A} = A + \zeta\, Dy \;,
\end{equation}
where $\zeta$ is a $\mathfrak{g}$-valued 5d zero-form. The gravitational analogue of this relation 
consists of the expression for the spin connection components, which can be computed from \eqref{KK_vielbein}:
\beq
\hat{\omega}_{ab} = \tilde{\omega}_{ab} + \goa_{ab}  Dy \ ,\qquad \qquad
\hat{\omega}_{a 5} = \gob_a + \goc_a Dy \; ,
\eeq
where $\tilde{\omega}$ is the 5d spin connection determined by $\tilde{e}^a_\mu$. 
The zero-forms $\goa_{ab},\goc_a$, and the one-form $\gob_a $ are given by
\begin{align}
& \goa_{ab} = \tfrac{1}{2} r^2 \tilde{e}^\mu_a \tilde{e}^\nu_b F^0_{\mu\nu}\, , &
& \gob_a =\tfrac{1}{2} r \tilde{e}^\lambda_a F^0_{\lambda\mu}\; dx^\mu\, , &
& \goc_a = -\tilde{e}^\lambda_a \tilde \nabla_\lambda r \;,
\end{align}
where $\tilde \nabla_\lambda$ is the 5d Levi-Civita connection before Weyl rescaling. 

We are now in a position to write down the Kaluza-Klein Ansatz for the two-forms 
$\hat{B}^\alpha$. Care has to be taken because the 6d transformation rule (\ref{6d_B_rule}) 
entangles the degrees of freedom encoded in $\hat{B}^\alpha$ with those of vectors and gravity. 
Thus, we set 
\begin{equation} \label{KK_2form}
 \hat{B}^\alpha = B^\alpha - \left[ A^\alpha -  \tfrac{1}{2} a^\alpha\, \tr (\goa \tilde{\omega})
 - 2 b^\alpha \, \tr (\zeta A)    \right] \wedge Dy\ .
\end{equation}
In this way $A^\alpha, B^\alpha$ have the simplest possible gauge transformations, 
\begin{align}
\delta A^\alpha &= d\mu^\alpha \\
\delta B^\alpha &= d\Lambda^\alpha +\mu^\alpha F^0 -\tfrac{1}{2} a^\alpha\, \tr(\ell d\tilde{\omega})
 -2b^\alpha\, \tr( \lambda dA) \; ,
\end{align}
where the infinitesimal parameters are a $\mathfrak{g}$-valued 5d zero-form $\lambda$, 
a $\mathfrak{so}(1,4)$-valued 5d zero-form $\ell$, 5d zero-, one-forms $\mu^\alpha, \Lambda^\alpha$. 
The first relation implies that $A^\alpha$ has a standard, Abelian field strength 
\begin{equation}
 F^\alpha = dA^\alpha \; .
\end{equation}
However, the na\"ive field strength $dB^\alpha$ is not gauge invariant, and must be improved by setting
\begin{equation} \label{defG5d}
G^\alpha = dB^\alpha - A^\alpha \wedge F^0
 + \tfrac{1}{2} a^\alpha \tilde{\omega}^{\mathrm{CS}}_{\mathrm{grav}} 
 + 2b^\alpha  {\omega}^{\mathrm{CS}} \;,
\end{equation}
where
\begin{align}
\tilde{\omega}^{\mathrm{CS}}_{\mathrm{grav}} &= 
\tr \left( \tilde{\omega} \wedge d\tilde{\omega}
 + \tfrac{2}{3}\tilde{\omega} \wedge \tilde{\omega} \wedge \tilde{\omega} \right)\; ,\\
 {\omega}^{\mathrm{CS}} &= \tr \left( A \wedge dA + \tfrac{2}{3}A \wedge A \wedge A \right)\;.
\end{align}
The corresponding non-standard Bianchi identity reads
\begin{equation} \label{5d_bianchi}
dG^\alpha = -F^\alpha \wedge F^0 + \tfrac{1}{2} a^\alpha \tr \tilde{\mathcal{R}} 
\wedge \tilde{\mathcal{R}} +2 b^\alpha \tr  F \wedge F \;.
\end{equation}

In the rest of this subsection, we will only focus on the two-derivative Lagrangian. 
As a consequence, we drop higher curvature terms from the 6d pseudo-action, and we also 
neglect gravitational contribution to the gauge transformation of $B^\alpha$ and to the field strength $G^\alpha$. A discussion of the higher curvature 
corrections can be found in subsection \ref{6dhighercurvature}.

Dimensional reduction of action (\ref{6d_action}) is performed in appendix \ref{appendix_6dto5d}, 
to which we refer the reader for more details. However, let us just stress here that the 
resulting 5d action is a proper action, without any need for auxiliary self-duality 
conditions. This is possible because the 6d two-forms $\hat B^\alpha$ dimensionally reduce 
to two-forms $B^\alpha$ and vectors $A^\alpha$ as seen in \eqref{KK_2form}.
At the same time, we also have to dimensionally reduce the self-duality constraint (\ref{6d_self}). 
Explicitly we find
\begin{equation} \label{5d_self_text}
r g_{\alpha\beta} \tilde{*} G^\beta = - \Omega_{\alpha\beta} \mathcal{F}^\beta  \;,
\end{equation}
where we have introduced the shorthand notation
\begin{equation} \label{def-cF}
\mathcal{F}^\alpha = F^\alpha -4 b^\alpha \tr( \zeta F) + 2 b^\alpha \tr(\zeta\zeta ) F^0\;.
\end{equation}
The key point is that the 5d duality condition \eqref{5d_self_text} now relates two-forms and 
vectors. Since it does not involve a self-duality, it can be imposed on the level of the action itself.
Hence, in computing the 5d action we proceed in the two steps:
\begin{itemize}
\item[1.] We rewrite the 5d pseudo-action $S^{(5)\rm F}_{\rm pseudo}$ resulting from reduction of 
(\ref{6d_action}) in a form such that $B^\alpha$ only 
appears through its field strength $G^\alpha$. Moreover, $G^\alpha$ can be
treated as an independent variable which enters the action only algebraically.
\item[2.] The 5d pseudo-action $S^{(5)\rm F}_{\rm pseudo}$ can be replaced by an actual action by adding terms
of the schematic form $\Omega_{\alpha\beta} dB^\alpha \wedge F^\beta$  to the 
action to impose the condition \eqref{5d_self_text}. More precisely, the modification is of the form
\begin{align}
\Delta S^{(5)\rm F} = -\int_{\cM_5} \tfrac{1}{2} \Omega_{\alpha\beta} ( G^\alpha + A^\alpha \wedge F^0 - 2 b^\alpha \omega^{\mathrm{CS}}) \wedge F^\beta \;.
\end{align}
The first term proportional to $G^\alpha$ acts as Lagrangian multiplier term to link 
$G^\alpha$ with its dual $\cF^\alpha$. The remaining two terms act as source terms which ensure 
compatibility with the modified Bianchi identity \eqref{5d_bianchi} of $G^\alpha$. 
Including these modifications, both the self-duality constraint 
and the Bianchi identity for $G^\alpha$ follow from the equations of motion.
We are thus able to integrate $G^\alpha$ out and obtain a 5d proper action $S^{(5)\rm F}$, written in 
terms of the vectors $A^\alpha$ only. 
\end{itemize}
The 5d action which results from this algorithm can be found in \eqref{5d_action_KK_nonabelian}.
It is interesting to note that these two steps can be performed even if we reintroduce the gravitational 
part of the generalized Green-Schwarz term, and all gravitational contributions to $G^\alpha$, as
discussed in section \ref{6dhighercurvature}. 

\subsection{Moving to the Coulomb branch} \label{56Coulomb}

In the following sections, we will explore the dynamics of F-theory in six dimensions by 
means of the duality with M-theory on a Calabi-Yau threefold, as introduced in section~\ref{F-theoryin6d}. 
In this framework, we can access
directly only the Coulomb branch of our non-Abelian gauge sector. The full gauge group $G$ 
is spontaneously broken down to $U(1)^{\mathrm{rank}(G)}$, which is spanned by the Cartan generators
$T_i$, $i=1,...,\mathrm{rank}(G)$. We take them 
to be normalized in such a way that
\begin{equation}
\tr (T_i T_j) = C_{ij}
\end{equation}
where $C_{ij}$ is the Cartan matrix of $G$.

The spontaneous break down of gauge symmetry 
is triggered by non-vanishing VEVs of some adjoint scalars $\zeta$ in the vector multiplets.
In particular, inspection of the terms
\begin{equation}
 -2 r^{2/3} \Omega_{\alpha\beta} j^\alpha b^\beta \tr F \wedge {*}F
-2 r^{-2} \Omega_{\alpha\beta} j^\alpha b^\beta \tr D\zeta \wedge {*} D\zeta 
\end{equation}
in the non-Abelian 5d action \eqref{5d_action_KK_nonabelian} shows that the usual
Higgs mechanism originates a mass term for the vectors lying outside of the
Cartan subalgebra. We refer to these massive vectors as W-bosons. Their scalar partners 
acquire a mass, as well. From an effective field theory perspective, 
we are thus left only with the massless fields $A^i, \zeta^i$ associated to the 
Cartan subalgebra of the full gauge algebra. As a result, replacements such as
\begin{gather}
\tr (F \wedge *F) \rightarrow C_{ij} F^i \wedge *F^j \, , \qquad
\tr (D\zeta \wedge * D\zeta) \rightarrow C_{ij} d\zeta^i \wedge * d\zeta^j \nn \\
{\omega}^{\mathrm{CS}}  \rightarrow C_{ij} A^i \wedge F^j \;
\end{gather}
have to be made in \eqref{5d_action_KK_nonabelian} to get the relevant 5d action.

In a similar fashion, charged hypermultiplets acquire a mass through the 5d scalar potential
\begin{equation}
 V = r^{-1} \hat V + r^{-8/3} h_{U V} \zeta^I \zeta^J (T^{\bf R}_I q)^U (T^{\bf R}_J q)^{V} 
\end{equation}
given in the last line of \eqref{5d_action_KK_nonabelian}. Note that the second term originates
directly from dimensional reduction of the 6d kinetic 
term $h_{U V }\hat\cD q^U \wedge \hat * \hat \cD {q}^{V}$. It is quadratic
in the scalars of the charged hypermultiplets and is the source for their masses once gauge symmetry is spontaneously broken. 
Following the effective field theory paradigm, one should integrate out the massive hypermultiplets 
and only keep neutral hypermultiplets in the 5d action
in the Coulomb branch. We use lower-case indices $u,v = 1,...,4 n_H^{\rm neutral}$ to enumerate them.
Hence, we have the replacement rule
\begin{equation}
 h_{U V} \cD q^U \wedge * \cD q^{V} \rightarrow h_{u  v}dq^u \wedge * d q^{v} \; ,
\end{equation}
where $h_{uv}$ is a quantum corrected hypermultiplet metric. Determining 
$h_{uv}$ after integrating out the massive states is in general a complicated task,
but we will later give the M-theory expression for $h_{uv}$ where certain 
corrections have been taken into account implicitly via the geometry.   
In accord with supersymmetry we also drop the scalar potential from the effective 
action for the massless modes.

The interested
reader can find the explicit expression for the effective action in the Coulomb
branch in \eqref{5d_action_KK_coulomb}. However, it is crucial to 
recast this result in a more transparent form in order to implement the F-theory lift discussed in section 
\ref{F-theory-lift}. The aim of the following section is precisely the reformulation of the 5d action
in terms of new variables, in such a way to exploit the underlying supersymmetric structure. Hence, 
we begin our analysis with a concise review of 5d $\cN = 2$ supergravity. 

\subsection{The 5d effective action and its canonical form} \label{canonica5daction}

Let us briefly recall the field content of 5d $\mathcal{N} = 2$ (8 real supercharges) 
supersymmetry multiplets \cite{Ceresole:2000jd}:
\begin{itemize}
\item gravity multiplet: the graviton,  one vector (referred to as `graviphoton'), 
one Dirac \footnote{It is customary to replace one 
Dirac fermion by a $SU(2)$ doublet of Dirac fermions satisfying a symplectic 
Majorana condition. This explains the notation $\cN =2$.} gravitino;
\item vector multiplet: one vector, one scalar, one Dirac gaugino;
\item hypermultiplet: 2 complex scalars, one Dirac hyperino.
\end{itemize}

Let the spectrum consist of the gravity multiplet, $n_V^{(5)}$ vector multiplets, 
$n_H^{(5)}$ hypermultiplets, and let us focus on the bosonic sector. We are not 
going to study gauged supergravity models, and therefore the framework outlined 
in \cite{Gunaydin:1983bi} is general enough for our purposes \footnote{In order 
to compare formulae below with the reference, the reader should be aware that 
we have changed notation, should recall our conventions on Riemann tensor contractions (cf.~appendix \ref{appendix_conventions}), 
and should also note that $C_{\mathcal{I} \mathcal{J} \mathcal{K}}^{\mathrm{there}} 
= \frac{\sqrt{6}}{8} C_{\mathcal{I} \mathcal{J} \mathcal{K}} ^{\mathrm{here}}$.}. 
As usual, each hypermultiplet contributes four real scalars to the spectrum, 
and we will use notation $q^u$ with $u=1,...,4 n_H^{(5)}$. The 
hypersector is entirely specified once a quaternionic structure with metric $h_{u v}$
 is given.  Since the graviphoton and the vectors from the vector multiplets are 
naturally entangled by the dynamics of the theory, let us denote them collectively 
as $A^{\mathcal{I}}$ where $\mathcal{I} = 0, ..., n_V^{(5)}$. The scalars coming 
from the vector multiplets parameterize a $n_V^{(5)}$ manifold which is most 
conveniently described in terms of so-called very special coordinates $M^{\mathcal{I}}$. 
These are $n_V^{(5)} + 1$ real coordinates which describe an auxiliary 
$(n_V^{(5)}+1)$-dimensional manifold in which the actual scalar manifold is embedded 
as an hypersurface, as explained below.

The dynamics of gravity-vector sector at two-derivative level is entirely specified 
once the cubic potential
\begin{equation} \label{canonical-cN}
\mathcal{N} = \tfrac{1}{3!} C_{\mathcal{I} \mathcal{J} \mathcal{K}} M^{\mathcal{I}} M^{\mathcal{J}} M^{\mathcal{K}}
\end{equation}
is given in terms of very special coordinates and of a constant 
symmetric tensor $C_{\mathcal{I} \mathcal{J} \mathcal{K}}$. First of all, 
the scalar manifold is identified with the hypersurface described by the 
so-called very special geometry constraint
\begin{equation} \label{very-special-geometry-constraint}
\mathcal{N} = 1 \;.
\end{equation}
Second of all, the gauge coupling function and the metric on the scalar manifold 
coincide and are constructed out of second derivatives of the cubic potential,
\begin{equation} \label{gauge-coupling-function}
 G_{\cI \cJ} = \left[ -\tfrac{1}{2} \partial_{M^\cI} \partial_{M^\cJ} \log \mathcal{N} \right]_{\mathcal{N}=1} 
= \left[ -\tfrac{1}{2}  \cN_{\cI \cJ}  + \tfrac{1}{2}  \cN_\cI \cN_\cJ \right]_{\cN =1}  \;.
\end{equation}
In this expression, and in the following, downstairs indices $\cI, \cJ, ...$ denote partial
derivative with respect to coordinates $N^\cI, M^\cJ, ...$

Finally, the constant tensor $C_{\mathcal{I} \mathcal{J} \mathcal{K}}$ itself appears 
in the action as Chern-Simons coupling. Indeed, the action is given by
\begin{align} \label{5d_action_canonical}
S^{(5)\rm can} = \int_{\cM_5} & + \tfrac{1}{2} R \,*1
 - \tfrac{1}{2} G_{\mathcal{I} \mathcal{J}} dM^{\mathcal{I}} \wedge * dM^{\mathcal{J}}
-h_{u v} dq^u \wedge * d q^{v} \nn \\
& -\tfrac{1}{2} G_{\mathcal{I} \mathcal{J}} F^{\mathcal{I}} \wedge * F^{\mathcal{J}}
 - \tfrac{1}{12} C_{\mathcal{I} \mathcal{J} \mathcal{K}} A^{\mathcal{I}} \wedge F^{\mathcal{J}} \wedge F^{\mathcal{K}} \;.
\end{align}

Let us now discuss the relation between the spectrum of a 6d supergravity model and the spectrum 
of its Kaluza-Klein reduction on a circle. Suppose the numbers of 6d tensor, vector and hypermultiplets
are $n_T, n_V, n_H$ respectively.  To begin with, we note that the bosonic part of a hypermultiplet 
behaves trivially under dimensional reduction on $S^1$. Hence, we can conclude that 
the number $n_H^{(5)}$ of 5d hypermultiplets is given simply by
\begin{equation} \label{nH}
n_H^{(5)} = n_H^{\rm neutral} \; ,
\end{equation}
where the label `neutral' has been added to remind the reader that charged 6d hypermultiplets
are integrated out and do not appear in the 5d effective theory.

As far as 5d vectors are concerned,  they are generated by three different 
mechanisms. First of all, one vector $A^0$ is introduced by the off-diagonal component 
of the Kaluza-Klein Ansatz for the 6d metric. Second of all, $n_T+1$ vectors $A^\alpha$ 
come from the (anti)-self-dual two-forms in six-dimensions. Finally, reduction 
of 6d vectors gives us $n_V$ additional $A^i$. We thus have a total of $1+(n_T+1)+n_V$ vectors, which we 
denote collectively as $ A^{\mathcal{I}} = (A^0, A^\alpha, A^i) $. They fit into
\begin{equation} \label{nV}
n^{(5)}_V = n_V + n_T +1
\end{equation}
5d vector multiplets, because one linear combination of $\{A^0,A^\alpha\}$ has to be 
identified with the graviphoton and sits in the gravity multiplet\footnote{We include 
$A^\alpha$ because we cannot exclude a contribution from the 6d anti-self-dual two-form 
in the gravity multiplet.}. The corresponding scalar degrees of freedom are provided 
by $j^\alpha, \zeta^i,r$ for a total of $(n_T +1)+n_V +1$ variables. However, they are subject 
to one constraint, which in 6d language is given by (\ref{j_squared}). This counting is consistent
with the existence of very special coordinates $M^\cI = (M^0, M^\alpha, M^i)$ satisfying
\eqref{very-special-geometry-constraint}.

In the remaining part of this section we discuss in which way, and to which extent, the
results of the dimensional reduction performed in \ref{KKreduction} can be expressed in canonical
form \eqref{5d_action_canonical}. 
The first step towards this direction is provided by the correct identification of the very special
coordinates $M^\cI$ on the vector multiplet scalar manifold. It turns out that these new coordinates are defined 
in terms of the old coordinates $(r,j^\alpha, \zeta^i)$ by relations 
\begin{align} \label{coord_shifts}
 M^0 &= r^{-4/3}\nn \;, \\
 M^\alpha &= r^{2/3} \left( j^\alpha + 2 b^\alpha r^{-2} C_{ij} \zeta^i \zeta^j \right) \nn \;, \\
 M^i &= r^{-4/3} \zeta^i \;. 
\end{align}
Next, let us define
\begin{equation} \label{def-cN}
 \mathcal{N}^{\rm F} = \Omega_{\alpha \beta} M^0 M^\alpha M^\beta - 4 \Omega_{\alpha \beta} b^\alpha C_{ij}  M^\beta M^i M^j 
+ 4 \Omega_{\alpha\beta} b^\alpha b^\beta C_{ij} C_{kl} \frac{M^i M^j M^k M^l}{M^0} \;.
\end{equation}
Expressions \eqref{coord_shifts} and \eqref{def-cN} are engineered in such a way that
\begin{equation}
   \mathcal{N}^{\rm F}= \Omega_{\alpha \beta} j^\alpha j^\beta = 1\ 
\end{equation}
holds identically. In particular, note that this identity depends on the non-trivial
interplay of the non-linear $b^\alpha$-shifted redefinition 
of the coordinates $M^\alpha$ \eqref{coord_shifts}
and the fact that there is a non-polynomial term in the 
definition \eqref{def-cN} of $\cN^{\rm F}$, including an inverse power of $M^0$. This non-polynomial
term in $\cN$ is a significant deviation from the canonical case, in which $\cN$ is a cubic polynomial, and 
will be discussed further in the following. However, 
note that $\cN^{\rm F}$ is still a homogeneous function of degree three in the coordinates $M^\cI$. 

Once the new coordinates $M^\cI$ are introduced, the 5d effective action takes the form
\begin{align} \label{5d_action_KK}
 S^{(5)\rm F} = \int_{\cM_5} &+\tfrac{1}{2} R\; *1 -h_{uv} dq^u \wedge * dq^{v} - \tfrac{1}{2} G_{\cI \cJ} dM^\cI \wedge *dM^\cJ \nn \\
&-\tfrac{1}{2} G_{\cI \cJ} F^\cI \wedge * F^\cJ -\tfrac{1}{12} X_{\cI  \cJ \cK } A^\cI \wedge F^\cJ \wedge F^\cK \;.
\end{align}
where the metric $G_{\cI \cJ}$ and the coefficients $X_{\cI  \cJ \cK } = X_{\cI (\cJ \cK) } $ are functions of the scalar fields $M^\cI$. Note that the gauge coupling function and the metric in the kinetic term for scalars $M^\cI$ coincide, as expected for a 5d $\cN=2$ theory. Moreover, both $G_{\cI \cJ}$ and $X_{\cI \cJ \cK}$ are completely determined by the function $\cN^{\rm F}$ introduced above, 
as explained in the following.

As far as the metric $G_{\cI \cJ}$ is concerned, it is given precisely by \eqref{gauge-coupling-function}. 
It is interesting to point out that the non-polynomial term in the definition of $\cN^{\rm F}$
is crucial for \eqref{gauge-coupling-function} to hold for the Kaluza-Klein reduced action.

The Chern-Simons term in \eqref{5d_action_KK},
\begin{equation} \label{CS-from-KK}
  S^{(5)\rm F}_{\mathrm{CS}} =-\tfrac{1}{12} \int_{\cM_5} X_{\cI \cJ \cK} A^\cI \wedge F^\cJ \wedge F^\cK \;,
\end{equation}
deserves more discussion. Its variation under an Abelian gauge transformation $\delta A^\cI = d\lambda ^\cI$
can be written as a boundary term, plus
\begin{equation} \label{CS-variation}
\delta S^{(5)\rm F}_{\mathrm{CS}} = -\tfrac{1}{12}  \int_{\cM_5} \lambda^\cI dX_{\cI  \cJ \cK} \wedge F^\cJ \wedge F^\cK \;.
\end{equation}
For each value of indices $\cI,\cJ,\cK$, two possibilities may occur:
\begin{enumerate}
 \item $X_{\cI\cJ\cK}$ is constant: the corresponding contribution to the Chern-Simons term is gauge invariant in five dimensions;
 \item $X_{\cI \cJ \cK}$ depends non-trivially on the scalars $M^\cI$: the corresponding contribution to the Chern-Simons term breaks 5d gauge invariance explicitly.
\end{enumerate}
Usually, only the first case is encountered in supergravity models. As a consequence, only the totally symmetric
 part of $X_{\cI \cJ \cK}$ effectively enters the action, because we are allowed to integrate by parts and permute indices on 
the vector and the field strengths in \eqref{CS-from-KK}. This symmetry argument breaks down if
 some components of $X_{\cI \cJ\cK}$ are non-constant. In fact, the first slot of this tensor plays a 
distinguished role: exactly those gauge symmetries are broken, whose gauge vector has index $\cI$ such 
that not all components $\{ X_{\cI \cJ \cK} \}_{\cJ,\cK}$
are constant, as can be see from \eqref{CS-variation}.

As mentioned above, all data needed to construct \eqref{CS-from-KK} can be extracted 
 from the function $\cN^{\rm F}$ introduced above. To this end, 
it is useful to note that $\cN^{\rm F}$ naturally splits in a polynomial part $\cN^{\rm F}_{\rm p}$ and a non-polynomial part
$\cN^{\rm F}_{\rm np}$,
\begin{align}
 \cN^{\rm F}_{\rm p} &= \Omega_{\alpha \beta} M^0 M^\alpha M^\beta - 4 \Omega_{\alpha \beta} b^\alpha C_{ij}  M^\beta M^i M^j \nn \\
 \cN^{\rm F}_{\rm np} &= 4 \Omega_{\alpha\beta} b^\alpha b^\beta C_{ij} C_{kl} \frac{M^i M^j M^k M^l}{M^0} \;.
\end{align}
 On the one hand, since $\cN^{\rm F}_{\rm p}$ is a homogeneous polynomial of degree three, its third derivatives with respect to
coordinates $M^\cI$ are constants. In fact, they turn out to be simply related to the coefficients of the gauge invariant part of \eqref{CS-from-KK}.
On the other hand, third derivatives of $\mathcal{N}^{\rm F}_{\rm np}$ are non-constant, and indeed they are proportional 
to the coefficient functions appearing in the gauge-anomalous contributions to \eqref{CS-from-KK}.
More precisely, we have
\begin{equation} \label{CS-12-16}
 S^{(5)\rm F}_{\mathrm{CS}} =  -\tfrac{1}{12} \int_{\cM_5} (\mathcal{N}^{\rm F}_{\rm p})_{\cI \cJ \cK} A^\cI \wedge F^\cJ \wedge F^\cK
-\tfrac{1}{16} \int_{\cM_5} (\mathcal{N}^{\rm F}_{\rm np})_{i \cJ \cK} A^i \wedge F^\cJ \wedge F^\cK \;.
\end{equation}
Two remarks are due at this point. Firstly, observe that the first term fits into the canonical form discussed above, since for 
a cubic polynomial as \eqref{canonical-cN} one has precisely $\mathcal{N}_{\cI \cJ \cK} = C_{\cI \cJ \cK}$. Secondly, note that
 in the second term the first index never takes values $0,\alpha$. This means that the $U(1)$ gauge symmetries 
associated to vectors $A^0,A^\alpha$ are unbroken, while those associated to vectors $A^i$ are broken.

It may be considered questionable, if not inconsistent, to construct a 5d effective action which
fails to be gauge invariant. However, this should not come as a surprise. Our starting point in six dimensions \eqref{6d_action}
is not gauge invariant as well, because of the introduction of the Green-Schwarz terms. As discussed in section \ref{6dgeneralities}, these terms 
are needed in order to implement the anomaly cancellation mechanism: they introduce tree-level gauge violations which counterbalance
one-loop anomalous diagrams generated by the chiral matter content of the theory. As a result, the sum of the tree-level and one-loop contributions to the 6d effective action 
is gauge invariant, while the two summands are not invariant separately. This suggests that a gauge invariant 5d effective action
could be obtained supplementing the computation of this section with the reduction of the one-loop 6d effective action. However, we do not
need to address this ambitious task, since we will show that all relevant data about the effective action of F-theory in six dimensions
can already be extracted from the reduction of the tree-level action only.  

It is worth mentioning a crucial distinction between anomalous terms in six and five dimensions.
It is well known that 5d theories do not develop quantum anomalies. Indeed, possible non-gauge invariant terms
can always be cancelled by adding suitable local counter-terms to the tree level action, in such a way that 
the full effective action at one-loop is gauge-invariant. This kind of anomalies is referred to as `irrelevant'.
The aforementioned counterterms in 5d take the form $\int A \wedge *J$,  
where $A$ is one of the vectors whose gauge invariance is anomalous, and $J$ is a gauge invariant 5d current, such that $*J \propto F \wedge F$. It is precisely the gauge invariance of this current which makes the anomaly irrelevant. 
If we were to implement a similar mechanism to treat 6d anomalies, we would have $*J \propto A \wedge F \wedge F$, which 
is manifestly non gauge invariant. 

From this point of view, the non-gauge invariant Chern-Simons term which appears in \eqref{CS-12-16} has the same form as
the counterterms discussed above. More precisely, the corresponding gauge invariant current reads
\begin{equation}
*J_i = -\tfrac{1}{16} (\cN^{\rm F}_{\rm np})_{i\cJ \cK} F^\cI \wedge F^\cK \; .
\end{equation}
Note that all scalar fields in $(\cN^{\rm F}_{\rm np})_{i\cJ \cK}$ are neutral under the gauge group $U(1)^{\mathrm{rank}(G)}$
after spontaneous symmetry breaking to the Coulomb branch.  

In summary, we are able to cast the Kaluza-Klein reduced action in canonical form, even though some subtle points
 have to be stressed:
\begin{itemize}
\item $\cN$ has to be promoted from a cubic polynomial to a homogeneous function $\cN^{\rm F}$ of degree three; 
the very special geometry constraint $\cN^{\rm F}=1$ and the metric $G_{\cI \cJ}$ are formulated in terms of this non-polynomial $\cN^{\rm F}$;
\item the Chern-Simons term coming from Kaluza-Klein reduction and the Chern-Simons term obtained through the
canonical prescription $C_{\cI \cJ \cK} = (\cN^{\rm F})_{\cI \cJ \cK}$ share the same gauge-invariant part,
and differ only for non gauge-invariant terms; these can be interpreted as local counterterms 
which make 5d anomalies irrelevant.  
\end{itemize}   
Since counterterms are completely specified by the classical data of the model, all information about the 
effective 5d action is encoded in the polynomial part of $\cN^{\rm F}$ and the corresponding gauge-invariant  Chern-Simons terms.

\subsection{Higher order curvature corrections} \label{6dhighercurvature}

As we have seen in subsection \ref{6dgeneralities}, anomaly cancellation requires 
the introduction of a higher curvature term in the 6d action,
\begin{equation}
\hat{S}^{(6)}_{\cR^2} = -\tfrac{1}{4}\int_{\cM_6}  \Omega_{\alpha \beta} a^\alpha \hat{B}^\beta \wedge
 \tr \hat{\mathcal{R}} \wedge \hat{\mathcal{R}} \; .
\end{equation}
Furthermore, local Lorentz transformations act non-trivially on the two-forms 
$\hat{B}^\alpha$, in such a way that the corresponding field strength $\hat{G}^\alpha$ 
receives a gravitational contribution. Even if we are not going to perform the 
dimensional reduction of the complete, higher-derivative action, we can make general 
remarks about some interesting feature of the resulting 5d action.

First of all, as stated in subsection \ref{KKreduction}, inclusion 
of gravitational contributions does not interfere with the possibility to get rid 
of 5d two-forms $B^\alpha$ in favour of vectors $A^\alpha$. Indeed, 
gravitational terms modify the action in such a way that $F^\beta$ in 
\begin{equation}
\Delta S^{(5)\rm F} = - \tfrac{1}{2}\int_{\cM_5} \Omega_{\alpha\beta} dB^\alpha \wedge F^\beta
\end{equation}
is replaced by a more complicated expression, which is nonetheless exact. 
$\Delta S^{(5)\rm F}$ is still a total derivative, and the elimination of 
$B^\alpha$ can proceed along the same line as in the two-derivative case.

Secondly, it can be verified that all possible non-gauge invariant terms in the final 
5d action are proportional to
\begin{equation}
 \Omega_{\alpha \beta} a^\alpha a^\beta \qquad \mathrm{or} \qquad  
\Omega_{\alpha \beta} a^\alpha b^\beta  \qquad \mathrm{or} \qquad 
\Omega_{\alpha \beta} b^\alpha b^\beta \; .
\end{equation}
This observation will be relevant for the discussion of F-theory lift, in section \ref{F-theory-lift}.

Finally, let us present one particular higher curvature contribution to the 5d action, 
which will play a prominent role in the matching with M-theory on a Calabi-Yau threefold. 
It is the $A\mathcal{R} \mathcal{R}$ term coming from dimensional reduction of the 
$\hat{B} \hat{\mathcal{R}} \hat{\mathcal{R}}$ 6d term written above. In order to 
extract this term from the total 5d action, we can effectively set $A^0$ to zero and 
treat $r$ as a constant:\footnote{The Weyl rescaling 
$\tilde{g}_{\mu\nu} = r^{-2/3} g_{\mu\nu}$ has no effect on the leading, moduli-independent terms 
in the expression of the curvature two-form.}
\begin{align*}
 \hat{\mathcal{R}}_{ab} &= {\mathcal{R}}_{ab} + \dots\\
 \hat{\mathcal{R}}_{a5} &= 0 + \dots \; ,
\end{align*}
where $a,b,=0,\dots,4$ are 5d flat spacetime indices, and `5' refers to the compact direction.
As a consequence, we have
\begin{equation}
 \tr \hat{\mathcal{R}} \wedge  \hat{\mathcal{R}} = \tr \mathcal{R} \wedge \mathcal{R} + \dots \; .
\end{equation}
A first contribution to the term we are looking for is then given by
\begin{equation} \label{final6dcorrection}
  \tfrac{1}{4} \int_{\cM_5}  \Omega_{\alpha \beta} a^\alpha A^\beta \wedge 
\tr \mathcal{R} \wedge \mathcal{R} \; ,
\end{equation}
in which the change of sign comes from the Ansatz (\ref{KK_2form}). Note however that an addition
contribution arises when $\Delta S^{(5)\rm F}$ is added in order to eliminate tensors from the 5d action,
as can be seen recalling the definition of $G^\alpha$ \eqref{defG5d}:
\begin{equation}
 - \tfrac{1}{2}\int_{\cM_5} \Omega_{\alpha\beta} dB^\alpha \wedge F^\beta \supset + \tfrac{1}{4} \int_{\cM_5} \Omega_{\alpha\beta} a^\alpha \omega_{\rm CS}^{\rm grav} \wedge F^\beta =  \tfrac{1}{4} \int_{\cM_5}  \Omega_{\alpha \beta} a^\alpha A^\beta \wedge 
\tr \mathcal{R} \wedge \mathcal{R} \; .
\end{equation}
In summary, we find the 5d higher curvature term
\begin{equation}
  S^{(5)\rm F}_{A \cR\cR} =\tfrac{1}{2} \int_{\cM_5}  \Omega_{\alpha \beta} a^\alpha A^\beta \wedge 
\tr \mathcal{R} \wedge \mathcal{R} \; .
\end{equation}

We conclude this subsection describing the effect of higher curvature terms on the 
canonical form of 5d supergravity. As done in \cite{Hanaki:2006pj}, superconformal 
techniques can be used to construct the 5d supersymmetric completion of the 
$A\mathcal{R}\mathcal{R}$ term. In this formalism, the supersymmetry algebra closes 
off-shell, at the expense of introducing auxiliary fields in the gravity, vector and 
hypermultiplets. The scalar manifold associated to vector multiplets is still 
described by constrained coordinates $M^\mathcal{I}$. However, the constraint is no longer 
\begin{equation}
 \tfrac{1}{3!} C_{\mathcal{I} \mathcal{J} \mathcal{K}} M^{\mathcal{I}} M^{\mathcal{J}} M^{\mathcal{K}} = 1
\end{equation}
but gets corrected by terms proportional to the constants $c_{2\mathcal{I}}$ appearing 
in front of $A^\mathcal{I} \wedge \tr \mathcal{R} \wedge \mathcal{R}$ in the higher 
derivative Lagrangian \cite{Cremonini:2008tw}:
\begin{equation}
  \tfrac{1}{3!} C_{\mathcal{I} \mathcal{J} \mathcal{K}}
 M^{\mathcal{I}} M^{\mathcal{J}} M^{\mathcal{K}} =
 1 - \tfrac{1}{72} c_{2\mathcal{I}} \left( D M^\mathcal{I} + v^{\mu\nu} F^\mathcal{I}_{\mu\nu} \right) \; ,
\end{equation}
where $D, v_{\mu\nu}$ are the auxiliary bosonic fields in the gravity multiplet. 
It is possible to integrate them out iteratively in a small $c_{2\mathcal{I}}$ expansion;
 the result reads schematically $C M^3 = 1 + c F^2$.


\section{M-theory on a Calabi-Yau threefold} \label{Mtheory_on_threefold}

In this section we recall the dimensional reduction of M-theory 
on a Calabi-Yau threefold and adapt it to the case of elliptic 
fibrations with resolved singularities. The basics of the Kaluza-Klein 
reduction restricted to the zero-modes are presented in subsection \ref{Mreduction}, 
with more details summarized in appendix \ref{appendix_11d5dreduction}. We also discuss the 
the specification to a resolved elliptically fibred Calabi-Yau threefold. 
In subsection \ref{11dhighercurvature} we perform the dimensional reduction of a higher curvature 
correction to 11d supergravity focusing on the terms needed in the matching with 
the 5d higher curvature terms of section \ref{6dhighercurvature}.

\subsection{M-theory action on an elliptic Calabi-Yau threefold} \label{Mreduction}

In this subsection, we start with the unique two-derivative action for 11d supergravity \cite{Cremmer:1978km}, whose 
 purely bosonic part is
\begin{equation} \label{Mtheorytree}
\hat{S}^{(11)} =\int_{\cM_{11}} \tfrac{1}{2} \hat{R} \hat{*} 1 
- \tfrac{1}{4} \hat{F}_4 \wedge \hat{*} \hat{F}_4 
- \tfrac{1}{12} \hat{C}_3 \wedge \hat{F}_4 \wedge \hat{F}_4
\end{equation}
where $\hat{F}_4 = d\hat{C}_3$. In this subsection a hat will always denote 11d fields. 
Supergravity compactified from eleven to five dimensions is discussed 
e.g.~in \cite{Cadavid:1995bk}. 

Following the standard recipe for dimensional reduction on a Calabi-Yau threefold $Y_3$, we expand 
11d fields on a basis of zero-modes of the appropriate differential operator on the internal 
manifold. We refer the reader to appendix \ref{appendix_calabiyau} for an account on our 
notations for Calabi-Yau threefolds. The background metric has a line element
\begin{equation} \label{backgroundmetric}
ds_{11}^2 = \tilde{g}_{\mu\nu}(x) dx^\mu dx^\nu + 2 g_{\ib j}(y) d\bar y^\ib dy^j
\end{equation}
where the external metric $\tilde{g}_{\mu\nu}$ describes a maximally symmetric spacetime (Minkowski, dS, AdS) 
and a twiddle reminds us that a Weyl rescaling will be performed later. Fluctuations of 
the internal metric $g_{\ib j}$ are zero-modes of the Lichnerowicz operator and are expanded onto 
the $(1,1)$ and $(1,2)$ cohomologies,
\beq \label{deltag}
  \delta g_{i \jb} = -i  (\omega_\Lambda)_{i\jb} \, \delta v^\Lambda \ ,\quad\qquad
  \delta g_{ij} = (\bar{b}_{\bar{\kappa}})_{ij} \, \delta \bar{z}^{\bar{\kappa}}\ ,
\eeq
in which
\begin{equation}
(\bar{b}_{\bar{\kappa}})_{ij} = \tfrac{i}{\| \Omega \|^2} (\bar{\chi}_{\bar{\kappa}})_{i \bar{k}\bar{l}}
 \Omega^{\bar{k}\bar{l}}_{\phantom{kl}j}\ ,
\end{equation}
where $\Omega$ is the holomorphic $(3,0)$-form, and $\chi_{\kappa}$ is a basis of $(2,1)$-forms on $Y_3$.
The variations \eqref{deltag} are parameterized by the complex structure moduli $z^\kappa$, and the  
K\"{a}hler moduli $v^\Lambda$ which are obtained
in the expansion of the K\"ahler form 
\beq \label{Kahlerexpand}
J = v^\Lambda \omega_\Lambda\ . 
\eeq
The excitations of the three-form $\hat C_3$ are zero-modes of the internal Laplacian, and are therefore 
expanded on a cohomology basis of the internal manifold,
\begin{equation} \label{3formAnsatz}
  \hat{C}_3 = \xi^K \alpha_K - \tilde{\xi}_K \beta^K + A^\Lambda \wedge \omega_\Lambda + C_3 \;,
\end{equation}
where $(\alpha_K,\beta^K)$ is a real symplectic basis of the middle cohomology of $Y_3$. The 
zero-modes $(\xi^K,\tilde \xi_K)$ are scalars, $A^\Lambda$ are vectors, and $C_3$ is a three-form in five dimensions.

Let us now discuss how these fields fit into 5d $\mathcal{N}=2$ supersymmetry multiplets. 
As a preliminary remark, recall that in five dimensions three-forms can be dualized into 
scalars, so that we are allowed to trade $C_3$ for a real scalar field $\Phi$. The gravity 
multiplet consists of $\tilde{g}_{\mu\nu}$ and of one (linear combination) of the $A^\Lambda$ 
vectors. The remaining vectors fit into
\begin{equation}
n_V^{(5)} = h^{1,1}(Y_3) -1
\end{equation}
vector multiplets, along with the K\"{a}hler moduli $v^\Lambda$. It seems like there is a 
mismatch of degrees of freedom, since we have $h^{1,1}(Y_3)$ scalars. This seeming difficulty 
is overcome by the following observation.  We introduce the total volume of the Calabi-Yau threefold as 
\beq 
  \mathcal{V}=\frac{1}{3!} \int_{Y_3} J \wedge J\wedge J = \frac{1}{3!}  \cV_{\Lambda \Sigma \Theta} v^\Lambda v^\Sigma v^\Theta\ , 
\eeq 
where $\cV_{\Lambda \Sigma \Theta}$  are the intersection numbers of the 
Calabi-Yau threefold introduced in \eqref{def-intersectionnumbers}. Then, $\cV$ actually sits in the universal hypermultiplet,  
leaving effectively $h^{1,1}(Y_3)-1$ scalar degrees of freedom in the vector sector. 

To discuss hypermultiplets we need to recall the decomposition of the third cohomology 
into complex cohomologies,
\begin{equation}
H^3(Y_3) = \left[ H^{1,2}(Y_3) \oplus H^{2,1}(Y_3) \right] \oplus \left[ H^{0,3}(Y_3) \oplus H^{3,0}(Y_3) \right] \;.
\end{equation}
Real scalars $\xi^K, \tilde{\xi}_K$ provide $h^{1,2}(Y_3)+1$ complex degrees of freedom: 
$h^{1,2}(Y_3)$ of these correspond to the $ H^{1,2}(Y_3) \oplus H^{2,1}(Y_3)$ component and 
combine with the complex structure moduli $z^\kappa$ to give $h^{1,2}(Y_3)$ hypermultiplets; the 
remaining complex degree of freedom lives in $H^{0,3}(Y_3) \oplus H^{3,0}(Y_3)$ and combines 
with $\mathcal{V}, \Phi$ in the universal hypermultiplet. In conclusion, we have found
\begin{equation}
n_H^{(5)} = h^{1,2}(Y_3) + 1
\end{equation}
hypermultiplets, which will be collectively denoted by $q^u$.

The dimensional reduction is carried out in detail in appendix \ref{appendix_11d5dreduction}. 
Since the overall volume sits in the universal hypermultiplet it is natural to define scalar fields
\begin{equation}
L^\Lambda =\mathcal{V}^{-\tfrac{1}{3}}  v^\Lambda \; ,
\end{equation}
which are the real scalars in the vector multiplets. They only parameterize $h^{1,1}(Y_3) -1$
degrees of freedom, since due to their definition they are subject to the constraint
\begin{equation} \label{veryspecial}
\tfrac{1}{3!} \mathcal{V}_{\Lambda\Sigma\Theta} L^\Lambda L^\Sigma L^\Theta = 1\ .
\end{equation}
We are naturally led to interpret $L^\Lambda$ as 5d very special 
coordinates, in term of which the cubic potential reads
\begin{equation}
\mathcal{N} = \tfrac{1}{3!} \mathcal{V}_{\Lambda\Sigma\Theta} L^\Lambda L^\Sigma L^\Theta \;.
\end{equation}
Some additional details of this 5d formalism have been already given in section \ref{canonica5daction}.

Once the cubic potential $\mathcal{N}$ is known, the only missing ingredient to specify 
the model is the quaternionic metric $h_{uv}$ on the hypermultiplet scalar manifold: 
its expression in terms of $\mathcal{V},\Phi,\xi^K, \tilde{\xi}_K, z^\kappa$ can be found in 
appendix \ref{appendix_11d5dreduction}. In summary, the reduced bosonic action reads
\begin{align} \label{5d-action-M}
S^{(5)\rm M} = \int_{\cM_5} & + \tfrac{1}{2} R*1 
- \tfrac{1}{2} G_{\Lambda\Sigma} dL^\Lambda \wedge * dL^\Sigma 
-h_{uv} dq^u \wedge * dq^v \\
& - \tfrac{1}{2} G_{\Lambda \Sigma} F^\Lambda \wedge *F^\Sigma 
- \tfrac{1}{12} \mathcal{V}_{\Lambda \Sigma \Theta} A^\Lambda \wedge F^\Sigma \wedge F^\Theta \;, \nn
\end{align}
where, as expected,
\begin{equation}
G_{\Lambda \Sigma} =\left[ -\tfrac{1}{2} \partial_{L^\Lambda} \partial_{L^\Sigma} \log \mathcal{N}  \right]_{\mathcal{N}=1} \;.
\end{equation}
When restricted to elliptic fibrations as discussed next, it will be this form 
of the 5d action which can be matched to the circle reduced action of section \ref{canonica5daction}. 

Let us now specify this result to the elliptically 
fibred geometry introduced in subsection \ref{elliptic_basics}. We first split the 
index $\Lambda$ into $(0,\alpha,i)$ and write
\beq
L^\Lambda = (R, L^\alpha, \xi^i) \ , \qquad \quad A^\Lambda = (A^0, A^\alpha, A^i) \;.
\eeq
Combining this notation with the intersection numbers \eqref{ell_intersections} of an elliptic fibration we get
\begin{align} \label{N-beforeshift}
\mathcal{N} = & \tfrac{1}{2} \eta_{\alpha\beta} R L^\alpha L^\beta + \tfrac{1}{2}\eta_{\alpha\beta} K^\alpha R^2 L^\beta 
+ \tfrac{1}{6} \eta_{\alpha\beta} K^\alpha K^\beta R^3 \nn \\
&- \tfrac{1}{2} \eta_{\alpha\beta} C^\alpha C_{ij} L^\beta \xi^i \xi^j
 + \tfrac{1}{6} \mathcal{V}_{ijk} \xi^i \xi^j \xi^k \;.
\end{align}

As we will discuss in section \ref{F-theory-lift} couplings of the form $R^2 L^\alpha$ 
in \eqref{N-beforeshift} are not compatible with the 6d/5d lift. 
However, there is as simple field redefinition
which allows us to get rid of these $R^2 L^\alpha$ terms. More precisely, one introduces the shifted
fields \footnote{This field redefinition is also crucial in the 4d/3d treatment of F-theory on Calabi-Yau fourfolds as discussed in 
\cite{arXiv:1109.3191,arXiv:1008.4133}.} 
\beq \label{def-checkLA}
\check L^\alpha = L^\alpha + \tfrac{1}{2} K^\alpha R \ , \qquad \check A^\alpha = A^\alpha + \tfrac{1}{2} K^\alpha A^0\ ,
\eeq
where the shift of the vectors is required by supersymmetry. 
Clearly, the new $\check L^\alpha$, and new vectors can be obtained
by expanding $J$ and $C_3$ in a new basis of two-forms 
\beq \label{check-basis}
   \check \omega_0 = \omega_0 - \tfrac12 K^\alpha \omega_\alpha \ , \qquad \check \omega_\alpha = \omega_\alpha \ , \qquad \check \omega_i = \omega_i \ . 
\eeq
In fact, this new basis is better suited to identify the vectors $\check A^\alpha $ as dualizable into 5d tensors. 
The cubic potential in the new coordinates given by  
\begin{align} \label{cNM}
\mathcal{N}^{\rm M} = & \tfrac{1}{2} \eta_{\alpha\beta} R \check L^\alpha \check L^\beta  
+ \tfrac{1}{24} \eta_{\alpha\beta} K^\alpha K^\beta R^3 \nn \\
&- \tfrac{1}{2} \eta_{\alpha\beta} C^\alpha C_{ij} \check L^\beta \xi^i \xi^j
 +\tfrac{1}{4} \eta_{\alpha\beta} C^\alpha C_{ij} K^\beta R \xi^i \xi^j 
+ \tfrac{1}{6} \mathcal{V}_{ijk} \xi^i \xi^j \xi^k \;.
\end{align}
Using this expression of $\cN$ the Chern-Simons term takes the form
\begin{align}
S^{(5) \rm M}_{\rm CS} = \int_{\cM_5} & 
- \tfrac{1}{4} \eta_{\alpha\beta} A^0 \wedge \check F^\alpha \wedge \check F^\beta
 +\tfrac{1}{4} \eta_{\alpha\beta} C^\alpha C_{ij} \check A^\alpha \wedge F^i \wedge F^j   \nn \\
& - \tfrac{1}{48} \eta_{\alpha\beta} K^\alpha K^\beta A^0 \wedge F^0 \wedge F^0
- \tfrac{1}{8} \eta_{\alpha\beta} C^\alpha C_{ij} K^\beta A^0 \wedge F^i \wedge F^j \nn \\
&- \tfrac{1}{12} \mathcal{V}_{ijk} A^i \wedge F^j \wedge F^k \;,
\end{align}
where $\check F^\alpha$ is the usual field strength of the 
vectors $\check A^\alpha$ introduced in \eqref{def-checkLA}.

\subsection{Higher order curvature corrections} \label{11dhighercurvature}

Several higher-derivative corrections to the 11d M-theory action \eqref{Mtheorytree} are 
known \cite{Vafa:1995fj,Duff:1995wd}. In the following, we will focus on the mixed gauge-gravitational 
correction \footnote{As discussed in section \ref{6dgeneralities}, factors of $(2\pi)^{-1}$ are understood
in $\hat \cR$. Moreover, the relative normalization of this higher-derivative term and the two-derivative 
action \eqref{Mtheorytree} depends on the value of the 11d gravitational constant. It is suppressed everywhere,
adopting a convention
which is best suited to make contact with the 6d Green-Schwarz term, in which the 6d gravitational constant
has been equally suppressed.} 
\begin{equation} \label{I8correction}
  \hat S ^{(11)}_{C \cR^4} = \tfrac{1}{96} \int_{\mathcal{M}_{11}} \hat C_3 \wedge 
\left[ \tr \hat\cR^4  - \tfrac{1}{4} (\tr \hat\cR^2)^2   \right]
\end{equation}
because this terms allows us to make direct contact to the higher-derivative corrections 
studied in subsection \ref{6dhighercurvature}.

Rather than performing a complete dimensional 
reduction of \eqref{I8correction}, we will extract the relevant terms and we will 
systematically neglect all contributions which involve gradients of the K\"ahler and complex 
structure moduli. This means that we can effectively neglect fluctuations and compute 
curvature invariants on the background, which is the product space $\cM_{11} = \cM_5 \times Y_3$. 
As a result, we have simply \footnote{Just like in the reduction from six
to five dimensions, performing the Weyl rescaling on the 5d metric does not affect the moduli-independent  
terms in the expression of the curvature two-form.}
\begin{equation}
 \hat\cR = \cR + \cR_{Y_3} \;, 
\end{equation}
where $\mathcal{R}_{Y_3}$ is the curvature two-form on the Calabi-Yau threefold, and $\mathcal{R}$ 
is the 5d curvature two-form. A straightforward computation gives then
\begin{equation}
  (\tr \hat{\mathcal{R}}^2)^2 = 2 \tr \mathcal{R}^2 \wedge  \tr \mathcal{R}^2_{Y_3} + \dots \;, \qquad \qquad
 \tr \hat{\mathcal{R}}^4 =  0+\dots;, 
\end{equation}
where the dots are a reminder of the moduli-dependent, neglected 
terms. It is useful to recall the definition of the first Pontryagin class of the Calabi-Yau threefold $Y_3$,
\begin{equation}
 p_1(Y_3) = -\tfrac{1}{2} \tr \cR^2_{Y_3} \;,
\end{equation}
and its relation with the second Chern class,
\begin{equation}
 p_1(Y_3) = -2 c_2(Y_3) \;.
\end{equation}
Combining these equations with the three-form expansion \eqref{3formAnsatz},
we can deduce that the 11d correction \eqref{I8correction} yields, among 
other terms, the following 5d correction \cite{Antoniadis:1997eg}
\begin{equation} \label{Mcorrection_before_fibre}
 S^{(5) \rm M}_{A \cR \cR} = \tfrac{1}{48}\, \check c_\Lambda \int_{\cM_5} \check A^\Lambda \wedge \tr \mathcal{R}^2 \;,
\end{equation}
where we have defined
\begin{equation}
 \check c_\Lambda =  \;  \int_{Y_3}  \check \omega_\Lambda \wedge c_2(Y_3) \;.
\end{equation}

To make further progress it is crucial to specialize to the case of an elliptically fibred 
Calabi-Yau threefold $Y_3$.
Let us discuss a smooth fibration first. The second Chern class of the 
total space can then be expressed in term of Chern classes on the base space $B_2$, 
by means of \cite{Friedman:1997yq}
\begin{equation}
 c_2(Y_3) = c_2(B_2) + 11c_1^2(B_2) + 12\omega_0 \wedge c_1(B_2) \;.
\end{equation}
Making use of \eqref{omega0squaredidentity} we get
\begin{equation}
 \int_{Y_3} \omega_0 \wedge c_2(Y_3) = \int_{Y_3} \omega_0 \wedge [c_2(B_2) - c_1^2(B_2)] =
 \int_{B_2} c_2(B_2) - c_1^2(B_2)\;.
\end{equation}
This equation can be evaluated further by using the explicit expressions 
of the integrals of $c_2$ and $c_1^2$ on $B_2$ given in \eqref{chic12_base} as
\beq \label{intc2omega0}
   \int_{Y_3} \omega_0 \wedge c_2(Y_3) =  2 h^{1,1}(B_2) - 8\ .
\eeq
Furthermore, we can also evaluate the second Chern class on the basis $\omega_\alpha$ as
\begin{equation} \label{omegaa_c2}
\int_{Y_3} \omega_{\alpha} \wedge c_2(Y_3) = \int_{Y_3} \omega_\alpha \wedge [c_2(B_2) + 11c_1^2(B_2) 
+ 12\omega_0 \wedge c_1(B_2)]\ .
\end{equation}
Since the first two terms have all their indices on the base, only the last term provides a 
non-vanishing contribution. 
Using $c_1(B_2) = - K^\alpha \omega_\alpha$, as introduced in 
subsection \ref{elliptic_basics}, we compute
\begin{equation} \label{checkcalpha}
 \check c_{\alpha} = \int_{Y_3} \check \omega_{\alpha} \wedge c_2(Y_3) = - 12 \eta_{\alpha\beta} K^\beta \;,
\end{equation}
where we have used $\check \omega_\alpha = \omega_\alpha$.
In order to obtain $\check c_{0}$ from \eqref{intc2omega0}, \eqref{omegaa_c2} we have 
to recall the definition \eqref{check-basis} of $\check \omega_0$, and find 
\begin{equation} \label{checkc0}
   \check c_{0} =  52 - 4\, h^{1,1}(B_2)\ .
\end{equation}

So far we have worked on a smooth elliptic fibration. We now include the effects of singularities and 
their resolution. Clearly, the presence of resolved singularities induces new couplings 
\beq \label{def-ci}
   \check c_i = \int_{\tilde Y_3} \check \omega_i \wedge c_2(Y_3)\ .
\eeq  
One expects that this expression evaluated for a given gauge group has a 
group theoretic interpretation. Giving its precise form is beyond the scope of this 
work. However, let us note that also the other couplings $\check c_0$ and $\check c_\alpha$
could be corrected by the inclusion of blow-up divisors. Indeed,
a general shift of $c_2(\tilde Y_3)$ with the blow-up divisors induces
\beq \label{changec2}
 \int_{\tilde Y_3} \omega_0 \wedge \Delta c_2 (\tilde Y_3) =0 \ ,\qquad \int_{\tilde Y_3} \omega_\alpha \wedge \Delta c_2 (\tilde Y_3) = \cC^{ij} \int_{\tilde Y_3} \omega_{\alpha} \wedge \omega_i \wedge \omega_j \ ,
\eeq
where we have used the vanishing of the intersections \eqref{ell_intersections} 
with only one $\omega_i$ and two $\omega_\alpha$, and $\omega_i \wedge \omega_0 = 0$.
Note that  a shift in $\check c_0$ could still be induced due to the 
basis change \eqref{check-basis} inducing a term proportional to $\check c_{\alpha}$.
We claim that also $\check c_\alpha$ is uncorrected, and thus  
$\check c_0$ and $\check c_\alpha$ remain unchanged.
Despite that we do not have a general proof, we have checked for many examples that \eqref{checkcalpha} and \eqref{checkc0}
are still true:
\beq \label{def-calpha0}
 \check c_{\alpha} = - 12 \eta_{\alpha\beta} K^\beta \;, \qquad \check c_{0} =  52 - 4\, h^{1,1}(B_2) \ .
\eeq
As we will show later, the fact that $\check c_\alpha$ is not changed is consistent with 
the F-theory lift. The fact that $\check c_0$ does not change in this case follows from \eqref{changec2}.


\section{F-theory lift and one-loop corrections} \label{F-theory-lift}

In this section we compare the result of the circle reduction of the general 6d $(1,0)$ supergravity
theory with the M-theory reduction on an elliptically fibred Calabi-Yau threefold. We identify 
terms which appear at classical level on both sides and can be immediately matched as discussed in 
subsection \ref{match1}.  We also comment on the matching 
of certain higher derivative terms. It is crucial insight that both reductions contain additional terms which 
have no immediate analogue in the dual reduction. We suggest in 
subsection \ref{match2} that these terms arise at the 
quantum level and encode the same information about the underlying fully quantized theory. 
In particular, we argue that certain intersections on the M-theory side correspond in the 
6d/5d reduction on a circle to one-loop corrections with charged matter fermions and Kaluza-Klein modes 
of all 6d chiral fields running in the loop.  In conclusion this allows us to extract all data from M-theory required 
to specify the 6d action including the complete information about 6d anomalies.

\subsection{Classical action in the F-theory lift} \label{match1}

In order to extract information about F-theory in six dimensions, we have to compare the 5d 
action coming from Kaluza-Klein reduction from six dimensions with the 5d action of M-theory 
on an elliptically fibred Calabi-Yau threefold. Our strategy will be similar to 
the 4d/3d treatment of F-theory on Calabi-Yau fourfolds presented in \cite{arXiv:1008.4133}.

As a first step, we present the match of the number of multiplets in five dimensions in order to 
give the number of 6d multiplets in terms of the topological data of the F-theory 
compactification manifold $Y_3$. This was already implicit in our choice of indices 
in sections \ref{6d5dsection} and \ref{Mtheory_on_threefold}. More precisely, for the $\alpha$-index we 
find that the number of 6d tensors is given by
\beq \label{nTh11}
   n_T+1 = h^{1,1}(B_2)\ ,
\eeq   
where we recall that there are $n_T$ 6d tensor multiplets and $1$ tensor in the gravity multiplet. 
In the F-theory reduction the tensors arise from the reduction of the Type IIB RR four-form 
into a base of $H^{2}(B_2)$.
Since $A^i$ parameterize the Coulomb branch of the 6d/5d gauge theory, one finds 
\beq
  {\rm rank} (G) = h^{1,1}(\tilde Y_3) - h^{1,1}(B_2) -1 \ ,
\eeq
which counts the number of independent blow-up divisors induced to resolve the singular elliptic fibration to 
obtain $\tilde Y_3$. Note that for ADE gauge groups $G$ the number of 6d vector multiplets is then given by 
\beq
   n_{V} = (c_{G} +1) {\rm rank} (G) \ ,
\eeq
where $c_{G}$ is the dual Coxeter number of $G$. In F-theory these vectors arise from the 
seven-brane gauge potentials. 
Finally, one can match the number of hypermultiplets, simply by noting that a 6d hypermultiplet 
becomes a 5d hypermultiplet in the circle reduction. This leads to the following 
number of neutral 6d multiplets
\beq \label{nH_neutral}
   n^{\rm neutral}_{H} = h^{2,1}(\tilde Y_3) +1 \ .
\eeq
In F-theory on $Y_3$ these neutral hypermultiplets contain the complex deformations 
of the seven-branes and their Wilson line moduli.\footnote{See ref.~\cite{Braun:2009bh}, for a detailed matching with the 
orientifold picture with D7-branes.} The universal hypermultiplet in the F-theory reduction 
contains as one complex scalar the volume of the base together with the scalar of the Type 
IIB RR four-form expanded in the volume form of $B_2$. The remaining two real scalar 
degrees of freedom in the universal hypermultiplet arise in the expansion of the Type IIB RR and  
NSNS two-forms into the universal two-form mode present for any $B_2$.
The proof of the match \eqref{nTh11}-\eqref{nH_neutral} follows from the match 
of the effective theories presented in the following.

In order to systematically 
approach the match of the effective action, we would first like to identify the terms which 
are classical on both sides. This is not hard for the 6d/5d reduction. More complicated 
is the distinction of the various terms in the M-theory potential. We will address the 
two sides in turn.

In the 6d/5d reduction performed in section \ref{6d5dsection} we found that 
there is a potential $\cN^{\rm F}$ given in \eqref{def-cN} which encodes the kinetic 
terms of the gauge coupling functions and the Chern-Simons terms in 
the 5d reduced action. It is crucial to recall the natural decomposition 
of $\cN^{\rm F}$ in \eqref{def-cN} into a polynomial and a non-polynomial part:
\begin{align} \label{cNpnp}
 \cN^{\rm F}_{\rm p} &= \Omega_{\alpha \beta} M^0 M^\alpha M^\beta - 4 \Omega_{\alpha \beta} b^\alpha C_{ij}  M^\beta M^i M^j\ , \nn \\
 \cN^{\rm F}_{\rm np} &= 4 \Omega_{\alpha\beta} b^\alpha b^\beta C_{ij} C_{kl} \frac{M^i M^j M^k M^l}{M^0} \;.
\end{align}
The terms in  $\cN^{\rm F}_{\rm p}$ are cubic and hence encode a standard $\cN=2$ 5d action.
In contrast $\cN^{\rm F}_{\rm np}$ is only homogeneous of degree three, but non-polynomial. 
As argued in section \ref{canonica5daction} it can be interpreted as a counterterm of the 5d one-loop 
effective action. 
Its 6d origin is related to the classical 
lack of gauge invariance of the 6d action. 
In fact, it vanishes precisely when
\beq
  \Omega_{\alpha\beta} b^\alpha b^\beta = 0\ .
\eeq 
This corresponds to the case where the 6d action is gauge invariant as inferred from \eqref{classical_gaugeinvariance}, 
and is consistent with the absence of 6d anomalies as discussed in appendix \ref{appendix_anomalies}.

Let us now turn to the M-theory reduction. Here the identification of the 
classical terms is more subtle. We have worked on the resolved space with 
finite size elliptic fibre. As discussed in the introductory section \ref{F-theoryin6d},
the F-theory limit corresponds to both shrinking the blow-up divisors as 
well as the size of the elliptic fibre. One expects that this selects classical 
terms in the potential $\cN^{\rm M}$ of equation \eqref{cNM}. It turns out to be useful 
to introduce an $\epsilon$-scaling to distinguish various terms in $\cN^{\rm M}$. 
For the volumes $v^0,v^\alpha,v^i$ appearing in the K\"ahler form $J=v^\Lambda \omega_\Lambda$, we
make the formal replacements 
\beq
   v^0 \mapsto \epsilon v^0  \ ,  \qquad v^\alpha \mapsto \epsilon^{-1/2} v^\alpha  \ , \qquad v^i \mapsto \epsilon^{1/4} v^i \ .
\eeq
Note that these scalings satisfy some important consistency checks. Firstly, the size of the 
elliptic fibre $v^0$ and the blow-up fibres $v^i$ vanish for $\epsilon \rightarrow 0$. Secondly, 
the total volume $\cV$ of $Y_3$ is finite, which is required by the fact that $\cV$ sits in 
a 5d hypermultiplet. Translated into the variables $R,L^\alpha,\xi^i$ one finds the 
replacements 
\beq \label{scaling}
   R \mapsto \epsilon R  \ ,  \qquad L^\alpha \mapsto \epsilon^{-1/2} L^\alpha  \ , \qquad \xi^i \mapsto \epsilon^{1/4} \xi^i \ .
\eeq
Since the redefined scalars $\check L^\alpha$ contain $L^\alpha$ linearly, they obey the same rescaling as $L^\alpha$.
In the limit $\epsilon \rightarrow 0$ two terms in \eqref{cNM} survive which we collect in $\mathcal{N}^{\rm M}_{\rm class}$. We thus divide the terms in 
\eqref{cNM} into  
\begin{align} \label{cNM-split}
\mathcal{N}^{\rm M}_{\rm class} &\, =\,  \tfrac{1}{2} \eta_{\alpha\beta} R \check L^\alpha \check L^\beta 
                                  - \tfrac{1}{2} \eta_{\alpha\beta} C^\alpha C_{ij} \check L^\beta \xi^i \xi^j\ ,   \\
\mathcal{N}^{\rm M}_{\rm loop} &\, =\, \tfrac{1}{24} \eta_{\alpha\beta} K^\alpha K^\beta R^3 
 +\tfrac{1}{4} \eta_{\alpha\beta} C^\alpha C_{ij} K^\beta R \xi^i \xi^j 
+ \tfrac{1}{6} \mathcal{V}_{ijk} \xi^i \xi^j \xi^k \;. \nn
\end{align}
It is now straightforward to match $\mathcal{N}^{\rm M}_{\rm class}$ with $\cN^{\rm F}_{\rm p}$ given in \eqref{cNpnp}. 
Note that the second term $\mathcal{N}^{\rm M}_{\rm loop}$ in \eqref{cNM-split} will be later reinterpreted as 
a loop correction, which gives another justification of the split induced by the F-theory limit \eqref{scaling}.

Let us first start by matching the fields on the 6d/5d and the M-theory side. In order to do that we have 
to fix the normalization of the fields, which cannot be uniquely extracted by comparing \eqref{cNpnp} 
and \eqref{cNM-split}. Supersymmetry relates the normalization of the real scalars and vectors 
in the vector multiplets. Hence, given a fixed normalization of the vectors the complete match of the 
scalar components can be inferred. On the one hand, in the 6d/5d compactification the vectors are normalized by the 
Green-Schwarz term \eqref{Green-Schwarz-term}, and the fixed definition of the anomaly coefficients $b^\alpha,a^\alpha$.
On the other hand, in M-theory the normalization of the vectors is fixed by a choice of integral basis in 
the expansion \eqref{3formAnsatz} of $\hat C_3$. Appropriately rescaling the 6d vectors to also 
adopt to an integral basis, one can infer the map  
\beq
  M^0 = 2R \ ,\qquad M^\alpha  = \tfrac{1}{2}\check L^\alpha \ ,\qquad M^i = \tfrac12 \xi^i \ ,
\eeq
while the constants are identified as
\beq \label{etaOmega_match}
  \Omega_{\alpha \beta} =  \eta_{\alpha \beta}\ , \qquad b^\alpha = C^\alpha\ .
\eeq
Note that our result are consistent with the findings of \cite{Sadov:1996zm,KumarTaylor}.

So far we have only discussed the vector and gravity sectors of the M-theory to F-theory 
matching. Clearly, both the 6d/5d reduction as well as the M-theory reduction contain a hypermultiplet 
sector. As discussed in section \ref{56Coulomb}, we found that in the dimensional reduction from six to five dimensions
the charged hypermultiplets are massive in the Coulomb branch. Therefore, they are not 
visible in the effective action of the massless modes of M-theory. We will include them in 
the study of loop corrections in the next subsections. However, the neutral hypermultiplets are massless
and their moduli space could be matched straightforwardly also leading to \eqref{nH_neutral}. 

Let us close this subsection by also comparing the classical parts of the higher curvature 
terms dimensionally reduced in sections \ref{6dhighercurvature} and \ref{11dhighercurvature}. 
We have focussed on the terms involving the 5d vectors and two 5d curvature forms $\cR$. 
In \eqref{final6dcorrection} and \eqref{Mcorrection_before_fibre} we found that such couplings are given by 
\begin{equation} 
   S^{(5) \rm F}_{A \cR \cR} = - \tfrac{1}{2}\, \Omega_{\alpha \beta} a^\beta \int_{\cM_5}  A^\alpha \wedge \tr \mathcal{R}^2 \;, \qquad   
   S^{(5) \rm M}_{A \cR \cR} = \tfrac{1}{48}\, \check c_\Lambda \int_{\cM_5} \check A^\Lambda \wedge \tr \mathcal{R}^2 \;.
\end{equation}
Recall that the coefficients $\check c_\Lambda$ have been determined in \eqref{def-calpha0}, and \eqref{def-ci}. Since 
in the 6d/5d reduction only the $A^\alpha$ appears, one suspects that, similar to 
the F-theory limit discussed above, that these are the only classical terms in 
the reduction. Using $\check c_\alpha = -12 \eta_{\alpha \beta} K^\beta$, as given in \eqref{def-calpha0}, we can 
apply the identification \eqref{etaOmega_match} to infer
\beq
    a^\alpha = K^\alpha\ .
\eeq 
Note that this is precisely, the identification dictated by anomaly cancellation conditions 
as found in \cite{Sadov:1996zm,KumarTaylor}. On the M-theory side we also found the non-vanishing 
couplings involving $\check c_i,\check c_0$. Similar to the split found for $\cN^{\rm M}$ we believe 
that these couplings are induced by one-loop corrections on the 
F-theory side. The remainder of this paper is devoted to the discussion of 
such one-loop quantum corrections.

\subsection{Completing the duality using one-loop corrections} \label{match2}

As we have seen in the previous subsection, only some terms of the 5d cubic potential $\cN^{\rm M}$ of
M-theory compactified on a Calabi-Yau threefold admit a straightforward dual in the potential $\cN^{\rm F}$
arising from circle compactification of 6d supergravity. In this subsection, we will provide 
a framework for the interpretation of the remaining terms in $\cN^{\rm M}$, which we record here again
for the ease of the reader,
\begin{equation} \label{loop-terms}
\mathcal{N}^{\rm M}_{\rm loop} \, =\, \tfrac{1}{24} \eta_{\alpha\beta} K^\alpha K^\beta R^3 
 +\tfrac{1}{4} \eta_{\alpha\beta} C^\alpha C_{ij} K^\beta R \xi^i \xi^j 
+ \tfrac{1}{6} \mathcal{V}_{ijk} \xi^i \xi^j \xi^k \;. 
\end{equation}
Recall that 5d $\cN=2$ supersymmetry ensures that exactly the same amount of information is contained in the
cubic potential $\cN$ and in the Chern-Simons couplings of vectors. The following discussion is
conveniently formulated in terms of the latter. 
As already anticipated, we relate these couplings to one-loop effects in the 6d/5d dual description.

In order to clarify the precise meaning of this statement, let us analyse in more detail the origin of
Chern-Simons couplings in the effective 5d theory arising from 6d supergravity on a circle.
A possible source of this kind of interactions is of course provided by dimensional reduction of the
Green-Schwarz term in the classical 6d action. These interactions are precisely the ones which we have considered 
in the previous subsection. However, additional contributions arise, which are understood in the framework
of effective quantum field theory. In fact, from a quantum perspective, the 5d effective action 
resulting from compactification on a circle of 6d supergravity encodes all information about the
low-energy dynamics, including interactions induced by massive fields which have
to be integrated out when we restrict our attention to the lightest states of the theory.

In the case under examination, we identify two different families of massive fields which
can alter 5d effective couplings:
\begin{itemize}
 \item {\it Kaluza-Klein modes}. All 6d fields can be schematically expanded into Kaluza-Klein modes as
\begin{equation} \label{general-KK}
\hat{\varphi}(x,y) = \sum_{n\in\mathbb{Z}} \varphi^{(n)}(x) e^{iny} \ .
\end{equation}
 The modes $\varphi^{(n)}$ with non-zero $n$ appear in the 5d theory as massive fields, with mass inversely
 proportional to the radius $r$ of the compactification 
circle \footnote{This holds before possible Weyl rescalings are taken into account.}, 
$m^{(n)} \sim |n|/r$. As argued in the introductory section \ref{FviaM}, zero-modes only are 
 sufficient to fix all data needed to specify the 6d model we are compactifying, and this is why we have systematically
 neglected excited modes so far. Nonetheless, Kaluza-Klein modes can run in 5d loop diagrams. 
\item {\it Fields which are given a mass by gauge symmetry breaking}. Recall that F-/M-theory duality 
can be applied in a geometric regime only if the 
5d gauge symmetry is spontaneously broken down to the Coulomb phase and the compactification 
threefold is resolved. This amounts to giving non-vanishing VEVs
to some scalars in the vector multiplets. As described in subsection \ref{56Coulomb}, these VEVs provide mass terms
for the W-bosons and the scalars in charged hypermultiplets. Supersymmetry implies that their fermionic partners, gaugini and hyperini, get massive as well.
We claim that these fields can run in 5d loops in
such a way as to induce effective Chern-Simons couplings.
\end{itemize}

We are able to provide a geometric picture for these families in the F-theory set-up. 
As recalled in the previous subsection, F-theory is conveniently analysed in a phase with 
finite size of the elliptic fibre and of the exceptional divisors introduced by resolution 
of singularities. However, F-/M-theory duality holds only in the limit in which these 
cycles are shrunk to zero size. In the M-theory picture, M2-branes can wrap these shrinking submanifolds. By means of the 
chain of dualities described in section \ref{FviaM}, it is possible to identify the states of 
M2-branes wrapping the elliptic fibre as Kaluza-Klein modes in the 6d/5d picture. 
Furthermore, 5d Higgsing to the Coulomb branch is dual to the blowing-up
of singularities provided by exceptional divisors. M2-branes wrapping such divisors provide the degrees of freedom of
both W-bosons and charged hypermultiplets, whose mass vanishes as the divisor is blown-down. 

We now turn to a more detailed description of the mechanism responsible for Chern-Simons couplings 
in the effective 5d theory. We follow closely reference \cite{Witten:1996qb}. A term of the form
\begin{equation}
 A \wedge F \wedge F
\end{equation}
in the Lagrangian corresponds to an amplitude with three external vectors. If these carry momenta $p,q,-p-q$ and 
polarizations $\alpha, \beta, \gamma$, the amplitude will be proportional to
\begin{equation}
 \epsilon^{\alpha\beta\gamma\mu\nu} p_\mu q_\nu \;.
\end{equation}
Suppose we compute a three-vector amplitude in the 5d theory with massive fields of the kind
listed above. General arguments imply that only one-loop diagrams provide
corrections to the classical Chern-Simons interactions. It is crucial to observe that the structure of the Chern-Simons coupling we are interested in
can be extracted unambiguously by looking at the parity violating terms with quadratic dependence on the external
momenta $p,q$. In particular, a Chern-Simons effective coupling can arise only if a totally antisymmetric
tensor $\epsilon^{\alpha\beta\gamma\mu\nu}$ is found in the computation of the three-vector amplitude.

We argue that this tensorial structure can be generated 
if massive modes of 6d chiral fields run in the loop. First of all, vertices between fermions and vectors are 
 able to give this kind of parity
violating term. From a Feynman diagram perspective, this can be seen as follows. In the computation of a one-loop
amplitude with fermions running in the loop, the trace of a string of 5d gamma matrices is involved. However, 
5d Clifford algebra implies, e.g. 
\begin{align}
 \tr \Gamma_a \Gamma_b \Gamma_c \Gamma_d \Gamma_e &= 4 \epsilon_{abcde} \;, \nn \\
 \tr \Gamma_a \Gamma_b \Gamma_c \Gamma_d  \Gamma_e \Gamma_f \Gamma_g &= 4  \epsilon_{abcde} \eta_{fg} + \text{other terms} \;.
\end{align}
Indeed, as explained in \cite{Witten:1996qb}, whenever a 5d fermion $\psi$ runs in the loop, with standard propagator and 
coupling to vectors of the form $A^\mu \bar\psi \Gamma_\mu \psi$, a contribution to the effective Chern-Simons coupling
is found.
Second of all, we claim that massive Kaluza-Klein modes of tensors can contribute to 
the parity violating part of the loop amplitude. On very general grounds, an electric coupling 
to the graviphoton $A^0$ is expected for all excited Kaluza-Klein modes. Moreover, the epsilon
tensor can enter the diagram by means of a term of the form $B \wedge dB$ in the 5d effective action.

We are now in a position to state our claim about the quantum origin of terms \eqref{loop-terms}: they are generated 
by 5d one-loop diagrams with three external vectors and massive chiral modes running in the loop. In order for this
mechanism to work, we have to show that the fields in the three massive families listed above interact with 5d vectors
in the correct way such that the result of \cite{Witten:1996qb} can be applied. A thorough derivation of \eqref{loop-terms} from one-loop
calculation in 5d dimensions is beyond the scope of this paper, and is left for further investigation in future work. 
Nonetheless, we can give a schematic illustration of the source of the relevant couplings and mass terms for the massive fermions
in the two families listed above. Massive modes of tensors would deserve further discussion, and the authors
hope to come back soon to this subject.

\subsection{Origin of the one-loop Chern-Simons couplings}

We start discussing fermionic Kaluza-Klein modes. Let $\hat\psi_{(\pm)}$ denote a general 6d spinor of given chirality. It is 
an 8-component spinor with complex entries, but the number of degrees of freedom is halved by restriction to
definite chirality. This counting agrees with the number of degrees of freedom of the (off-shell) 5d reduced spinor $\psi$, 
which can be represented as a 4-component vector with complex entries.

We can be more explicit. A representation of 6d gamma matrices $\hat\Gamma_{\hat a}$, $\{ \hat\Gamma_{\hat a}, \hat\Gamma_{\hat b}\} = 2 \hat\eta_{\hat a \hat b}$, $\hat a, \hat b=0,1,...,5$ can be found, such that
\beq
\hat\Gamma_a = \sigma_1 \otimes \Gamma_a \;, \qquad \hat\Gamma_5 = \sigma_2 \otimes \mathbb{I}_4 \; .
\eeq
In these equations, $\sigma_i$ are the usual Pauli matrices, while $\Gamma_a$, $\{\Gamma_a, \Gamma_b\} = 2 \eta_{ab}$, $a,b=0,1,...,4$ are 5d gamma matrices, satisfying
\beq
i \Gamma_0 \Gamma_1 \Gamma_2 \Gamma_3 \Gamma_4 = \mathbb{I}_4 \; .
\eeq
As a result, the 6d chirality matrix is simply given by
\beq
\hat \Gamma = \hat\Gamma_0 \hat\Gamma_1 \hat\Gamma_2 \hat\Gamma_3 \hat\Gamma_4 \hat\Gamma_5 = \sigma_3 \otimes \mathbb{I}_4 \; .
\eeq
We can thus write $\hat\psi_{(\pm)}$ in the factorized form
\beq
\hat\psi_{(\pm)} = \iota_{(\pm)} \otimes \psi \; ,
\eeq
where $\iota_{(\pm)}$ is a unit vector in $\mathbb{C}^2$, such that $\sigma_3 \iota_{(\pm)} =\pm \iota_{(\pm)}$, and $\psi$ is a 5d spinor.

Using these conventions, dimensional reduction of the 6d standard kinetic
term for $\hat \psi_{(\pm)}$ yields \footnote{In order to keep the argument simple, we work in a flat background and we do not Weyl rescale the 5d metric.} 
\begin{equation}
 \int d^6\hat x \; \hat{\bar\psi}_{(\pm)} \hat \Gamma^{\hat\mu} \hat\partial_{\hat\mu} \hat\psi_{(\pm)} =
2\pi \sum_{n\in \mathbb{Z}}\int d^5x\; r \big\{ \bar\psi^{(n)} \Gamma^\mu \partial_\mu \psi^{(n)} 
\mp \tfrac{n}{r} \bar\psi^{(n)} \psi^{(n)} + i n A^0_\mu \bar\psi^{(n)} \Gamma^\mu \psi^{(n)} \big\}\;.
\end{equation}
On the left hand side, a hat denotes 6d gamma matrices, indices, and coordinates. The modes $\psi^{(n)}$ of the fermion $\psi$ are defined 
as in \eqref{general-KK}. On the right hand side, we find
a result consistent with the general features of Kaluza-Klein models on a circle. In fact, the $n$-th excited Kaluza-Klein mode 
has a mass proportional to $n$ and is electrically charged with respect to the vector $A^0$. The charge is 
proportional to $n$ as well.

We can now turn to fermions in the vector multiplets. Let $\hat\lambda$ be a 6d spinor in the adjoint representation 
of the simple gauge group $G$. Its gauge-covariant derivative is given by
\begin{equation}
 \hat D \hat\lambda = d \hat\lambda + [\hat A, \hat\lambda] \;,
\end{equation}
where $\hat A$ are the non-Abelian 6d vectors introduced in section \ref{6dgeneralities}. In order to keep the discussion as simple 
as possible, we restrict our attention to Kaluza-Klein zero-modes only in this paragraph. As a consequence, dimensional reduction
of the 6d kinetic term for $\hat \lambda$ is of the form
\begin{equation} \label{gaugino-reduction}
 \int d^6 \hat x \tr \big( \hat{\bar\lambda}\hat\Gamma^{\hat\mu} \hat D_{\hat\mu} \hat\lambda \big) =
 2\pi \int d^5 x \; r \big\{ \tr \big( \bar\lambda \Gamma^\mu D_\mu \lambda \big)+ \tfrac{i}{r} \tr \big(\bar\lambda [\zeta, \lambda] \big) \big\} \;.
\end{equation}
On the right hand side, $D\lambda = d\lambda + [A,\lambda]$ is the 5d gauge-covariant derivative, while $\zeta$ is the adjoint scalar 
introduced in the Ansatz \eqref{KK_vector}. Note that the sign of the last term is determined by 
the requirement of left-handedness for the gaugini, and that no $A_0$-coupling emerges for the Kaluza-Klein zero-modes precisely 
thanks to the shift of 5d vectors described by \eqref{KK_vector}. When the gauge symmetry is spontaneously broken to the Coulomb branch,
the scalars $\zeta$ acquire a non-vanishing VEV orthogonal to the Cartan subalgebra. Furthermore, 
commutators $[A,\lambda],[\zeta,\lambda]$ vanish for the components of $\lambda$ lying in this subalgebra. However, they are 
non-trivial for the components orthogonal to it. 
These components receive a mass from the second term in \eqref{gaugino-reduction}, while the first term in the same equation 
provides electric coupling to the Abelian vectors $A^i$ associated to the generators of the Cartan subalgebra.
We can thus see that Higgsed gaugini have the correct coupling to generate the effective Chern-Simons 
interaction under examination.

A similar argument can be used to conclude that charged hyperini
can run in the loop and furnish a non-vanishing contribution. More precisely, dimensional reduction of their kinetic term gives
\begin{equation} \label{hyperino-reduction}
 \int d^6 \hat x \tr \big[ h_{UV}\hat{\bar{\psi}}^U\hat\Gamma^{\hat\mu}  (\hat{\cD}_{\hat\mu} \hat\psi)^V \big] =
2\pi \int d^5 x \; r \big\{  h_{UV}{\bar{\psi}}^U \Gamma^{\mu}  ({\cD}_{\hat\mu} \psi)^V - \tfrac{i}{r} h_{UV}  \bar \psi^U \zeta^I (T^{\bf R}_I \psi)^V \big\} \nn\;.
\end{equation}
In this expression, the 6d covariant derivative of the hyperino is defined as
\begin{equation}
 (\hat{\cD}_{\hat\mu} \hat\psi)^U = \hat\nabla_{\hat\mu} \hat\psi^U + \hat A^I_{\hat\mu} (T^{\bf R}_I \hat\psi)^U \;,
\end{equation}
and an analogous expression is understood for the 5d covariant derivative on the right hand side. Note that
the sign of the last term has changed with respect to the gaugino reduction, because hyperini are right-handed. 
Upon spontaneous gauge symmetry breaking to the Coulomb branch, this term provides a mass for charged hyperini, 
while neutral hyperini are unaffected and remain in the massless 5d spectrum.

The reader might wonder whether there are massive fermions which are electrically coupled to
vectors $A^\alpha$. Our analysis suggests that this is not the case. A thorough explanation would require
dimensional reduction of the full 6d pseudo-action, including fermionic terms. Such a pseudo-action can be 
found e.g.~in \cite{Riccioni:1997ik}. However, it is crucial to recall that 5d vectors $A^\alpha$ are obtained by
dimensional reduction of 6d two-forms $\hat B^\alpha$. Such two-forms enter the 6d action in a qualitatively 
different way as 6d vectors. Geometrically, they are not connection forms, and cannot be used to
build 6d covariant derivatives. Therefore, the reduced 5d action lacks electric couplings of vectors
$A^\alpha$ to fermions. Nonetheless, different couplings are possible, which can be referred to as magnetic.
They read schematically $m_\alpha \bar\psi \Gamma^{\mu\nu} F^\alpha_{\mu\nu} \psi$ where $\psi$ stands
for a 5d fermion. Even though these interactions may play a role in the full one-loop 5d effective action,  
in the absence of electric vertices they are not able to generate contributions to the Chern-Simons couplings.

It is interesting to point out the connection between this argument and the shift of vectors
performed in \eqref{def-checkLA}. As explained in section \ref{Mreduction}, this shift is crucial
to identify properly  5d vectors coming from 6d two-forms. As we can see by comparing \eqref{N-beforeshift}
 and \eqref{cNM}, the field redefinition \eqref{def-checkLA} is such that 
in the cubic potential $\cN^{\rm M}$ the term $R^2 L^\alpha$ gets replaced by the term $R\xi^i \xi^j$. 
As argued in the previous paragraph, it would be impossible to generate the former term 
using 5d fermion loops, while in the following we will show how the latter term can emerge from such Feynman diagrams.

After these general remarks about massive fermions in the 5d theory, let us discuss in more detail each 
term in \eqref{loop-terms}. The first term corresponds to a Chern-Simons coupling of the form $A^0 \wedge F^0 \wedge F^0$. As we argued
above, Kaluza-Klein modes are the fields which are electrically charged under $A^0$. We therefore claim 
that this 5d interaction is generated by diagrams in which Kaluza-Klein excited modes 
coming from reduction of all chiral 6d fields can run in the loops. In order to get a finite result, the sum over modes has to be suitably
regularized, e.g. by means of the Riemann zeta function. We expect the outcome of the computation to
be independent of the specific regularization scheme chosen, since it describes a physical observable.
It is intriguing to recall at this point the interplay between the 6d anomaly coefficients and the numbers of
multiplets in 6d the theory. In particular, we can consider equations \eqref{eq:hv} and \eqref{eq:aa-condition}, which we record here again,
\begin{equation}
n_H-n_V =  273-29 n_T \ , \qquad \Omega_{\alpha \beta} a^\alpha a^\beta =  9 - n_T  \ .
\end{equation}
These relations can be combined with the identification $K^\alpha = a^\alpha$ found in the previous subsection and strongly 
suggest that the prefactor of the first term in \eqref{loop-terms} can be extracted from 5d loop computations
involving all species of  chiral fields of the theory. Each species gives a contribution proportional to number 
of the corresponding 5d multiplets. Note that the relation between 5d and 6d multiplets has been worked out
in section \ref{canonica5daction}, see \eqref{nH}, \eqref{nV}. 

The next term in \eqref{loop-terms} corresponds to a Chern-Simons vertex of the form $A^0 \wedge F^i \wedge F^j$. In order to 
reproduce this effective coupling using 5d one-loop diagrams, we need fermions which are electrically coupled both 
to the Kaluza-Klein vector $A^0$ and to the Abelian vectors $A^i$ in the Coulomb branch. Our discussion above singles out 
Kaluza-Klein modes of Higgsed gaugini and charged hyperini as natural candidates to run in the loop.

Finally, we focus our attention on the last term in \eqref{loop-terms}, which gives rise to a 
Chern-Simons term $A^i \wedge F^j \wedge F^k$. We identify the source of this coupling in the
 Higgsed gaugini and the massive charged hyperini. The one-loop effect due to these fermions has 
been computed \cite{Intriligator:1997pq} for a 5d $\cN=2$ supersymmetric gauge theory decoupled from 
gravity. The full result for the purely gauge part 
of the 5d cubic potential $\cN$, including quantum corrections, reads 
\begin{equation} \label{intriligator-eq}
 \cN^{\rm gauge} =\tfrac{1}{2}m_0 C_{ij} \xi^i \xi^j + \tfrac{1}{6}c_{\rm class} d_{ijk} \xi^i \xi^j \xi^k+
 \tfrac{1}{12} \bigg( \sum _\mathbf{R} |\mathbf{R} \cdot \xi|^3 
- \sum_f \sum_{\mathbf{w}\in \mathbf{W}_f} |\mathbf{w} \cdot \xi + m_f|^3   \bigg) \;.
\end{equation}
In this equation $\xi$ is a vector whose component are the scalar fields $\xi^i$ associated to
vectors $A^i$. In $\xi\cdot {\bf R}$ it is contracted with a root of the simple gauge group $G$,
while in $\xi\cdot {\bf w}$ it contracts with a weight of a the representation 
in which the charged fermions transform. The first sum in \eqref{intriligator-eq}
runs over all the roots of $G$, and arises from integrating out the Higgsed gaugini, i.e.~the 
fermionic partners of massive W-bosons. The second sum in \eqref{intriligator-eq} runs over all 
massive charged fermions $f$ and all weights in $\mathbf{W}_f$, i.e.~all elements of 
the set of weights of the representation in which the fermion $f$ transforms.
$m_f$ is the classical mass of the fermion $f$. Finally, the group theoretical invariants $C_{ij}$ and $d_{ijk}$ are given by
\begin{equation}
 C_{ij} = \tr T_i T_j \ , \qquad d_{ijk} = \tfrac{1}{2} \tr T_i(T_j T_k + T_k T_j) \ .
\end{equation}

To apply the formula \eqref{intriligator-eq} to our 6d/5d compactification, we recall the 
classical expression \eqref{cNpnp} for $\cN^{\rm F}$. This leads to the identification 
\beq
    m_0  = - 8 M^\alpha b^\beta \Omega_{\alpha \beta} \ ,  \qquad c_{\rm class} = 0\ ,
\eeq
where we have used the fact that upon decoupling gravity the $M^\alpha$ are simply parameters. Following 
the discussion of section \ref{match1} this matches the classical M-theory result. 
A careful comparison of the loop terms in \eqref{intriligator-eq} and the intersection numbers $\mathcal{V}_{ijk}$ of the resolved
Calabi-Yau threefold $\tilde Y_3$ would require the introduction of new technical tools and lies out of the main line 
of development of this section. However, let us stress that  the reader can find a detailed discussion
of this point in \cite{Grimm:2011fx}, appendix A: as explained there, the match can be performed successfully
in many examples of Calabi-Yau threefolds with $SU(N)$ singularities. The classical mass $m_f$ is zero in this case.

In summary, we are confident that all terms in the M-theory expression \eqref{loop-terms} arise from one-loop quantum corrections
in the 6d/5d dual picture. Moreover, it is tempting to extend this analysis to some higher-derivative
couplings which appear naturally in the M-theory reduction on a Calabi-Yau threefold, but seem to be absent 
in the reduction of 6d supergravity on a circle. Since we have not addressed the problem of the full reduction of 
higher-derivative actions, we limit ourselves to an example. In section \ref{11dhighercurvature} 
we have seen that M-theory higher-curvature correction induce a  term \eqref{Mcorrection_before_fibre} which has a
non-vanishing contribution
involving the Kaluza-Klein vector $A^0$. It is proportional to the shifted component $\check c_0$ of the second
Chern class of the Calabi-Yau threefold $c_2(Y_3)$ and reads schematically
\begin{equation}
 A^0 \wedge \tr \cR \wedge \cR \;,
\end{equation}
and corresponds to an amplitude with one Kaluza-Klein vector $A^0$ and two 5d gravitons.
It is impossible to extract such a coupling from the higher-curvature Green-Schwarz term \eqref{Green-Schwarz-term} in the 6d
pseudo-action. Hence, we are led to claim that on the 6d/5d side this interaction emerges as quantum effect, in
a similar fashion as the $A^0 \wedge F^0 \wedge F^0$ coupling analysed above. In particular, since the tensorial structure 
of this vertex involves the totally antisymmetric symbol $\epsilon^{\mu\nu\rho\sigma\lambda}$, we can apply the same
argument used above and infer that  the only
non-vanishing contributions to this coupling are due to massive modes  
from the reduction of 6d chiral fields. Given the universality
of gravitational interactions and Kaluza-Klein couplings involving $A^0$, it is natural to expect that all 
species contribute to this amplitude.  A more systematic treatment of this issue is not possible in the context
of the present paper, and the authors regard it as possible subject for further research.


\section{Conclusions}


In this paper we derived the 6d $(1,0)$ effective action 
of F-theory compactified on a singular elliptically fibred 
Calabi-Yau manifold $Y_3$. Our strategy was to use an M-theory compactification 
on the resolved space $\tilde Y_3$, and compare the effective 5d action with a general
6d action reduced on a circle. We included an extensive discussion of 5d one-loop 
corrections to the Chern-Simons term and their interplay with the 6d anomaly conditions.

In the first part of this work we performed 
the circle reduction of a general 6d $(1,0)$ supergravity theory 
with a non-Abelian gauge group $G$. We performed the Kaluza-Klein reduction in 
the non-Abelian phase and later discussed the modifications when 
the effective 5d theory is considered on the Coulomb branch. We argued 
that the charged hypermultiplets and the vector multiplets containing the 
W-bosons are massive in this phase and need to be integrated out when comparing with 
an M-theory reduction on $\tilde Y_3$. Moreover, we presented a 
careful treatment of the self-dual and anti-self-dual tensors present 
in a general 6d theory. While we used a 6d pseudo-action, which 
has to be accompanied by the self-duality conditions on the 
level of the equations of motion, we showed in detail that in the Kaluza-Klein reduced 
theory the self-duality can be imposed on the level of the action now 
relating 5d vectors and tensors. However, due to the fact that the 6d theory 
requires an anomaly cancelling Green-Schwarz term, the resulting 5d theory 
is also classically non-gauge invariant. We showed that its vector 
sector can nevertheless be encoded by a single real function $\cN^{\rm F}$ which 
is homogeneous of degree three. However, $\cN^{\rm F}$ contains a non-polynomial term which 
is required to encode the complete 5d metric for the vectors and enforce $\cN^{\rm F}=1$
consistent with the 6d supergravity constraint $\Omega_{\alpha \beta} j^\alpha j^\beta =1$ imposing 
a condition on the real scalars $j^\alpha$ in the tensor multiplets. 
The non-polynomial term is not present in a standard 5d $\cN=2$ supergravity theory 
and induces a non-gauge invariant term. We identify this term as a one-loop counterterm.
The 6d Green-Schwarz term also contains a higher curvature coupling 
and we presented a partial dimensional 5d reduction of this term.

In the second part of this paper we compared the circle 
reduced action with the 5d effective action of M-theory on a 
Calabi-Yau threefold $\tilde Y_3$. 
To extract the 5d $\cN=2$ characteristic 
data in a geometric regime one has to work with the resolved threefold $\tilde Y_3$, where 
both the gauge group singularities at co-dimension one in $B_2$, as well as the matter singularities 
at co-dimension two in $B_2$ are resolved. Accordingly all M2-brane states wrapped on cycles 
in $\tilde Y_3$ are massive and do not arise as dynamical degrees of freedom in the 5d effective theory.
However, the 5d effective action of M-theory on $\tilde Y_3$ contains terms which arise 
by consistently integrating out these massive states. To disentangle these from the 
terms present in the classical 6d/5d reduction we introduced a scaling limit corresponding 
to the F-theory limit. The finite terms in the M-theory reduction are readily matched 
with the general 6d/5d result. This enabled us to determine the core characteristic data 
required to evaluate the 6d $(1,0)$ F-theory effective action in terms of the geometric 
data of $\tilde Y_3$. Also dimensionally reducing the known M-theory higher curvature terms 
we were able to extract from a 5d comparison the integral vectors $(a^\alpha,b^\alpha)$ encoding 
all 6d anomalies. 

In the treatment of the massive states we have discovered an intriguing interplay 
of 5d one-loop corrections and 6d anomalies. In fact, since the M-theory reduction is 
on the resolved $\tilde Y_3$, all M2-brane states wrapped 
on the resolving $\mathbb{P}^1$-fibres are massive. These M2-brane states are 
dual in the F-theory limit of M-theory to the 6d charged hypermultiplets, and 
6d vector multiplets containing the W-bosons. Accordingly, one can only 
compare the 5d theories if these massive 
states are consistently integrated out also in the circle reduced theory. 
This is equally true for the M2-brane states 
on the elliptic fibre itself which are massive for a finite fibre volume. 
Using the M-theory to F-theory lift we identify these modes as certain 
Kaluza-Klein modes. More generally, this implies that also massive Kaluza-Klein modes have 
to be integrated out consistently in the circle reduction to compare the 5d result with the 
M-theory reduction. We have focused in this work on the investigation of the 
5d Chern-Simons couplings which only receive corrections due to massive 
5d modes of 6d chiral fields in one-loop diagrams.
The investigation of the various 
 couplings allowed us to identify the one-loop diagrams 
generating the classically absent couplings. In 
particular, we argued that the couplings $A^0 \wedge F^0 \wedge F^0$ and $A^0 \wedge \tr \cR \wedge \tr \cR$ 
are generated by integrating out massive  Kaluza-Klein modes.
We expect that both fermions and tensors can run in this loop diagram. 
A detailed account of possible fermionic coupling has been given, while
we leave a  proper
discussion of massive tensors for future investigation. 
More familiar, are the couplings $A^i \wedge F^i \wedge F^j$, which are generated 
by integrating out massive hyperini and gaugini. The mixed terms, such as 
$A^0 \wedge F^i \wedge F^j$, are induced by combining the vertices and 
propagators of both sets of massive fermionic modes. 
We believe that comparing the resulting coefficient functions to the geometric M-theory
result leads to a 5d derivation of the 6d anomaly cancellation conditions. While we have  
summarized the necessary tools to perform these one-loop integrals, we leave the explicit evaluation 
of all Chern-Simons coefficients to future work. 

There are various interesting directions for further research. Firstly, 
one can extend the classical reduction on both the 6d/5d action and the 
M-theory side to more then one non-Abelian gauge group. Also the 
extension to include Abelian $U(1)$ gauge groups is desirable. Additional $U(1)$ gauge 
fields will modify the 6d anomaly constraints and lead to new insights about the interplay 
of resolved geometries and 6d gauge theories.\footnote{See ref.~\cite{arXiv:1110.5916}\cite{arXiv:1111.2351} for recent progress in this direction.}
Beyond the classical analysis it would be important to extend the study of loop corrections
to all terms in the 5d action obtained by circle reduction. This includes 
a detailed study of the metric for the neutral hypermultiplets.
Also a evaluation of the coefficient of the 5d higher curvature corrections, generated 
at the quantum level, will be desirable. Comparing the results with the coefficients
$\check c_i,\check c_0$ predicted by the geometry of $\tilde Y_3$ on the M-theory side, 
will be a non-trivial test of the F-theory limit and its consistency 
with 6d anomaly cancellation. Reversely, one might also be able to 
use known 6d higher curvature terms to infer additional terms in the 
11d supergravity action. This is particularly interesting 
since the 6d/5d Kaluza-Klein vector is part of the M-theory three-form. 

Let us close by noting that in this work we have only dealt with Abelian 
tensor fields in the 6d action. We have found that in this case the couplings 
of the form $A^\alpha \wedge F^\beta \wedge F^\gamma$ are not generated in the 5d effective 
theory. In a future project we hope to generalize the transdimensional 
treatment of tensors to the non-Abelian case. It will be interesting 
to investigate how the various terms expected for non-Abelian 
tensors are generated in the M-theory picture.

\vspace*{1cm}
\noindent
{\bf Acknowledgments}: 
We would like to thank Hirotaka Hayashi, Stefan Hohenegger, Denis Klevers, Albrecht Klemm, Noppadol Mekareeya, Daniel Park, Raffaele Savelli, 
Maximilian Schmidt-Sommerfeld and Wati Taylor for interesting discussions. 
This work was supported by a research grant of the 
Max Planck Society.

\vspace*{2cm}


\appendix

\noindent {\bf \LARGE Appendices}

\section{Notations and conventions} \label{appendix_conventions}

For every spacetime dimension $d$, we adopt the mostly plus convention for the metric $g_{\mu\nu}$,
and the $(+++)$ conventions of \cite{Misner} for the Riemann tensor: explicitly,
\begin{align}
\Gamma^\rho_{\phantom{a}\mu\nu} &= \tfrac{1}{2} g^{\rho\sigma} \left( \partial_{\mu} g_{\nu\sigma} 
+  \partial_{\nu} g_{\mu\sigma} -  \partial_{\sigma} g_{\mu\nu}\right) \;, \nn \\
R^{\lambda}_{\phantom{a}\tau \mu\nu} &= \partial_{\mu} \Gamma^\lambda_{\phantom{a}\nu\tau} 
- \partial_{\nu} \Gamma^\lambda_{\phantom{a}\mu\tau}
 + \Gamma^\lambda_{\phantom{a}\mu\alpha} \Gamma^\alpha_{\phantom{a}\nu\tau} 
- \Gamma^\lambda_{\phantom{a}\nu\alpha} \Gamma^\alpha_{\phantom{a}\mu\tau} \;,\nn  \\
R_{\mu\nu} &= R^{\lambda}_{\phantom{a}\mu\lambda\nu} \;, \quad R = R_{\mu\nu} g^{\mu\nu} \;.
\end{align}

We use $\epsilon_{\mu_1 \dots \mu_d}$ to denote the Levi-Civita tensor, and use the 
metric to raise its indices. It is defined in such a way that, in any coordinate system $(x^0, x^1, \dots ,x^{d-1})$,
\begin{equation}
\epsilon_{01 \dots (d-1)} = +\sqrt{-\det g_{\mu\nu}} \;.
\end{equation}
Note that the following identity holds for arbitrary $k=0,...,d$:
\begin{align}
\epsilon_{\mu_1 \dots \mu_k \lambda_{k+1} \dots \lambda_d}  \epsilon^{\nu_1 \dots \nu_k \lambda_{k+1} \dots \lambda_{d}} 
= - k!(d-k!) \delta^{\nu_1}_{[\mu_1} \dots \delta^{\nu_k}_{\mu_k]} \;.
\end{align}

Differential $p$-forms are expanded on the basis of differential of the coordinates as
\begin{equation}
\lambda = \tfrac{1}{p!} \lambda_{\mu_1 \dots \mu_p} \; dx^{\mu_1} \wedge \dots \wedge dx^{\mu_p} \;,
\end{equation}
so that the wedge product of a $p$- and a $q$-form satisfies
\begin{equation}
(\alpha \wedge \beta)_{\mu_1 \dots \mu_{p+q}} 
= \tfrac{(p+q)!}{p!q!} \alpha_{[\mu_1 \dots \mu_p} \beta_{\mu_{p+1} \dots \mu_{p+q}]} \;.
\end{equation}
Exterior differentiation of a $p$-form is given by
\begin{equation}
(d\alpha)_{\mu_0 \dots \mu_p} = (p+1) \partial_{[\mu_0} \alpha_{\mu_1 \dots \mu_p]} \; .
\end{equation}
The Hodge dual of $p$-form in real coordinates and arbitrary spacetime dimension $d$ 
is defined by expression
\begin{equation}
(*\alpha)_{\mu_1 \dots \mu_{d-p}} = \tfrac{1}{p!} \alpha^{\nu_1 \dots \nu_p} 
\epsilon_{\nu_1 \dots \nu_p \mu_1 \dots \mu_{d-p}} \;.
\end{equation}
As a consequence,
\begin{equation}
\alpha \wedge *\beta = \tfrac{1}{p!} \alpha_{\mu_1 \dots \mu_p} \beta^{\mu_1 \dots \mu_p} \; *1
\end{equation}
holds identically for arbitrary $p$-forms $\alpha, \beta$.


\section{Anomalies in 6d supergravity} \label{appendix_anomalies}

In subsection \ref{6dgeneralities} we mentioned generalized Green-Schwarz mechanism 
\cite{Green:1984sg, Sagnotti:1992qw, Sadov:1996zm} for anomaly cancellation in a 6d supergravity 
model with simple gauge group $G$. In this appendix we review this mechanism in the more 
general case in which the gauge group is the direct product of several simple groups $G_i$. 
Possible Abelian factors are not take into account.

In 6d models, tree-level exchange of $\hat{B}^\alpha$ quanta can counterbalance one-loop 
anomalous diagrams. For this to be possible, the total anomaly polynomial must be of the 
form
\begin{equation} \label{eq:anomaly_polynomial_factorization}
\hat{I}_8 = \tfrac{1}{2} \Omega_{\alpha \beta} \hat{X}^\alpha_4 \wedge \hat{X}^\beta_4 \;,
\end{equation}
where 
\begin{equation}
\hat{X}^\alpha_4 = \tfrac{1}{2} a^\alpha \tr \hat{\mathcal{R}} \wedge \hat{\mathcal{R}} 
+ \sum_i 2 b^\alpha_i \, \lambda_i^{-1} \tr\!_f \hat{F}_i \wedge \hat{F}_i \;.
\end{equation}
In these expressions  $a^\alpha, \ b_i^\alpha$ transform as vectors in the space
$\mathbb{R}^{1,T}$ with symmetric inner product $\Omega_{\alpha\beta}$. Furthermore,
$\tr\!_f$ of $\hat{F}_i^2$ denotes the trace in the fundamental representation, and
$\lambda_i$ are normalization constants depending on the type of each
simple group factor. In the main text, this constant is always reabsorbed in
the normalization of the trace of field strengths, $\tr\! = \lambda^{-1}\tr\!_f$. We refer the reader to \cite{KumarTaylor} for 
the value of $\lambda$ for various simple gauge groups.

If condition (\ref{eq:anomaly_polynomial_factorization}) is met, 
the theory can be made anomaly-free by introduction of the generalized Green-Schwarz term
\begin{equation}
\hat{S}^{\mathrm{GS}} = -\int_{\cM_5} \tfrac{1}{2} \Omega_{\alpha\beta} \hat{B}^\alpha \wedge \hat{X}^\beta_4 \;.
\end{equation}

By computation of the anomaly polynomial $\hat{I}_8$ in terms of the chiral matter content 
and comparison with the factorized form (\ref{eq:anomaly_polynomial_factorization}), 
the following necessary conditions for anomaly cancellation are found:
\begin{align}
n_H-n_V &=  273-29 n_T \label{eq:hv}\\
0 &= B^i_{\rm adj} - \sum_{\bf R}
x^i_{\bf R} B^i_{\bf R} \label{eq:f4-condition}\\
\Omega_{\alpha \beta} a^\alpha a^\beta &=  9 - n_T  \label{eq:aa-condition}\\
-\Omega_{\alpha \beta} a^\alpha  b^\beta_i &=  \frac{1}{6} \lambda_i  \Big(  \sum_{\bf R}
x^i_{\bf R} A^i_{\bf R}-
A^i_{\rm adj} \Big)  \label{eq:ab-condition}\\
\Omega_{\alpha \beta} b^\alpha_i b^\beta_i &= \frac{1}{3} \lambda_i^2 \Big(  \sum_{\bf R} x_{\bf
  R}^i C^i_{\bf R}  -C^i_{\rm adj}\Big) & &\text{(no sum over $i$)} \label{eq:bb-equation}\\
\Omega_{\alpha \beta} b^\alpha_i  b^\beta_j &=   \lambda_i \lambda_j \sum_{\bf R S} x_{\bf R S}^{ij} A_{\bf R}^i
A_{\bf S}^j &  &\text{($i \neq j$).}   \label{eq:bij-condition}
\end{align}
In these equations, $n_H,n_V,n_T$ are the numbers of hyper-, vector and tensor multiplets 
in the model, $A_{\bf R},
B_{\bf R}, C_{\bf R}$ are group theory coefficients defined through
\begin{align}
\tr_{\bf R} \hat{F}^2 & = A_{\bf R}  \tr\!_f \hat{F}^2 \\
\tr_{\bf R} \hat{F}^4 & = B_{\bf R} \tr\!_f \hat{F}^4+C_{\bf R} (\tr\!_f\hat{F}^2)^2 \label{eq:bc-definition}\,,
\end{align}
and 
$x_{\bf R}^i$, $x_{\bf R S}^{ij}$
denote the number of matter fields that transform in the irreducible
representation ${\bf R}$ of gauge group factor $G_i$,
and $({\bf R} , {\bf S})$ of $G_i \times G_j$, respectively.
Note that for groups such as $SU(2)$ and $SU(3)$, which lack a fourth
order invariant, $B_{\bf R} = 0$ and there is no condition
\ref{eq:f4-condition}.


\section{Two-derivative 6d (1,0) supergravity on a circle} \label{appendix_6dto5d}

In this appendix we discuss the dimensional reduction of 6d $(1,0)$ supergravity at two-derivative level.
 Our starting point is therefore 
(\ref{6d_action}), which we write down again for convenience,
\begin{align}
\hat{S}^{(6)} = \int_{\cM_6} &+ \tfrac{1}{2} \hat{R} \hat{*} 1
 - \tfrac{1}{4} g_{\alpha \beta} \hat{G}^\alpha \wedge \hat{*} \hat{G}^\beta 
-\tfrac{1}{2} g_{\alpha\beta} dj^\alpha \wedge \hat{*} dj^\beta 
-h_{U  V} \hat \cD q^U \wedge \hat{*} \hat\cD  q^{V} \nn\\
&- 2 \Omega_{\alpha\beta} {j}^\alpha b^\beta \tr \hat{F} \wedge \hat{*} \hat{F} 
-  \Omega_{\alpha \beta} b^{\alpha} \hat{B}^\beta \wedge \tr \hat{F} \wedge \hat{F} - \hat V \hat * 1 \;.
\end{align}
The Kaluza-Klein Ansatz for the metric was given in (\ref{KK_metric}), while vectors 
and two-forms are expanded in 5d fields according to (\ref{KK_vector}), 
(\ref{KK_2form}). Consistently with our two-derivative approximation, 
we omit the gravitational contribution proportional to $a^\alpha$ in eq. (\ref{KK_2form}). 
This implies that the gravitational part is dropped in $G^\alpha$, too.

Standard dimensional reduction techniques can be applied to this pseudo-action, considered 
as a functional of both $A^\alpha$ and $B^\alpha$ independently. One computes
\begin{align} \label{5d_naive}
S^{(5)\rm F}_{\rm pseudo} = \int_{\cM_5} &+ \tfrac{1}{2} r \tilde{R} \; \tilde{*}1
 - \tfrac{1}{4} r^3 F^0 \wedge \tilde{*} F^0  
-\tfrac{1}{2} r g_{\alpha\beta} dj^\alpha \wedge \tilde{*} dj^\beta  
-r h_{U V} \cD q^U \wedge \tilde{*} \cD  q^{ V} \nn \\
& -2 r \Omega_{\alpha\beta} j^\alpha b^\beta \tr (F-\zeta F^0)\wedge \tilde{*}(F - \zeta F^0) 
 - 2 r^{-1} \Omega_{\alpha\beta} j^\alpha b^\beta \tr D\zeta \wedge \tilde{*} D\zeta \nn \\
& - \tfrac{1}{4} rg_{\alpha \beta} G^\alpha \wedge \tilde{*} G^\beta
 - \tfrac{1}{4} r^{-1} g_{\alpha \beta} \mathcal{F}^\alpha \wedge \tilde{*} \mathcal{F}^\beta \nn \\
& -\tfrac{1}{2} \Omega_{\alpha\beta} G^\alpha \wedge (\mathcal{F}^\beta - F^\beta) 
+  \Omega_{\alpha\beta} b^\alpha A^\beta \wedge \tr F \wedge F   \nn \\
& - 2 \Omega_{\alpha\beta} b^\alpha b^\beta \omega^{\mathrm{CS}} \wedge
 \left(2\tr \zeta F - \tr \zeta\zeta F^0 \right) \nn \\
& - 2 \Omega_{\alpha\beta} b^\alpha b^\beta \tr \zeta A \wedge 
\left( \tr F\wedge F - 2 \tr \zeta F \wedge F^0 + \tr \zeta \zeta F^0 \wedge F^0 \right) \nn \\
& - \big[ r \hat V + r^{-1} h_{U V} \zeta^I \zeta^J (T^{\bf R}_I q)^U (T^{\bf R}_J q)^{ V} \big] \tilde * 1 \;.
\end{align}
In this expression, $D\zeta = d\zeta + [A,\zeta]$ is the gauge covariant derivative for
the adjoint scalars $\zeta$, while $\cD q^U = dq^U + A^I (T^{\bf R}_I q)^U$ are the 5d gauge 
covariant derivatives for the scalars $q^U$ in the hypermultiplets. Furthermore,  
 we have introduced the shorthand notation
\begin{equation}
\mathcal{F}^\alpha = F^\alpha - 4 b^\alpha \tr \zeta F + 2 b^\alpha \tr \zeta\zeta F^0 \;.
\end{equation}

Dimensional reduction of the the self-duality constraint (\ref{6d_self}) gives
\begin{equation} \label{5d_self}
r g_{\alpha\beta} \tilde{*} G^\beta =- \Omega_{\alpha\beta} \mathcal{F}^\beta  \;,
\end{equation}
where the minus sign comes from our Ansatz \eqref{KK_2form}. This relation
means that  $A^\alpha$ and $B^\alpha$ encode the same physical degrees of freedom. 
Let us now discuss in detail how we can obtain a proper 5d action written in 
terms of vectors $A^\alpha$ only. The first step amounts to adding 
 a total derivative to the action above: $S^{(5)\rm F} = 
S^{(5)\rm F}_{\rm pseudo} + \Delta S^{(5)\rm F}$, where
\begin{align}
\Delta S^{(5)\rm F} &=\int_{\cM_5}- \tfrac{1}{2} \Omega_{\alpha\beta} dB^\alpha \wedge F^\beta \\
&= \int_{\cM_5} -\tfrac{1}{2} \Omega_{\alpha\beta} G^\alpha \wedge F^\beta
 +\tfrac{1}{2} \Omega_{\alpha\beta} (-A^\alpha F^0 + 2 b^\alpha \omega^{\mathrm{CS}}) \wedge F^\beta \;.
\end{align}
If we now consider $S^{(5)\rm F}$ as a functional of $G^\alpha$, $A^\alpha$, 
the equations of motion ensure both the self-duality condition (\ref{5d_self}) and the 
non-standard Bianchi identity (\ref{5d_bianchi}). Moreover, $G^\alpha$ enters 
$S^{(5)\rm F}$ only quadratically, and is therefore readily integrated out:
\begin{align} \label{5d_before_weyl}
S^{(5)\rm F} = \int_{\cM_5} &+ \tfrac{1}{2} r \tilde{R} \; \tilde{*}1 
- \tfrac{1}{4} r^3 F^0 \wedge \tilde{*} F^0  
-\tfrac{1}{2} r g_{\alpha\beta} dj^\alpha \wedge \tilde{*} dj^\beta 
 -r h_{U V} \cD q^U \wedge \tilde{*} \cD q^{V} \nn \\
& -2 r \Omega_{\alpha\beta} j^\alpha b^\beta \tr (F-\zeta F^0)\wedge \tilde{*}(F - \zeta F^0) 
 - 2 r^{-1} \Omega_{\alpha\beta} j^\alpha b^\beta \tr D\zeta \wedge \tilde{*} D\zeta \nn \\
& - \tfrac{1}{2} r^{-1} g_{\alpha \beta} \mathcal{F}^\alpha \wedge \tilde{*} \mathcal{F}^\beta 
-\tfrac{1}{2} \Omega_{\alpha\beta} A^0 \wedge F^\alpha \wedge F^\beta 
 +  2 \Omega_{\alpha\beta} b^\alpha A^\beta \wedge \tr F\wedge F  \nn \\
& - 2 \Omega_{\alpha\beta} b^\alpha b^\beta \omega^{\mathrm{CS}} \wedge 
\left(2\tr \zeta F - \tr \zeta\zeta F^0 \right) \nn \\
& -2  \Omega_{\alpha\beta} b^\alpha b^\beta \tr \zeta A \wedge 
\left( \tr F\wedge F - 2 \tr \zeta F \wedge F^0 + \tr \zeta \zeta F^0 \wedge F^0 \right) \nn \\
& - \big[ r \hat V + r^{-1} h_{U V} \zeta^I \zeta^J (T^{\bf R}_I q)^U (T^{\bf R}_J q)^{V}  \big] \tilde * 1 \;.
\end{align}
It is worth pointing out that $- \tfrac{1}{4} rg_{\alpha \beta} G^\alpha \wedge \tilde{*} G^\beta
 - \tfrac{1}{4} r^{-1} g_{\alpha \beta} \mathcal{F}^\alpha \wedge \tilde{*} \mathcal{F}^\beta$ 
vanishes identically after elimination of $G^\alpha$, and that the kinetic term for 
vectors $- \tfrac{1}{2} r^{-1} g_{\alpha \beta} \mathcal{F}^\alpha \wedge \tilde{*} \mathcal{F}^\beta$ 
comes from the Chern-Simons term $-\tfrac{1}{2} \Omega_{\alpha\beta} G^\alpha \wedge \mathcal{F}^\beta$. 
Moreover, the term $+ 2 \Omega_{\alpha\beta} b^\alpha A^\alpha \wedge \tr F\wedge F$ has a 
different prefactor because two different contributions must be taken into account: one was 
already present in (\ref{5d_naive}), the other one is found in $\Delta S^{(5)\rm F}$.

The last step consists of the Weyl rescaling $\tilde{g}_{\mu\nu} = r^{-2/3} g_{\mu\nu}$, 
which brings the Einstein-Hilbert term in (\ref{5d_before_weyl}) into canonical form:
\begin{align} \label{5d_action_KK_nonabelian}
S^{(5)\rm F} = \int_{\cM_5} &+ \tfrac{1}{2} {R} \; {*}1 -\tfrac{2}{3} r^{-2} dr \wedge *dr
 -\tfrac{1}{2}  g_{\alpha\beta} dj^\alpha \wedge {*} dj^\beta  \nn \\
&- 2 r^{-2} \Omega_{\alpha\beta} j^\alpha b^\beta \tr D\zeta \wedge {*} D\zeta 
 - h_{U  V} \cD q^U \wedge {*} \cD q^{V} \nn \\
& - \tfrac{1}{4} r^{8/3} F^0 \wedge {*} F^0 
 - \tfrac{1}{2} r^{-4/3} g_{\alpha \beta} \mathcal{F}^\alpha \wedge {*} \mathcal{F}^\beta  \nn \\
& -2 r^{2/3} \Omega_{\alpha\beta} j^\alpha b^\beta \tr (F-\zeta F^0) \wedge {*}(F - \zeta F^0)  \nn \\
& -\tfrac{1}{2} \Omega_{\alpha\beta} A^0 \wedge F^\alpha \wedge F^\beta 
 + 2 \Omega_{\alpha\beta} b^\alpha A^\beta \wedge \tr F \wedge F \nn \\
& - 2 \Omega_{\alpha\beta} b^\alpha b^\beta \omega^{\mathrm{CS}} \wedge 
\left(2\tr \zeta F - \tr \zeta\zeta F^0 \right) \nn \\
& - 2 \Omega_{\alpha\beta} b^\alpha b^\beta \tr \zeta A \wedge 
\left( \tr F\wedge F - 2 \tr \zeta F \wedge F^0 + \tr \zeta \zeta F^0 \wedge F^0 \right) \nn \\
&- \big[ r^{-1} \hat V + r^{-8/3} h_{U V} \zeta^I \zeta^J (T^{\bf R}_I q)^U (T^{\bf R}_J q)^{V}  \big]  * 1 \;.
\end{align}

As explained in subsection \ref{56Coulomb}, we are interested in the broken phase of the theory
corresponding to the Coulomb branch of the gauge sector. The fields which acquire a mass 
during the spontaneous breaking of gauge symmetry are omitted from the final 5d effective action. 
These include W-bosons and charged hypermultiplet scalars. As a consequence, 
the lower-case indices $u,v$ now only run over neutral hypermultiplets. For the same reason, 
the scalar potential is omitted.
The final form of the effective action in the Coulomb branch thus reads
\begin{align} \label{5d_action_KK_coulomb}
S^{(5)\rm F} = \int_{\cM_5} &+ \tfrac{1}{2} {R} \; {*}1 -\tfrac{2}{3} r^{-2} dr \wedge *dr
 -\tfrac{1}{2}  g_{\alpha\beta} dj^\alpha \wedge {*} dj^\beta  \nn \\
&- 2 r^{-2} \Omega_{\alpha\beta}  j^\alpha b^\beta C_{ij} d\zeta^i \wedge {*} d\zeta^j 
 - h_{u  v} d q^u \wedge {*} d q^{ v} \nn \\
& - \tfrac{1}{4} r^{8/3} F^0 \wedge {*} F^0 
 - \tfrac{1}{2} r^{-4/3} g_{\alpha \beta} \mathcal{F}^\alpha \wedge {*} \mathcal{F}^\beta  \nn \\
& -2 r^{2/3} \Omega_{\alpha\beta} C_{ij} j^\alpha b^\beta (F^i-\zeta^i F^0) \wedge {*}(F^j - \zeta^j F^0)  \nn \\
& -\tfrac{1}{2} \Omega_{\alpha\beta} A^0 \wedge F^\alpha \wedge F^\beta 
 + 2 \Omega_{\alpha\beta} C_{ij} b^\alpha A^\beta \wedge F^i \wedge F^j \nn \\
& -2 (\Omega_{\alpha \beta} b^\alpha b^\beta) (C_{kl}\zeta^k \zeta^l) C_{ij}  \zeta^i A^j \wedge F^0 \wedge F^0 \nn \\
& +2 (\Omega_{\alpha \beta} b^\alpha b^\beta) (C_{ij} C_{kl} \zeta^k \zeta^l + 2 C_{ik}C_{jl} \zeta^k \zeta^l ) A^i \wedge F^j \wedge F^0 \nn \\
& -6 (\Omega_{\alpha \beta} b^\alpha b^\beta) C_{(ij} C_{k)l} \zeta^l  A^i \wedge F^j \wedge F^k \;.
\end{align}



\section{Calabi-Yau reference formulae} \label{appendix_calabiyau}

The main purpose of this appendix is fixing some notation about Calabi-Yau threefolds. 
Therefore, it is not meant to be complete nor self-contained. 
We refer the reader to e.g.~\cite{UTTG-07-90,hep-th/9702155}
for a more detailed account of the material covered hereafter.

A Calabi-Yau threefold $Y_3$ can be described locally either by means of six real coordinates $\{\xi^{\hat\imath}\}_{\hat\imath = \hat 1 ,\dots \hat 6}$,
or by means of three complex coordinates $\{y^i\}_{i=1,2,3}$, defined as
\begin{align}
y^1 &= \tfrac{1}{\sqrt{2}} \left( \xi^{\hat{1}} + i \xi^{\hat{2}} \right)\,, &
y^2 &= \tfrac{1}{\sqrt{2}} \left( \xi^{\hat{3}} + i \xi^{\hat{4}} \right)\,, &
y^3 &= \tfrac{1}{\sqrt{2}} \left( \xi^{\hat{5}} + i \xi^{\hat{6}} \right) \;.
\end{align}
In the following, we will make use of complex coordinates, and their associated holomorphic
indices $i,j,... = 1,2,3$ and antiholomorphic indices  $\ib, \jb,... = \bar{1}, \bar{2}, \bar{3}$.
Accordingly, a complex differential form of degree $(r,s)$ is expanded on the basis of differentials 
of complex coordinates as
\begin{equation}
\alpha = \tfrac{1}{r!s!} \alpha_{i_1 \dots i_r \jb_1 \dots \jb_s} \; dy^{i_1} \wedge \dots
 \wedge dy^{i_r} \wedge d\bar{y}^{\jb_1} \wedge \dots \wedge d\bar{y}^{\jb_s} \;.
\end{equation}

Being a K\"ahler threefold, $Y_3$ is endowed with an Hermitian metric $g_{i\jb}$, whose
K\"{a}hler $(1,1)$-form $J = i g_{i\jb} \; dy^i \wedge d\bar{y}^{\jb}$
is closed. The Calabi-Yau condition ensures the existence of a globally defined, non-vanishing,
holomorphic $(3,0)$-form, which we denote by $\Omega$. 
The volume form, the K\"{a}hler form and the holomorphic $(3,0)$-form are related by
\begin{equation}
*1 = \tfrac{1}{3!} J\wedge J \wedge J = \tfrac{i}{\| \Omega \|^2} \Omega \wedge \bar{\Omega} \;, \qquad
\text{where} \qquad \| \Omega \| ^2 = \tfrac{1}{3!} \Omega_{i j k} \bar{\Omega}^{ijk} \;.
\end{equation}

Since our definition of a Calabi-Yau threefold implies strict $SU(3)$ holonomy, the only
independent Hodge numbers of $Y_3$ are $h^{1,1}(Y_3)$, $h^{1,2}(Y_3)$. Let us
fix our notations for the corresponding cohomology basis.

First of all, we choose an integral cohomology basis
$\{ \omega_\Lambda \}_\Lambda$ for $H^{1,1}(Y_3)$, with $\Lambda=1,\dots,h^{1,1}(Y_3)$. 
The intersection
numbers associated to this basis $\{\omega_\Lambda\}_{\Lambda}$ are
\begin{equation}
 \mathcal{V}_{\Lambda \Sigma \Theta} =\int_{Y_3} \omega_\Lambda \wedge \omega_\Sigma \wedge \omega_\Theta \;.
\end{equation}

Second of all, we take $H^{2,1}(Y_3)$ to be generated 
by the complex cohomology basis $\{\chi_\kappa \}_{\kappa}$, where $\kappa = 1,\dots,h^{1,2}(Y_3)$. 
It is also useful to consider an integral basis $\{ \alpha_K, \beta^K\}_{K}$ 
for the middle cohomology $H^3(Y_3)$, with $K=1,\dots, h^{1,2}(Y_3)+1$. Since three-forms anticommute, it is natural to introduce a symplectic 
structure on $H^3(Y_3)$ choosing $\alpha_K, \beta^K$ in such a way that
\begin{equation}
\int_{Y_3} \alpha_K \wedge \beta^L = \delta^K_L   \;.
\end{equation}
In order to discuss the metric on the moduli space of neutral hypermultiplets, we need to
 introduce matrices $A_K^{\phantom{K}L},B_{KL},C^{KL}$, such that
\begin{equation}
*\alpha_K  = A_K^{\phantom{K}L} \alpha_L + B_{KL} \beta^L \;, \qquad
* \beta ^K = C^{KL} \alpha_L - A_{L}^{\phantom{L}K} \beta^L \; ,
\end{equation}
where $*$ represents the Hodge star in $Y_3$. These matrices can be conveniently expressed in terms of a symmetric, complex matrix $\mathcal{M}$:
\begin{align}
A_K^{\phantom{K}L} & = (\mathrm{Re}\mathcal{M})_{KH} ( \mathrm{Im} \mathcal{M} )^{-1\, HL} \;, \\
B_{KL} &= - ( \mathrm{Im} \mathcal{M} )_{KL} 
- (\mathrm{Re}\mathcal{M})_{KH} ( \mathrm{Im} \mathcal{M} )^{-1\, HM} (\mathrm{Re}\mathcal{M})_{ML} \;, \\
C^{KL} &= ( \mathrm{Im} \mathcal{M} )^{-1\, KL} \;.
\end{align}

Let us now give a brief account on the moduli space of Calabi-Yau threefold $Y_3$. It is known that it
 can be written locally in a 
factorized form as the product of the K\"{a}hler moduli space and the complex 
structure moduli space: $\mathcal{M} = \mathcal{M}_{\text{K}} \times \mathcal{M}_{\text{cs}}$. 
We discuss each factor in turn.

On the one hand, the K\"ahler moduli space $\mathcal{M}_{\text{K}}$ is parameterized by the K\"ahler moduli $v^\Lambda$ which appear
in the expansion of the K\"ahler form $J$ on the basis $\{\omega_\Lambda\}_{\Lambda}$,
\begin{equation}
J = v^\Lambda \omega_\Lambda \;.
\end{equation}
Fluctuations of the K\"ahler moduli encode information about the variation of the mixed components of the Ricci-flat metric
as we move around in the moduli space of $Y_3$, as specified by
\begin{equation} \label{mixed_indices}
 \delta g_{i \jb} = -i  (\omega_\Lambda)_{i\jb} \, \delta v^\Lambda \;.
\end{equation}
The K\"ahler moduli $v^\Lambda$ also appear in the expression of the volume $\mathcal{V}$ of $Y_3$,
\begin{equation}
 \mathcal{V} = \tfrac{1}{3!}\int_{Y_3} J\wedge J \wedge J = \tfrac{1}{3!} \mathcal{V}_{\Lambda \Sigma \Theta} v^\Lambda v^\Sigma v^\Theta \;.
\end{equation}
For convenience, we introduce the shorthand notation
\begin{equation}
 \mathcal{V}_\Lambda = \tfrac{1}{2!} \mathcal{V}_{\Lambda \Sigma \Theta} v^\Sigma v^\Theta = 
\partial_{v^\Lambda} \mathcal{V} \;, \qquad 
\mathcal{V}_{\Lambda \Sigma} = \mathcal{V}_{\Lambda \Sigma \Theta}  v^\Theta = 
\partial_{v^\Lambda} \partial_{v^\Sigma} \mathcal{V} \;.
\end{equation}

On the other hand, the complex structure moduli space $\mathcal{M}_{\text{cs}}$ is described by suitable complex coordinates $z^\kappa$.
They are obtained as periods of the holomorphic $(3,0)$-form $\Omega$, and their 
fluctuations correspond to variations of the components of the Ricci-flat metric
with the same kind of indices. More precisely, $\bar\chi_{\bar \kappa} \in H^{1,2}(Y_3)$
are used to construct $\bar b_{\bar \kappa} \in H^{0,1}(Y_3;TY_3^{1,0})$, where $TY_3^{1,0}$ is
the holomorphic tangent bundle to $Y_3$, and the $\bar b_{\bar \kappa}$ encode the metric fluctuations. In our conventions, we have 
\begin{equation} \label{same_indices}
  \delta g_{ij} = (\bar{b}_{\bar{\kappa}})_{ij} \, \delta \bar{z}^{\bar{\kappa}} \;, \qquad
(\bar{b}_{\bar{\kappa}})_{i}^{\phantom{a}\jb} = \tfrac{i}{\| \Omega \|^2} (\bar{\chi}_{\bar{\kappa}})_{i \bar{k}\bar{l}}
 \Omega^{\bar{k}\bar{l} \jb} \;.
\end{equation}

Both moduli spaces $\mathcal{M}_{\rm K}$ and $\cM_{\rm cs}$ are equipped with a natural metric, which can be derived 
from a potential. These potentials are determined by
\begin{equation}
e^{\mathcal{K}_{\text{K}}} = \int_{Y_3} \tfrac{1}{3!} J\wedge J \wedge J \;, \qquad
e^{\mathcal{K}_{\text{cs}}} = i \int_{Y_3} \Omega \wedge \bar{\Omega} \;,
\end{equation}
and yield the metrics
\begin{align} 
G_{\Lambda \Sigma}(v) &= -\tfrac{1}{2} \partial_{v^\Lambda} \partial_{v^\Sigma} \mathcal{K}_{\text{K}}(v) = 
\tfrac{1}{2\mathcal{V}} \int_{Y_3} \omega_\Lambda \wedge * \omega_\Sigma = 
\tfrac{1}{2} \tfrac{\mathcal{V}_\Lambda \mathcal{V}_\Sigma }{\mathcal{V}^2} -
 \tfrac{1}{2} \tfrac{\mathcal{V}_{\Lambda \Sigma}}{\mathcal{V}} \;, \nn \\
\label{kaeler-metrics}
g_{\kappa \bar{\kappa}}(z,\bar{z}) &= 
\partial_{z^\kappa} \partial_{\bar{z}^{\bar{\kappa}}} \mathcal{K}_{\text{cs}}(z,\bar{z}) =
 - \tfrac{\int_{Y_3} \chi_\kappa \wedge \bar{\chi}_{\bar\kappa} }{\int_{Y_3} \Omega \wedge \bar{\Omega}} \;.
\end{align}


\section{11d supergravity on a Calabi-Yau threefold} \label{appendix_11d5dreduction}

This appendix is devoted to the presentation of the key points of the Kaluza-Klein
reduction of 11d supergravity on a Calabi-Yau threefold $Y_3$. Zero-modes only are taken into account.
For ease of reference we record again the 11d supergravity action
\begin{equation}
\hat{S}^{(11)} =\int_{\cM_11} \tfrac{1}{2} \hat{R} \hat{*} 1 - \tfrac{1}{4} \hat{F}_4 \wedge \hat{*} \hat{F}_4
 - \tfrac{1}{12} \hat{C}_3 \wedge \hat{F}_4 \wedge \hat{F}_4 \;,
\end{equation}
the Kaluza-Klein Ansatz for the three-form 
\begin{equation}
  \hat{C}_3 = \xi^K \alpha_K - \tilde{\xi}_K \beta^K + A^\Lambda \wedge \omega_\Lambda + C_3 \;,
\end{equation}
and the background line element
\begin{equation}
\langle d\hat{s}^2 \rangle = \langle \tilde{g}_{\mu\nu}(x) \rangle dx^\mu dx^\nu + 2 \langle g_{i \jb}(y) \rangle dy^i d\bar y^\jb \;.
\end{equation}

Let us starts discussing the reduction of the Einstein-Hilbert term. The full internal metric $g_{i \jb}$, background and fluctuations,
depends on the external coordinates through the K\"ahler moduli $v^\Lambda$ and the complex structure moduli $z^\kappa$, as 
can be seen from \eqref{mixed_indices}, \eqref{same_indices}. Note that the off-diagonal $dx\; dy$ components of the background metric
cannot fluctuate since a Calabi-Yau threefold has no continuous isometries. In order to get a 5d Einstein-Hilbert term with canonical
normalization, we have to perform the Weyl rescaling
\begin{equation}
\tilde{g}_{\mu\nu} = \mathcal{V}^{-2/3} g_{\mu\nu} \;.
\end{equation}
Straightforward calculation gives then
\begin{equation}
  \int_{\cM_{11}} \tfrac{1}{2} \hat{R} \hat{*} 1 = \int_{\cM_{5}} \tfrac{1}{2} R * 1 - \tfrac{1}{2} H_{\Lambda\Sigma}(v) 
dv^\Lambda \wedge * dv^\Sigma - g_{\kappa \bar\kappa} dz^\kappa \wedge * d\bar{z}^{\bar\kappa}
\end{equation}
where $g_{\kappa\bar\kappa}$ is the metric in the complex moduli space, defined in \eqref{kaeler-metrics}, and
\begin{equation}
H_{\Lambda \Sigma}(v) = -G_{\Lambda\Sigma}(v) - \cV^{-1} \mathcal{V}_{\Lambda\Sigma} \;.
\end{equation}
We have singled out the expression
\begin{equation} \label{preliminary-metric}
G_{\Lambda \Sigma}(v) = -\tfrac{1}{2} \partial_{v^\Lambda} \partial_{v^\Sigma} \log \mathcal{V}(v) =
-\tfrac{1}{2 } \cV(v)^{-1} \cV_{\Lambda \Sigma \Theta} v^\Theta + \tfrac{1}{8}\cV(v)^{-2} \cV_{\Lambda \Omega \Theta} 
\cV_{\Sigma \Psi \Xi}  v^\Omega v^\Theta v^\Psi v^\Xi
\end{equation}
because it is the natural metric on the K\"{a}hler moduli space. It is useful to define 
\begin{equation}
 L^\Lambda = \mathcal{V}^{-1/3} v^\Lambda
\end{equation}
since, as we shall see, the kinetic term for $v^\Lambda$ expressed in these coordinates takes a particularly
simple form. It is crucial to observe that the $L^\Lambda$'s parameterize one degree of freedom less than the $v^\Lambda$'s, since they
obey identically $\tfrac{1}{3!} \mathcal{V}_{\Lambda \Sigma \Theta} L^\Lambda L^\Sigma L^\Theta =1$. The kinetic term we are interested in reads
\begin{equation}
 - \tfrac{1}{2} H_{\Lambda\Sigma}(v) 
dv^\Lambda \wedge * dv^\Sigma = - \tfrac{1}{2} G_{\Lambda\Sigma}(L) dL^\Lambda \wedge * dL^\Sigma 
- dD \wedge *dD  \;.
\end{equation}
In this expression, we have defined
\begin{equation} \label{dilaton}
D= -\tfrac{1}{2} \log \mathcal{V}
\end{equation}
for future convenience, and we have introduced the symbol $G_{\Lambda\Sigma}(L)$ to denote the metric obtained by 
replacing $v^\Lambda$ by $L^\Lambda$ everywhere in \eqref{preliminary-metric}. It is easily checked that
$G_{\Lambda\Sigma}(L)$ can be written in a compact form as
\begin{equation} 
G_{\Lambda\Sigma}(L) = \left[ -\tfrac{1}{2} \partial_{L^\Lambda} \partial_{L^\Sigma} \log \mathcal{N} \right]_{\mathcal{N}=1} 
= \left[ -\tfrac{1}{2}  \cN_{\Lambda \Sigma}  + \tfrac{1}{2}  \cN_\Lambda \cN_\Sigma \right]_{\cN =1}
\end{equation}
provided that we introduce $\mathcal{N} = \tfrac{1}{3!} \mathcal{V}_{\Lambda \Sigma \Theta} L^\Lambda L^\Sigma L^\Theta$.

We are now in a position to describe the reduction of the other terms in the 11d Lagrangian.
As far as the three-form kinetic term is concerned, a 
straightforward computation shows that
\begin{align}
\int_{\cM_{11}} - \tfrac{1}{4} \hat{F}_4 \wedge \hat{*} \hat{F}_4 = \int_{\cM_5} & + \tfrac{1}{4} 
(d\tilde{\xi}_{K} - \mathcal{M}_{KM} d\xi^M)(\mathrm{Im} \mathcal{M})^{-1\, KL } \wedge 
\tilde{*}(d\tilde{\xi}_{L} - \mathcal{M}_{LN} d\xi^N) \nn \\
&-\tfrac{1}{2} \mathcal{V} G_{\Lambda \Sigma}(v) F^\Lambda \wedge \tilde{*} F^\Sigma 
-\tfrac{1}{4} \mathcal{V} F_4 \wedge \tilde{*} F_4  \;.
\end{align}
See appendix \ref{appendix_calabiyau} for the definition of $\mathcal M$.
For the Chern-Simons term, we find
\begin{equation}
\int_{\cM_{11}}- \tfrac{1}{12} \hat{C}_3 \wedge \hat{F}_4 \wedge \hat{F}_4  = \int_{\cM_5}
 -\tfrac{1}{12} \mathcal{V}_{\Lambda \Sigma \Theta} A^\Lambda \wedge F^\Sigma \wedge F^\Theta
 + \tfrac{1}{4} (\xi^K d\tilde{\xi}_K - \tilde{\xi}_Kd\xi^K) \wedge F_4 \;.
\end{equation}

As mentioned in the main text, we can dualize the three-form $C_3$ into a real 
scalar $\Phi$. To this end we add to the 5d action the term
\begin{equation}
\Delta S^{(5)\rm M} = \int_{\cM_5} \tfrac{1}{4} d\Phi \wedge F_4
\end{equation}
which implements Bianchi identity $dF_4=0$ if we consider $F_4$ rather than $C_3$ as 
independent variable. After elimination of $F_4$ via its equation of motion, we get
\begin{align}
S^{(5)\rm M}_{\text{non-grav}} = \int_{\cM_5} &+ \tfrac{1}{4} (d\tilde{\xi}_{K} 
- \mathcal{M}_{KM} d\xi^M)(\mathrm{Im} \mathcal{M})^{-1\, KL } \wedge \tilde{*}(d\tilde{\xi}_{L}
 - \mathcal{M}_{LN} d\xi^N) \\
&-\tfrac{1}{2} \mathcal{V} G_{\Lambda \Sigma}(v) F^\Lambda \wedge \tilde{*} F^\Sigma
 -\tfrac{1}{12} \mathcal{V}_{\Lambda \Sigma \Theta} A^\Lambda \wedge F^\Sigma \wedge F^\Theta \nn \\
& -\tfrac{1}{16\mathcal{V}} \left[\xi^K d\tilde{\xi}_K - \tilde{\xi}_Kd\xi^K 
+ d\Phi \right] \wedge \tilde{*} \left[\xi^K d\tilde{\xi}_K - \tilde{\xi}_Kd\xi^K + d\Phi \right] \;.\nn
\end{align}

Let us stress here  that we still have to take into account the Weyl rescaling of the metric $\tilde{g}_{\mu\nu}$.
 It is interesting to note that it is crucial to get the equality between the inverse gauge coupling function and
the metric of the moduli space of scalars $L^\Lambda$, since
\begin{equation}
-\tfrac{1}{2} \mathcal{V} G_{\Lambda \Sigma}(v) F^\Lambda \wedge \tilde{*} F^\Sigma = 
-\tfrac{1}{2}\mathcal{V}^{\tfrac{2}{3}} G_{\Lambda \Sigma}(v) F^\Lambda \wedge * F^\Sigma =
 -\tfrac{1}{2} G_{\Lambda \Sigma}(L) F^\Lambda \wedge * F^\Sigma \;.
\end{equation}

The final action was given in the main text in \eqref{5d-action-M}. We only need to specify the 
quaternionic kinetic for hypermultiplets, which turns out to be
\begin{align}
  h_{u v} dq^u \wedge * d q^{v} = &+ dD \wedge *dD 
+ g_{\kappa\bar\kappa} dz^\kappa \wedge *d\bar{z}^{\bar\kappa} \\
&+\tfrac{1}{4}e^{4D} \left[ d \Phi + (\xi^K d\tilde{\xi}_K 
- \tilde{\xi}_K d\xi^{K}) \right]^2 \nn \\
& - \tfrac{1}{2} e^{2D} (d\tilde{\xi}_{K} 
- \mathcal{M}_{KM} d\xi^M)(\mathrm{Im} \mathcal{M})^{-1\, KL } (d\tilde{\xi}_{L} 
- \mathcal{M}_{LN} d\xi^N)\nn \;.
\end{align}



\begin{thebibliography}{99}

\bibitem{Reports}
 M.~R.~Douglas and S.~Kachru,
  ``Flux compactification,''
  Rev.\ Mod.\ Phys.\ \ {\bf 79} (2007) 733
  [hep-th/0610102];\\[.1cm]
  R.~Blumenhagen, B.~Kors, D.~Lust and S.~Stieberger,
  ``Four-dimensional String Compactifications with D-Branes, Orientifolds and Fluxes,''
  Phys.\ Rept.\ \ {\bf 445} (2007) 1
  [hep-th/0610327].

\bibitem{Denef:2008wq}
  F.~Denef,
  ``Les Houches Lectures on Constructing String Vacua,''
  [arXiv:0803.1194 [hep-th]].

\bibitem{Taylor:2011wt}
  W.~Taylor,
  ``TASI Lectures on Supergravity and String Vacua in Various Dimensions,''
  [arXiv:1104.2051 [hep-th]].


\bibitem{hep-th/9611100}
  P.~Pasti, D.~P.~Sorokin and M.~Tonin,
  ``On Lorentz invariant actions for chiral p forms,''
  Phys.\ Rev.\ D\ {\bf 55} (1997) 6292
  [hep-th/9611100].



\bibitem{hep-th/9703075}
  H.~Nishino and E.~Sezgin,
  ``New couplings of six-dimensional supergravity,''
  Nucl.\ Phys.\ B\ {\bf 505} (1997) 497
  [hep-th/9703075].

\bibitem{Riccioni:1997ik}
  S.~Ferrara, F.~Riccioni and A.~Sagnotti,
  ``Tensor and vector multiplets in six-dimensional supergravity,''
  Nucl.\ Phys.\ B\ {\bf 519} (1998) 115
  [hep-th/9711059];\\[.1cm]
  F.~Riccioni, A.~Sagnotti,
  ``Some properties of tensor multiplets in six-dimensional supergravity,''
  Nucl.\ Phys.\ Proc.\ Suppl.\  {\bf 67 } (1998)  68-73.
  [hep-th/9711077];\\[.1cm]
  F.~Riccioni,
  ``Abelian vector multiplets in six-dimensional supergravity,''
  Phys.\ Lett.\ B\ {\bf 474} (2000) 79
  [hep-th/9910246].

\bibitem{arXiv:1012.1818}
  M.~Gunaydin, H.~Samtleben and E.~Sezgin,
  ``On the Magical Supergravities in Six Dimensions,''
  Nucl.\ Phys.\ B\ {\bf 848} (2011) 62
  [arXiv:1012.1818 [hep-th]].

\bibitem{Samtleben:2011fj}
  H.~Samtleben, E.~Sezgin, R.~Wimmer,
  ``(1,0) superconformal models in six dimensions,''
  [arXiv:1108.4060 [hep-th]].


\bibitem{Douglas:2010iu}
  M.~R.~Douglas,
  ``On D=5 super Yang-Mills theory and (2,0) theory,''
  JHEP {\bf 1102}, 011 (2011).
  [arXiv:1012.2880 [hep-th]].

\bibitem{Lambert:2010iw}
  N.~Lambert, C.~Papageorgakis, M.~Schmidt-Sommerfeld,
  ``M5-Branes, D4-Branes and Quantum 5D super-Yang-Mills,''
  JHEP {\bf 1101 } (2011)  083.
  [arXiv:1012.2882 [hep-th]].


\bibitem{arXiv:1104.4040} %
  P.~-M.~Ho, K.~-W.~Huang and Y.~Matsuo,
  ``A Non-Abelian Self-Dual Gauge Theory in 5+1 Dimensions,''
  JHEP\ {\bf 1107} (2011) 021
  [arXiv:1104.4040 [hep-th]].


\bibitem{Green:1984sg}
  M.~B.~Green, J.~H.~Schwarz,
  ``Anomaly Cancellation in Supersymmetric D=10 Gauge Theory and Superstring Theory,''
  Phys.\ Lett.\  {\bf B149 } (1984)  117-122.

\bibitem{Sagnotti:1992qw}
  A.~Sagnotti,
  ``A Note on the Green-Schwarz mechanism in open string theories,''
  Phys.\ Lett.\  {\bf B294 } (1992)  196-203.
  [hep-th/9210127].

\bibitem{Schwarz:1995zw}
  J.~H.~Schwarz,
  ``Anomaly - free supersymmetric models in six-dimensions,''
  Phys.\ Lett.\  {\bf B371 } (1996)  223-230.
  [hep-th/9512053].

\bibitem{Vafa:1996xn}
  C.~Vafa,
  ``Evidence for F theory,''
  Nucl.\ Phys.\  {\bf B469 } (1996)  403-418.
  [hep-th/9602022].

\bibitem{Morrison:1996na}
  D.~R.~Morrison, C.~Vafa,
  ``Compactifications of F theory on Calabi-Yau threefolds. 1,''
  Nucl.\ Phys.\  {\bf B473 } (1996)  74-92.
  [hep-th/9602114];\\[.1cm]
  D.~R.~Morrison, C.~Vafa,
  ``Compactifications of F theory on Calabi-Yau threefolds. 2.,''
  Nucl.\ Phys.\  {\bf B476 } (1996)  437-469.
  [hep-th/9603161].

\bibitem{hep-th/9604097}
  S.~Ferrara, R.~Minasian and A.~Sagnotti,
  ``Low-energy analysis of M and F theories on Calabi-Yau threefolds,''
  Nucl.\ Phys.\ B\ {\bf 474} (1996) 323
  [hep-th/9604097].



\bibitem{Sadov:1996zm}
  V.~Sadov,
  ``Generalized Green-Schwarz mechanism in F theory,''
  Phys.\ Lett.\  {\bf B388 } (1996)  45-50.
  [hep-th/9606008].


\bibitem{KumarTaylor}
  V.~Kumar and W.~Taylor,
  ``String Universality in Six Dimensions,''
  arXiv:0906.0987 [hep-th];\\[.1cm]
  V.~Kumar, D.~R.~Morrison and W.~Taylor,
``Mapping 6D N = 1 supergravities to F-theory,''
  JHEP\ {\bf 1002} (2010) 099
  [arXiv:0911.3393 [hep-th]];\\[.1cm]
  V.~Kumar, D.~R.~Morrison, W.~Taylor,
  ``Global aspects of the space of 6D N = 1 supergravities,''
  JHEP {\bf 1011 } (2010)  118.
  [arXiv:1008.1062 [hep-th]];\\[.1cm]
  V.~Kumar, D.~S.~Park and W.~Taylor,
  ``6D supergravity without tensor multiplets,''
  JHEP\ {\bf 1104} (2011) 080
  [arXiv:1011.0726 [hep-th]].


\bibitem{arXiv:1110.5916}
  D.~S.~Park and W.~Taylor,
  ``Constraints on 6D Supergravity Theories with Abelian Gauge Symmetry,''
  arXiv:1110.5916 [hep-th].

\bibitem{arXiv:1111.2351}
  D.~S.~Park,
  ``Anomaly Equations and Intersection Theory,''
  arXiv:1111.2351 [hep-th].

\bibitem{Bershadsky:1996nh}
  M.~Bershadsky, K.~A.~Intriligator, S.~Kachru, D.~R.~Morrison, V.~Sadov, C.~Vafa,
  ``Geometric singularities and enhanced gauge symmetries,''
  Nucl.\ Phys.\  {\bf B481 } (1996)  215-252.
  [hep-th/9605200].


\bibitem{hep-th/9606086}
  S.~H.~Katz and C.~Vafa,
  ``Matter from geometry,''
  Nucl.\ Phys.\ B\ {\bf 497} (1997) 146
  [hep-th/9606086].

\bibitem{arXiv:1106.3563}
  D.~R.~Morrison and W.~Taylor,
  ``Matter and singularities,''
  arXiv:1106.3563 [hep-th].


\bibitem{arXiv:1008.4133} 
  T.~W.~Grimm,
  ``The N=1 effective action of F-theory compactifications,''
  Nucl.\ Phys.\ B\ {\bf 845}, 48  (2011)
  [arXiv:1008.4133 [hep-th]].


\bibitem{Grimm:2011fx}
  T.~W.~Grimm, H.~Hayashi,
  ``F-theory fluxes, Chirality and Chern-Simons theories,'' 
  [arXiv:1111.1232 [hep-th]].

\bibitem{Cadavid:1995bk}
  A.~C.~Cadavid, A.~Ceresole, R.~D'Auria, S.~Ferrara,
  ``Eleven-dimensional supergravity compactified on Calabi-Yau threefolds,''
  Phys.\ Lett.\  {\bf B357 } (1995)  76-80.
  [hep-th/9506144].


\bibitem{Antoniadis:1997eg}
  I.~Antoniadis, S.~Ferrara, R.~Minasian, K.~S.~Narain,
  ``R**4 couplings in M and type II theories on Calabi-Yau spaces,''
  Nucl.\ Phys.\  {\bf B507 } (1997)  571-588.
  [hep-th/9707013].

\bibitem{GrimmTaylor}
 T.~W.~Grimm, W.~Taylor, \textit{work in progress.}

\bibitem{Witten:1996qb}
  E.~Witten,
  ``Phase transitions in M theory and F theory,''
  Nucl.\ Phys.\  {\bf B471 } (1996)  195-216.
  [hep-th/9603150].

\bibitem{Intriligator:1997pq}
  K.~A.~Intriligator, D.~R.~Morrison, N.~Seiberg,
  ``Five-dimensional supersymmetric gauge theories and degenerations of Calabi-Yau spaces,''
  Nucl.\ Phys.\  {\bf B497 } (1997)  56-100.

\bibitem{hep-th/9704097}
  P.~Candelas, E.~Perevalov and G.~Rajesh,
  ``Toric geometry and enhanced gauge symmetry of F theory / heterotic vacua,''
  Nucl.\ Phys.\ B\ {\bf 507} (1997) 445
  [hep-th/9704097].

\bibitem{Ferrara:1989ik}
  S.~Ferrara, S.~Sabharwal,
  ``Quaternionic Manifolds for Type II Superstring Vacua of Calabi-Yau Spaces,''
  Nucl.\ Phys.\  {\bf B332 } (1990)  317.

\bibitem{Andrianopoli:1996cm}
  L.~Andrianopoli, M.~Bertolini, A.~Ceresole, R.~D'Auria, S.~Ferrara, P.~Fre, T.~Magri,
  ``N=2 supergravity and N=2 superYang-Mills theory on general scalar manifolds: Symplectic covariance, gaugings and the momentum map,''
  J.\ Geom.\ Phys.\  {\bf 23 } (1997)  111-189.
  [arXiv:hep-th/9605032 [hep-th]].


\bibitem{IC-84-218}
  S.~Randjbar-Daemi, A.~Salam, E.~Sezgin and J.~A.~Strathdee,
  ``An Anomaly Free Model in Six-Dimensions,''
  Phys.\ Lett.\ B\ {\bf 151} (1985) 351.

\bibitem{hep-th/0504033}
  S.~D.~Avramis, A.~Kehagias and S.~Randjbar-Daemi,
  ``A New anomaly-free gauged supergravity in six dimensions,''
  JHEP\ {\bf 0505} (2005) 057
  [hep-th/0504033].

\bibitem{hep-th/0508172}
  S.~D.~Avramis and A.~Kehagias,
  ``A Systematic search for anomaly-free supergravities in six dimensions,''
  JHEP\ {\bf 0510} (2005) 052
  [hep-th/0508172].

\bibitem{hep-th/0512019}
  R.~Suzuki and Y.~Tachikawa,
  ``More anomaly-free models of six-dimensional gauged supergravity,''
  J.\ Math.\ Phys.\ \ {\bf 47} (2006) 062302
  [hep-th/0512019].

\bibitem{Ceresole:2000jd}
  A.~Ceresole, G.~Dall'Agata,
  ``General matter coupled N=2, D = 5 gauged supergravity,''
  Nucl.\ Phys.\  {\bf B585 } (2000)  143-170.
  [hep-th/0004111].

\bibitem{Gunaydin:1983bi}
  M.~Gunaydin, G.~Sierra, P.~K.~Townsend,
  ``The Geometry of N=2 Maxwell-Einstein Supergravity and Jordan Algebras,''
  Nucl.\ Phys.\  {\bf B242 } (1984)  244.


\bibitem{Hanaki:2006pj}
  K.~Hanaki, K.~Ohashi, Y.~Tachikawa,
  ``Supersymmetric Completion of an R**2 term in Five-dimensional Supergravity,''
  Prog.\ Theor.\ Phys.\  {\bf 117 } (2007)  533.
  [hep-th/0611329].

\bibitem{Cremonini:2008tw}
  S.~Cremonini, K.~Hanaki, J.~T.~Liu, P.~Szepietowski,
  ``Black holes in five-dimensional gauged supergravity with higher derivatives,''
  JHEP {\bf 0912 } (2009)  045.
  [arXiv:0812.3572 [hep-th]].

\bibitem{Cremmer:1978km}
  E.~Cremmer, B.~Julia, J.~Scherk,
  ``Supergravity Theory in Eleven-Dimensions,''
  Phys.\ Lett.\  {\bf B76 } (1978)  409-412.

\bibitem{arXiv:1109.3191} 
  T.~W.~Grimm and R.~Savelli,
  ``Gravitational Instantons and Fluxes from M/F-theory on Calabi-Yau fourfolds,''
  arXiv:1109.3191 [hep-th].



\bibitem{Vafa:1995fj}
  C.~Vafa, E.~Witten,
  ``A One loop test of string duality,''
  Nucl.\ Phys.\  {\bf B447 } (1995)  261-270.
  [hep-th/9505053].

\bibitem{Duff:1995wd}
  M.~J.~Duff, J.~T.~Liu, R.~Minasian,
  ``Eleven-dimensional origin of string-string duality: A One loop test,''
  Nucl.\ Phys.\  {\bf B452 } (1995)  261-282.
  [hep-th/9506126].

\bibitem{Haack:2001jz}
  M.~Haack, J.~Louis,
  ``M theory compactified on Calabi-Yau fourfolds with background flux,''
  Phys.\ Lett.\  {\bf B507 } (2001)  296-304.
  [hep-th/0103068].


\bibitem{Friedman:1997yq}
  R.~Friedman, J.~Morgan, E.~Witten,
  ``Vector bundles and F theory,''
  Commun.\ Math.\ Phys.\  {\bf 187 } (1997)  679-743.
  [hep-th/9701162].

\bibitem{Braun:2009bh}
  A.~P.~Braun, S.~Gerigk, A.~Hebecker, H.~Triendl,
  ``D7-Brane Moduli vs. F-Theory Cycles in Elliptically Fibred Threefolds,''
  Nucl.\ Phys.\  {\bf B836}, 1-36 (2010).
  [arXiv:0912.1596 [hep-th]].


\bibitem{Misner}
  C.~Misner, K.S.~Thorne, J.A.~Wheeler,
  {\it Gravitation}.
  San Francisco: W.H.~Freeman And Company, 1973.



\bibitem{UTTG-07-90}
  P.~Candelas and X.~de la Ossa,
  ``Moduli Space Of Calabi-yau Manifolds,''
  Nucl.\ Phys.\ B\ {\bf 355} (1991) 455.

\bibitem{hep-th/9702155}
  B.~R.~Greene,
  ``String theory on Calabi-Yau manifolds,''
  hep-th/9702155.

\end{thebibliography}
\end{document}